\documentclass[aps,singleocolumn, 12pt]{revtex4}
\usepackage{epsfig,graphics,amssymb,amsmath,subeqnarray,bm}
\usepackage{color}
\usepackage{setspace}

\def\Sp{{\text{Sp}  }}\def\r{{\mathbf{r}}}\def\t{{\mathbf{t}}}
\def\f{{\mathbf{f}}}\def\U{{\mathbf{U}}}
\def\I{{\mathbf{I}}}\def\a{{\mathbf{a}}}\def\b{{\mathbf{b}}}
\def\c{{\mathbf{c}}}\def\v{{\mathbf{v}}}
\def\n{{\mathbf{n}}}\def\c{{\mathbf{c}}}\def\d{{\mathbf{d}}}
\def\e{{\mathbf{e}}}\def\x{{\mathbf{x}}}\def\G{{\mathbf{G}}}
\def\Q{{\mathbf{Q}}}\def\D{{\mathbf{D}}}\def\S{{\mathbf{S}}}
\def\F{{\mathbf{F}}}\def\R{{\mathbf{R}}}
\def\M{{\mathbf{M}}}\def\h{{\mathbf{h}}}

\def\btau{\boldsymbol \tau}

\def\BOmega{{\boldsymbol \Omega}}
\def\Bsigma{{\boldsymbol \sigma}}

\def\bomega{{\boldsymbol \omega}}

\def\bgam{\boldsymbol{\gamma}}
\def\gamd{\dot\bgam}
\def\De{{\rm De}}
\def\n{{\mathbf n}}
\def\dgamd{\stackrel{\triangledown}{\gamd}}

\def\dbtau{\stackrel{\triangledown}{\btau}}

\begin{document}
\singlespacing

\title{Theoretical models in low-Reynolds-number locomotion}

\author{On Shun Pak}
\affiliation{
Department of Mechanical Engineering, Santa Clara University, 500 El Camino Real, Santa Clara, CA, 95053, USA.}

\author{Eric Lauga}
\affiliation{
Department of Applied Mathematics and Theoretical Physics, Centre for Mathematical Sciences,
University of Cambridge, 
Wilberforce Road, 
Cambridge, 
CB3 0WA, 
United Kingdom.}

\begin{abstract}
Chapter to appear in  {\it Low-Reynolds-Number Flows: Fluid-Structure Interactions}, 
Camille Duprat and Howard A. Stone (Eds.), Royal Society of Chemistry Soft Matter Series, 2014.

\end{abstract}

\maketitle

The locomotion of microorganisms in fluids is ubiquitous and plays  an important role in numerous biological processes. Mammalian spermatozoa undergo a long journey to reach the ovum during reproduction; bacteria and algae display coordinated movements to locate better nutrient sources; single-cell eukaryotes such as \textit{Paramecium} self-propel to escape  predators. 

The physics of swimming  governing life under the microscope  is very different from the one we experience in the macroscopic world, due to the absence of inertia (the low Reynolds number regime). For a typical microorganism such as \textit{Escherichia coli} (\textit{E. coli}), with a size $L\approx 10 \ \mu$m and a speed $U\approx 30 \ \mu$m/s, swimming in water (density $\rho \approx 1000 $\ kg/m$^3$ and shear viscosity $\mu \approx 10^{-3} \ $Pa$\cdot$s), the Reynolds number, $Re=\rho U L / \mu$, is on the order of $Re\approx 3 \times 10^{-4}$, and is thus negligible. Unlike humans, fish, insects, or birds, which accomplish swimming and flying by imparting momentum to the fluid, viscous damping is paramount in the microscopic world and  
microorganisms need to adopt different swimming strategies. 
The past decades have seen a tremendous growth in the number of theoretical and experimental studies of cell motility, 
both in the biological and physical communities,  due in part to  advances in observation techniques, leading to discovery of many new physical phenomena in the world of microorganisms, especially in hydrodynamics. Comprehensive reviews focusing on the hydrodynamics of swimming are available \cite{lighthill75,Brennen1977,Fauci2006,Lauga2009}. In this chapter, we present a tutorial on mathematical modelling of swimming at low Reynolds number. Viewing this chapter both as an introduction to the field and as a pedagogical review on some of the fundamental hydrodynamic issues, we purposely   keep only  the essential ingredients of each calculation and readers are referred to the original papers for mathematical rigor and more details.

\section{Swimming at Low Reynolds Number}

\subsection{Kinematic reversibility}\label{sec:KinRev}

{Locomotion in the incompressible flow of Newtonian fluids at zero Reynolds number is governed by the Stokes equations
\begin{subequations} \label{eqn:StokesEqn}
\begin{align}
\nabla p &= \mu \nabla^2 \v, \label{eqn:StokesEqn1}\\ 
\nabla\cdot \v &= 0,  \label{eqn:StokesEqn2}
\end{align}
\end{subequations}
where $p$ and $\v$ are, respectively, the pressure and velocity fields.} The absence of inertia, mathematically manifested by the linearity and time independence of Eq.~(\ref{eqn:StokesEqn}), leads to kinematic reversibility, an important property associated with the motion at zero Reynolds number. In this regime, time appears only as a parameter through the boundary conditions. Consider, for example,  the motion of a solid body. If we reverse time ($t \rightarrow -t$) in the boundary conditions, we reverse the prescribed velocity $\U$ and rotational rate $\BOmega$ of the body, which instantaneously reverses the direction of the velocity and pressure fields ($\v \rightarrow -\v$ and $p \rightarrow -p$)  due to the linearity and time-independence of the Stokes equations. The flow streamlines are not modified but the direction of the flow along these streamlines is reversed. The fluid stresses scale linearly with the pressure and velocity fields, and hence the force $\F$ and torque $\M$ on the body undergo the same reversal, $\F \rightarrow -\F$ and $\M \rightarrow -\M$. 

This property of kinematic reversibility, combined with mirror reflection symmetry, often allows  to deduce useful dynamic properties of a given problem without performing any calculation. We will illustrate using three examples. 
\begin{figure}[t]
\begin{center}
\includegraphics[width=\textwidth]{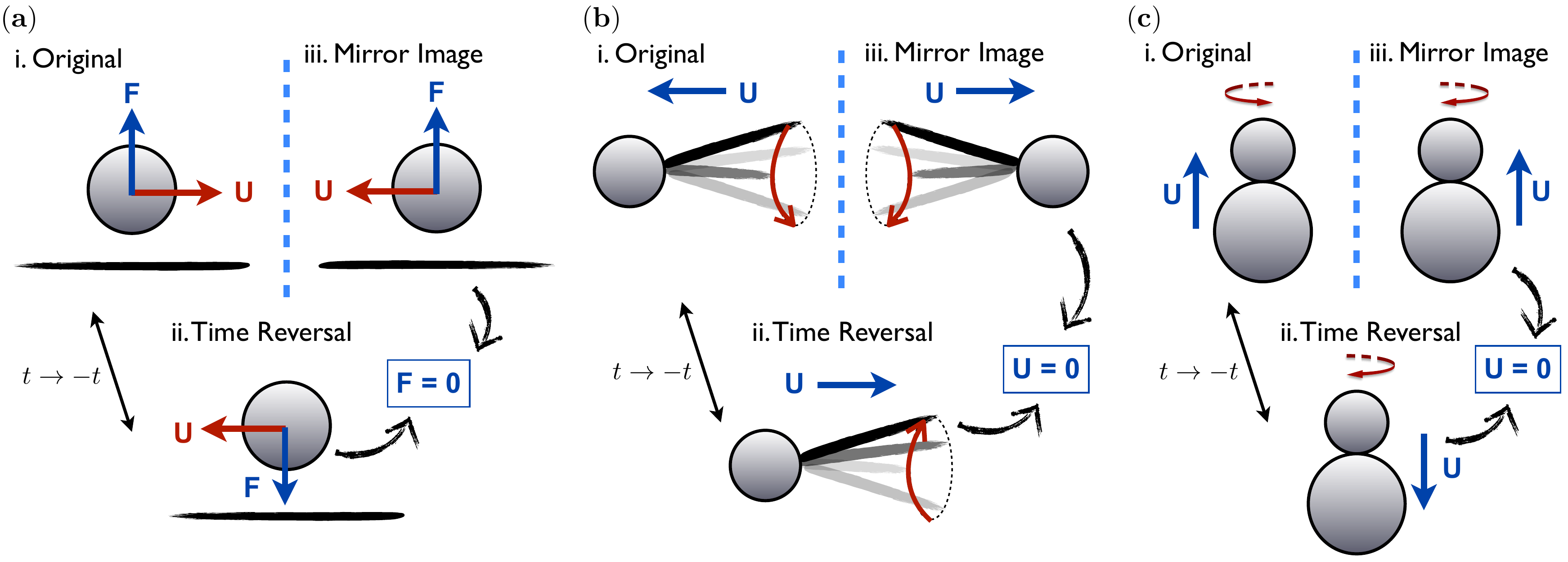}  
\caption{Illustrations of kinematic reversibility. $(\mathbf{a})$ The normal force on a sphere translating parallel to a  wall is zero ($\mathbf{F}=\mathbf{0}$). $(\mathbf{b})$ An organism which rotates its straight rigid tail sweeping out a cone is a non-swimmer ($\mathbf{U} = \mathbf{0}$) \cite{Childress1981}. $(\mathbf{c})$ The rotation of two unequal spheres about their line of centers does not lead to any translation ($\mathbf{U} =\mathbf{0}$).}
\label{fig:SA2}
\end{center}
\end{figure}
The first example considers a translating sphere of velocity $\U$ parallel to an infinite wall Fig.~\ref{fig:SA2}$\mathbf{a}$). The question of interest is whether or not the presence of a wall would induce a force (lift) normal to the wall. If so, is the force acting towards or away from the wall? To answer this using simple physical arguments, we first assume without loss of generality that there is a perpendicular force acting on the sphere away from the wall. We then construct the time-reversed kinematics by kinematic reversibility ($t \rightarrow - t$, $\v \rightarrow - \v$), where both the translational velocity $\mathbf{U}$ and the force $\mathbf{F}$ are reversed. Meanwhile, we can also construct a mirror image of the original solution. Such a mirror image also satisfies the Stokes equations and only the direction of the translation parallel to the wall is reversed in the mirror image solution. By comparing   the time-reversed and mirror image solutions, we observe that despite the same boundary conditions, the two solutions give opposite predictions on the direction of the force $\mathbf{F}$, hence the  force has to be  zero ($\mathbf{F}= \mathbf{0}$), \textit{i.e.}~there is no wall-induced lift.

Similar arguments are also useful to study swimming problems. For example, we can establish that a microorganism rotating a straight and rigid flagellum at an angle (as shown in Fig.~\ref{fig:SA2}$\mathbf{b}$), sweeping out a cone, cannot generate any propulsion \cite{Childress1981}. Again, without loss of generality, we assume the direction of rotation and propulsion speed to be as shown in Fig.~\ref{fig:SA2}$\mathbf{b}$. In the time-reversed solution, both the rotational direction of the flagellum and the swimming direction reverse. However, in the mirror image solution, the swimming direction is unchanged but the rotational direction of the flagellum reverses. Here again there are two solutions with the same boundary conditions but opposite predictions for the swimming direction, and thus  no swimming can  occur  for a rotating rigid and straight filament ($\mathbf{U} =\mathbf{0}$). The same result is true for any shape identical under a mirror image symmetry. Should instead the shape of the flagellum be chiral (\textit{e.g.}~a helix), the mirror-imaged geometry is no longer superposable with that in time-reversed solution, and the  arguments above  no longer hold. In addition, if the flagellum is not rigid (with some flexibility), a chiral deformation can develop as a result of the dynamic balance the bending and viscous forces, leading to propulsion \cite{Pak2011} (see Sec.~\ref{sec:ElasticityBody}). 

The final example considers the rotation of two unequal spheres connected as a rigid body (as shown in Fig.~\ref{fig:SA2}$\mathbf{c}$). Using similar arguments (left as an exercise for the readers), one can conclude that no propulsion can be generated upon imposing a rotation about the line of centers. Of course, this conclusion holds only for Stokes flows (and Newtonian fluids) such that we enjoy the property of kinematic reversibility. Should we remove this property by considering a viscoelastic (non-Newtonian) fluid, this  rigid body rotation does lead to propulsion along the line of centers \cite{Pak2012}. 

These simple physical arguments illustrate different geometrical constraints on low-Reynolds-number locomotion, and hence expose different methods to escape from them. We will also see the use of these  arguments in analyzing flagellar synchronization in Sec.~\ref{sec:Syn}.

\subsection{The scallop theorem}\label{sec:scallop}

\begin{figure}[t]
\begin{center}
\includegraphics[width=0.7\textwidth]{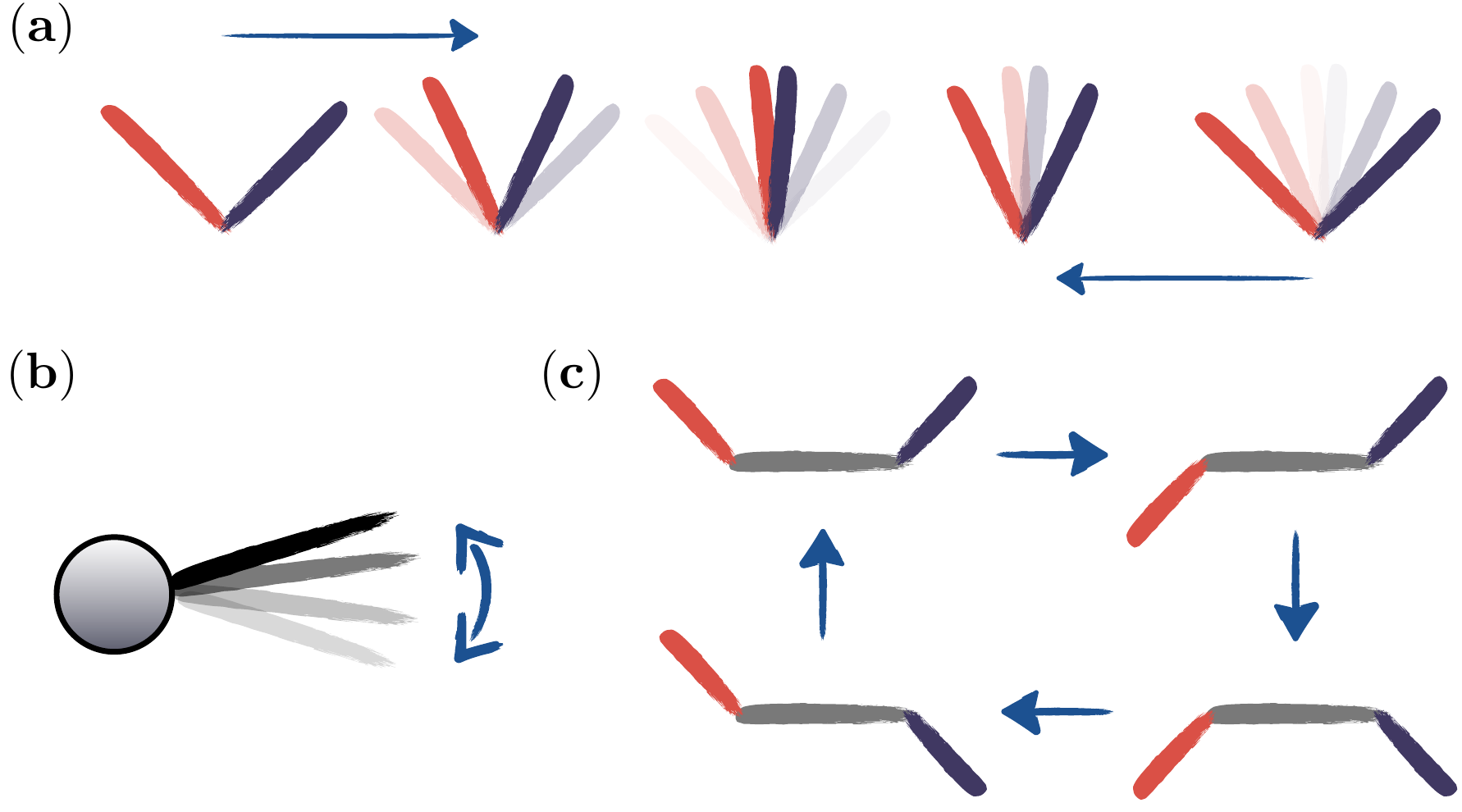}
\caption{$(\mathbf{a})$ A mathematical scallop periodically opening and closing its shell is a  nonswimmer in the Stokesian regime. The sequence of shapes is indistinguishable viewed forward or backward in time (reciprocal motion). $(\mathbf{b})$ An organism flapping its straight rigid tail (reciprocal motion)  cannot swim either. $(\mathbf{c})$ Purcell's three-link swimmer  is an example of a body undergoing non-reciprocal deformation and   swimming \cite{Purcell1977}.}
\label{fig:SA}
\end{center}
\end{figure}

As a direct application of kinematic reversibility, Purcell \cite{Purcell1977} put forward an important theorem for inertialess locomotion called the scallop theorem, stating that any reciprocal motion -- the sequence of shapes of a periodically deforming swimmer  identical under a time-reversal transformation -- cannot generate net propulsion (or fluid transport). A Stokesian scallop opening and closing its shell periodically is an example of reciprocal motion, and thus of a non-swimmer (Fig.~\ref{fig:SA}$\mathbf{a}$). Note that the scallop theorem does not concern the rates at which the forward or backward sequence is performed but only the sequence itself -- modulation of the opening and closing rate is ineffectual. The flapping motion of a rigid flapper, a common propulsion strategy in the macroscopic scale, is another example of reciprocal motion that is useless in the absence of inertia (Fig.~\ref{fig:SA}$\mathbf{b}$). A detailed mathematical proof of the theorem was given by Ishimoto and Yamada \cite{Ishimoto2012}.

Microorganisms and artificial micro-swimmers have thus to escape from the constraints of the scallop theorem in order to generate propulsion \cite{Lauga2011}. Purcell \cite{Purcell1977} proposed a simple mechanism, the three-link swimmer composed of two hinges connecting three rigid links rotating out of phase with each other, which performs the  non-reciprocal motion illustrated  in Fig.~\ref{fig:SA}$\mathbf{c}$ for propulsion.  A hydrodynamic analysis of Purcell's swimmer is given by Becker \textit{et al} \cite{Becker2003}. Other simple mechanisms were proposed and will be reviewed in Sec.~\ref{sec:Synthetic} while the next  section outlines the strategies employed by microorganisms.

\subsection{Propulsion of microorganisms}\label{sec:Microorganisms}

\begin{figure}[t]
\begin{center}
\includegraphics[width=\textwidth]{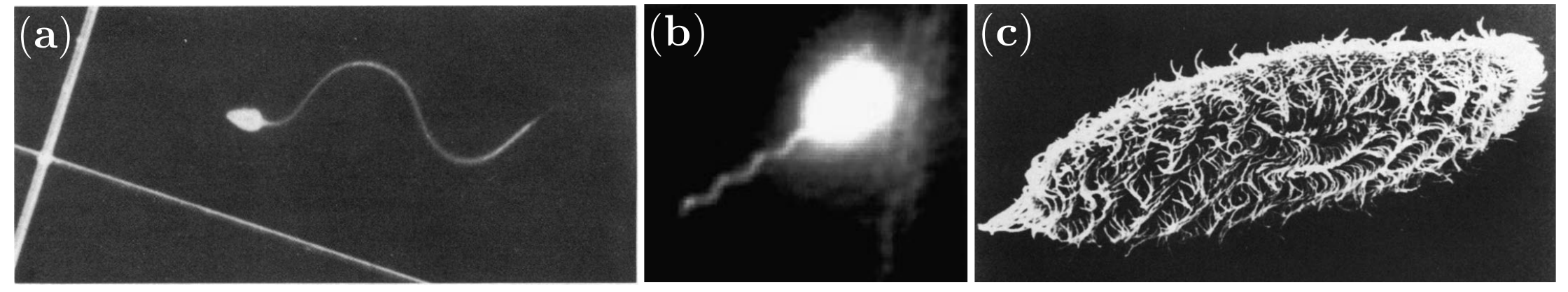}
\caption{$(\mathbf{a})$ A sea-urchin spermatozoon displaying a planar flagellar wave \cite{Rikmenspoel1985}. $(\mathbf{b})$ A bacterium (\textit{Vibrio alginolyticus}) swimming by rotating its helical flagellum, propagating an apparent helical flagellar wave \cite{Magariyamaa2001}. $(\mathbf{c})$ Ciliary motion in \textit{Paramecium} \cite{Hinrichsen1988}. All images were reprinted with permission: $(\mathbf{a})$ from Rikmenspoel and Isles \cite{Rikmenspoel1985}. Copyright \copyright 1985 Elsevier; $(\mathbf{b})$ from Magariyamaa \textit{et al.} \cite{Magariyamaa2001}. Copyright \copyright 2006 John Wiley and Sons; $(\mathbf{c})$ from Hinrichsen and Schultz \cite{Hinrichsen1988} Copyright \copyright 1988 Elsevier.}
\label{fig:microorganisms}
\end{center}
\end{figure}

Microorganisms adopt a variety of propulsion mechanisms\cite{Brennen1977}. Many of them use one or more appendages, called flagella and cilia, for propulsion (Fig.~\ref{fig:microorganisms}). {Eukaryotic flagella and cilia share a common structure, usually consisting of a core axoneme of nine doublet microtubules (long polymeric filaments) arranged around two inner microtubules. Molecular motors (dyneins) between adjacent doublet microtubules generate shear forces, which cause the sliding of the microtubules, leading to bending of the axoneme \cite{Bray2000}.} 

Some eukaryotic spermatozoa (such as sea-urchin spermatozoa) swim by propagating a planar travelling wave similar to a sinusoidal wave along the flagellum (Fig.~\ref{fig:microorganisms}$\mathbf{a}$).  Three-dimensional helical waves are also observed  {in some eukaryotic cells \cite{Baccetti1972} and bacteria \cite{Berg2003} (such as \textit{Escherichia coli}, \textit{Rhodobacter sphaeroides}, and \textit{Vibrio alginolyticus}). While the helical beating pattern observed in eukaryotic cells is again caused by the internal bending of the flagellum, the bacterial flagellum has a different structure and actuation mechanism from that of the eukaryotic flagellum. It is a  rigid and passive helical filament with a hook connecting to a rotary motor embedded in the cell wall, which rotates the flagellum (Fig.~\ref{fig:microorganisms}$\mathbf{b}$).} Some microorganisms such as ciliates (\textit{Opalina} and \textit{Paramecium}) and multicellular colonies of algae (\textit{Volvox}) swim by beating arrays of cilia (short flagella) covering their surfaces (Fig.~\ref{fig:microorganisms}$\mathbf{c}$). The cilia beat in a coordinated manner to produce a wave-like deformation of the envelope covering the cilia tips called a metachronal waves, similar to a  wave made by people standing then sitting in a stadium. %

Despite the diversity of propulsion mechanisms and flagellar waveforms among different cells, a common feature is the presence of wave propagation that breaks the time-reversal symmetry: {when time is reversed, so is the direction of wave propagation, the sequence of shapes is therefore different under time-reversal, making flagellar wave propagation a non-reciprocal deformation.} It should be noted here that breaking the time-reversal symmetry is a necessary but not  sufficient condition for propulsion at low Reynolds number. As a counter example, consider  a configuration formed by two identical flagellated cells arranged head-to-head as mirror-images from each other. Because of the flagellar wave propagation in both cells, the deformation of this mechanism is non-reciprocal. However, since the two cells are arranged head-to-head, their movements oppose each other and clearly they do not swim as a whole by symmetry.

\section{Flagellar Swimming}

In this section, we will introduce the framework for modelling flagellar swimming of microorganisms. Taylor \cite{Taylor1951} pioneered the hydrodynamic analysis of low-Reynolds-number swimming. By modelling the flagellum as a two-dimensional infinite waving sheet, Taylor showed that self-propulsion without inertia was possible as induced by the  propagation of a wave of deformation along the sheet. We revisit below this classical calculation (Sec.~\ref{sec:TaylorSheet}), which reveals many fundamental features of flagellar propulsion.  Next, we will consider another framework for analyzing flagellar swimming --  slender body theory (Sec.~\ref{sec:RFT}). In contrast to Taylor's analysis, slender body theory  allows the consideration of finite-size flagella and more complex geometries. This framework will allow us to revisit the propagation of a planar flagellar wave and compare with the results derived by Taylor. We will then apply it to helical flagellar waves as a model for the swimming of bacteria and other helically-propagating eukaryotic cells.

\subsection{Taylor's swimming sheet}\label{sec:TaylorSheet}

\begin{figure}[t]
\begin{center}
\includegraphics[width=0.5\textwidth]{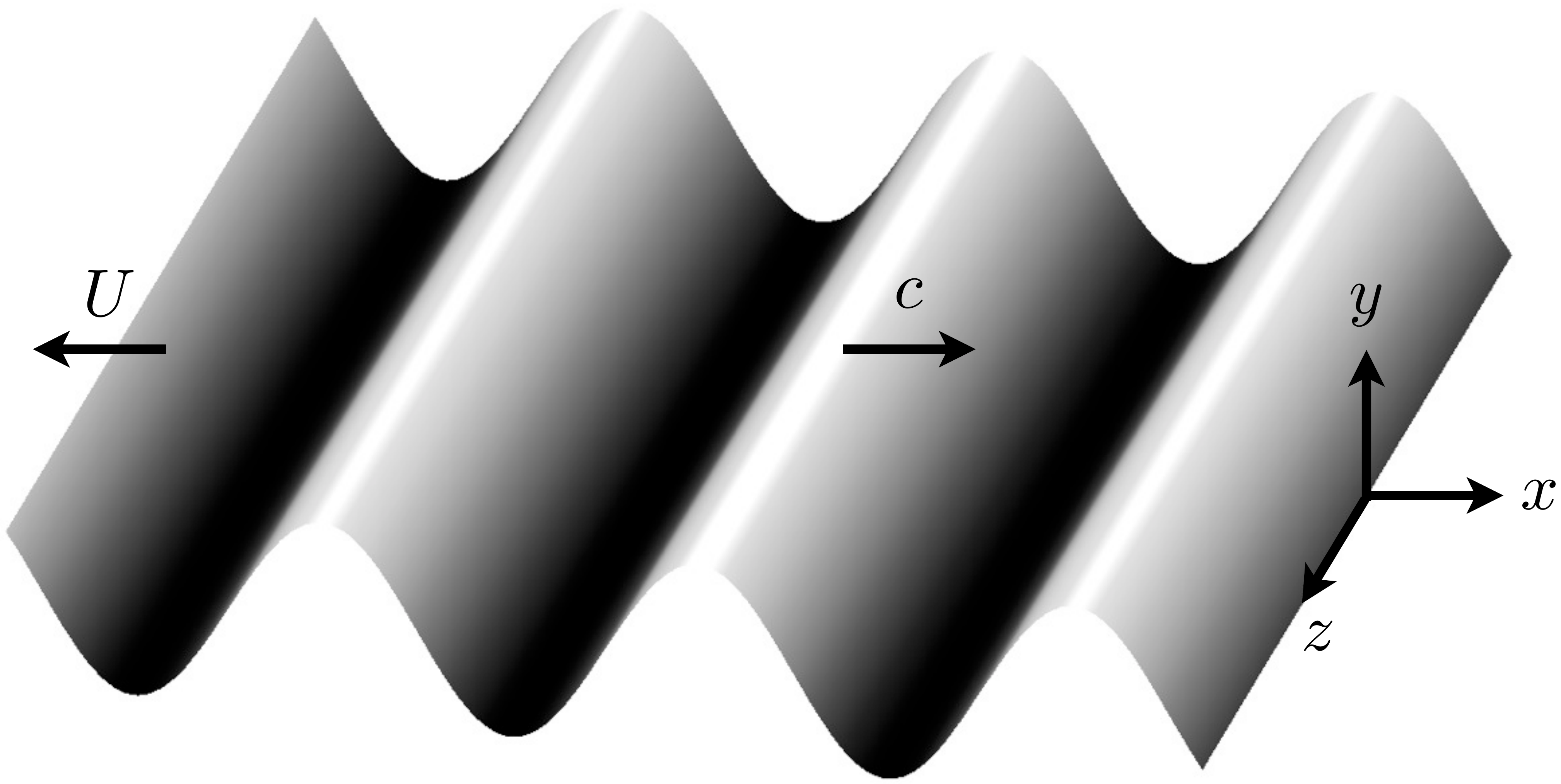}
\caption{Geometrical setup for Taylor's infinite waving sheet. A wave of transverse deformation is propagating  with phase speed $c$ along the $x$-direction inducing swimming at speed $U$ in the opposite direction.}
\label{fig:TaylorSheet}
\end{center}
\end{figure}

\subsubsection{Geometrical setup}
In Taylor's original model, the flagellum is approximated as a two-dimensional infinite waving sheet, which propagates a sinusoidal travelling wave in the positive $x$-direction (see notation in Fig.~\ref{fig:TaylorSheet}). From this waving action, a propulsion speed $U$ may develop, and the focus of this calculation is to compute its value. We assume the sheet swims in the direction opposite to the wave propagation, namely, the negative $x$-direction. Hence we denote the swimming velocity as $- U \e_x$ (Fig.~\ref{fig:TaylorSheet}).

Furthermore, we approach this problem by observing the motion in a frame moving with the (unknown) swimming velocity of the sheet ($- U \e_x$). In this frame, the vertical displacement of the material points is expressed as
\begin{align}
y = a \sin (kx - \omega t),
\end{align}
where the wave has an amplitude $a$, wavenumber $k$, angular frequency $\omega$, and hence phase speed $c = \omega/k$.

\subsubsection{Non-dimensionalization}
We first non-dimensionalize times by $1/\omega$, lengths by $1/k$, and hence speeds by $c$. The dimensionless position of material points on the sheet is therefore given by
\begin{align}
y^* = \epsilon \sin(x^*-t^*),
\end{align}
where $\epsilon = ak$ is the dimensionless wave amplitude compared with the wavelength. From the Stokes equations (Eq.~\ref{eqn:StokesEqn}), we then see that pressure scales as $\mu \omega$, giving the dimensionless Stokes equations as
\begin{subequations}
\begin{align}
\nabla^* p^* &= \nabla^{*2} \v^*, \label{eqn:TaylorStokes}\\
\nabla^* \cdot \v &= 0. \label{eqn:TaylorContinuity}
\end{align}
\end{subequations}
The stars represent dimensionless variables and are dropped hereafter for simplicity. All variables below are therefore dimensionless unless otherwise stated.

\subsubsection{Stream function formulation}
Since the problem is two-dimensional, it will be convenient to define the stream function $\psi(x,y,t)$ for the velocity field $\mathbf{v} = u \e_x+v \e_y$ such that $u = \partial \psi/ \partial y$ and $v = - \partial \psi/ \partial x$ thereby identically satisfying the continuity equation (Eq.~\ref{eqn:TaylorContinuity}). We  eliminate the pressure in Stokes equations by taking the curl of Eq.~(\ref{eqn:TaylorStokes}) (since $\nabla \times \nabla p = \mathbf{0}$), resulting in the equation
\begin{align}
\mathbf{0} = \nabla^2 \bomega,
\end{align}
where $\bomega$ is the vorticity field. For two-dimensional flows, vorticity is related to the stream function as $\bomega = - \nabla^2 \psi \e_z$. The stream function formulation of the Stokes equations is hence given by
\begin{align}
\nabla^4 \psi = 0, \label{eqn:TaylorBiharmonic}
\end{align}
which is the usual biharmonic equation for $\psi$ with analytical solutions  readily available \cite{Selvadurai2000}.

\subsubsection{Boundary conditions}
Since we live in a frame moving at the swimming velocity of the sheet ($- U \e_x$), the velocity field far from the sheet asymptotes to $U \e_x$, opposite to the swimming velocity of the sheet, leading to the boundary conditions
\begin{subequations} \label{eqn:TaylorBCFar}
\begin{align}
u |_{x, y \rightarrow \infty} &= \frac{\partial \psi}{\partial y} \bigg|_{x, y \rightarrow \infty}  =U,   \label{eqn:TaylorBCuFar}\\
v |_{x, y \rightarrow \infty}  &= -\frac{\partial \psi}{\partial x} \bigg|_{x, y \rightarrow \infty} =0. \label{eqn:TaylorBCvFar}
\end{align}
\end{subequations}
On the sheet, the velocity is given by a time derivative of the vertical displacement, leading to boundary conditions
\begin{subequations} \label{eqn:TaylorBC}
\begin{align}
u |_{x, y = \epsilon \sin(x-t)} &= \frac{\partial \psi}{\partial y} \bigg|_{x, y =\epsilon sin(x-t)} = 0, \label{eqn:TaylorBCu} \\
v |_{x, y = \epsilon \sin(x-t)} &= - \frac{\partial \psi}{\partial x} \bigg|_{x, y =\epsilon sin(x-t)} =-\epsilon \cos(x-t). \label{eqn:TaylorBCv}
\end{align}
\end{subequations}
In Eq.~(\ref{eqn:TaylorBC}) we see a typical technical difficulty of Stokesian locomotion: although the Stokes equations are linear, {the geometry of the boundary conditions can lead to nonlinearities. These nonlinearities are  usually addressed numerically or asymptotically.}

\subsubsection{Asymptotic expansions} 
To make analytical progress with the nonlinear boundary conditions (\textit{i.e.~}the fact that it is applied at $y=\epsilon \sin (x-t)$), the asymptotic limit $\epsilon \ll 1$ is considered here. Geometrically, we consider the scenario when the wave amplitude is much smaller than the wavelength. Regular perturbation expansions in powers of $\epsilon$ for both the stream function and the swimming speed in the form
\begin{subequations}
\begin{align}
\psi &= \epsilon \psi_1 + \epsilon^2 \psi_2 + ..., \label{eqn:TaylorPsiAsym}\\
U &= \epsilon U_1 + \epsilon^2 U_2 + ..., 
\end{align}
\end{subequations}
are sought. Substituting these expansions into the boundary conditions in the far field (Eq.~\ref{eqn:TaylorBCFar}), we have
\begin{subequations}\label{eqn:TaylorExpandBegin}
\begin{align}
\epsilon \frac{\partial \psi_1}{\partial y} \bigg|_{x,y \rightarrow \infty} + \epsilon^2 \frac{\partial \psi_2}{\partial y} \bigg|_{x,y \rightarrow \infty}  + ... &= \epsilon U_1 + \epsilon^2 U_2 + ..., \\
\epsilon \frac{\partial \psi_1}{\partial x} \bigg|_{x,y \rightarrow \infty} + \epsilon^2 \frac{\partial \psi_2}{\partial x} \bigg|_{x,y \rightarrow \infty} +...  &= 0.
\end{align}
\end{subequations}
The boundary conditions for $\psi_{1}$ and $\psi_{2}$ in the far field can then be obtained by balancing terms of the same order (Eqs.~\ref{eqn:TaylorBCFirst1}, \ref{eqn:TaylorBCFirst2}, \ref{eqn:TaylorBCSecond1}, and \ref{eqn:TaylorBCSecond2}). 

For the boundary conditions on the sheet (Eq.~\ref{eqn:TaylorBC}), since $\epsilon \ll 1$,  {the derivatives of the velocities on $y = \epsilon \sin(x-t)$ may be Taylor expanded about $y=0$ as: $\partial \psi/\partial y |_{x, y= \epsilon \sin(x-t)} =  \partial \psi/\partial y|_{x, y=0} + \epsilon \sin (x-t) \partial^2 \psi/\partial y^2 |_{x, y=0} + ...$, and $\partial \psi/\partial x |_{x, y= \epsilon \sin(x-t)} = \partial \psi/\partial x|_{x, y=0} + \epsilon \sin (x-t) \partial^2 \psi/\partial y \partial x |_{x, y=0} + ...$. Substituting these expansions together with the expansion of the stream function (Eq.~\ref{eqn:TaylorPsiAsym}) into Eq.~(\ref{eqn:TaylorBC}), the boundary conditions become} 
\begin{subequations}\label{eqn:TaylorExpandEnd}
\begin{align}
\epsilon \frac{\partial \psi_1}{\partial y} \bigg|_{y=0} +\epsilon^2 \frac{\partial \psi_2}{\partial y} \bigg|_{y=0} + \epsilon^2 \sin (x-t) \frac{\partial^2 \psi_1}{\partial y^2} \bigg|_{y=0} +...&= 0, \\
\epsilon \frac{\partial \psi_1}{\partial x} \bigg|_{y=0} +\epsilon^2 \frac{\partial \psi_2}{\partial x} \bigg|_{y=0} + \epsilon^2 \sin (x-t) \frac{\partial^2 \psi_1}{\partial y \partial x} \bigg|_{y=0} + ...&= \epsilon \cos(x-t). 
\end{align}
\end{subequations}
Grouping terms of the same order, we summarize the $O(\epsilon)$ boundary conditions as
\begin{subequations} \label{eqn:TaylorBCFirst}
\begin{align}
\frac{\partial \psi_1}{\partial y} \bigg|_{x, y \rightarrow \infty} &= U_1, \label{eqn:TaylorBCFirst1}\\
\frac{\partial \psi_1}{\partial x}\bigg|_{x, y \rightarrow \infty} &= 0,\label{eqn:TaylorBCFirst2}\\
\frac{\partial \psi_1}{\partial y}\bigg|_{x, y = 0} &= 0,\label{eqn:TaylorBCFirst3}\\
\frac{\partial \psi_1}{\partial x}\bigg|_{x, y = 0} &= \cos(x-t). \label{eqn:TaylorBCFirst4}
\end{align}
\end{subequations}

Similarly, the $O(\epsilon^2)$ boundary conditions are given by
\begin{subequations} \label{eqn:TaylorBCSecond}
\begin{align}
\frac{\partial \psi_2}{\partial y} \bigg|_{x, y \rightarrow \infty} &= U_2, \label{eqn:TaylorBCSecond1}\\
\frac{\partial \psi_2}{\partial x}\bigg|_{x, y \rightarrow \infty} &= 0, \label{eqn:TaylorBCSecond2}\\
\frac{\partial \psi_2}{\partial y}\bigg|_{x, y = 0} &= - \sin (x-t) \frac{\partial^2 \psi_1}{\partial y^2} \bigg|_{x, y=0}, \label{eqn:TaylorBCSecond3}\\
\frac{\partial \psi_2}{\partial x}\bigg|_{x, y = 0} &= - \sin(x-t) \frac{\partial^2 \psi_1}{\partial y \partial x} \bigg|_{x, y=0}. \label{eqn:TaylorBCSecond4}
\end{align}
\end{subequations}
Note that $U_1$ and $U_2$ here are, respectively, the  first-order and second-order swimming speeds, whose values are still  to be determined.

\subsubsection{First-order solution}
Substituting the expansion of $\psi$ into the governing equation (Eq.~\ref{eqn:TaylorBiharmonic}), the $O(\epsilon)$ governing equation is given by
\begin{align}
\nabla^4 \psi_1 =0,
\end{align}
which is a biharmonic equation subject to the $O(\epsilon)$ boundary conditions in Eq.~(\ref{eqn:TaylorBCFirst}). 
This can be solved by a repeated application of the method of separation of variables \cite{Selvadurai2000}. Analyzing the boundary conditions, the  solutions are given by 
\begin{align}
\psi_1 &= A x+ B y + ( C e^{-y} + D y e^{-y}) \sin(x-t), 
\end{align}
where $A, B, C, D$ are constants to be determined from the boundary conditions, Eq.~(\ref{eqn:TaylorBCFirst}). Specifically, Eq.~(\ref{eqn:TaylorBCFirst2}) gives $A = 0$; Eq.~(\ref{eqn:TaylorBCFirst4}) gives $C = 1$; Eq.~(\ref{eqn:TaylorBCFirst3}) gives $ D= C = 1$ and $B=0$; finally, Eq.~(\ref{eqn:TaylorBCFirst1}) gives that $B = U_1 = 0$. Therefore, the first-order solution is given by
\begin{align}
\psi_1  = (1+y) e^{-y} \sin(x-t), \label{eqn:Taylor1stSol}
\end{align}
and swimming does not occur at this order $(U_1 = 0)$. We then proceed to the second-order calculation.

\subsubsection{Second-order solution}
The governing equation at this order is again the biharmonic equation
\begin{align}
\nabla^4 \psi_2 &= 0,
\end{align}
subject to the $O(\epsilon^2)$ boundary conditions in Eq.~(\ref{eqn:TaylorBCSecond}). With the first-order solution $\psi_1$ determined, the boundary conditions now read explicitly as
\begin{subequations}\label{eqn:TaylorBCSecondb}
\begin{align}
\frac{\partial \psi_2}{\partial y} \bigg|_{x, y \rightarrow \infty} &= U_2, \label{eqn:TaylorBCSecond1b}\\
\frac{\partial \psi_2}{\partial x}\bigg|_{x, y \rightarrow \infty} &= 0, \label{eqn:TaylorBCSecond2b}\\
\frac{\partial \psi_2}{\partial y}\bigg|_{x, y = 0} &= \sin^2(x-t) = \frac{1}{2} - \frac{\cos 2(x-t)}{2}, \label{eqn:TaylorBCSecond3b}\\
\frac{\partial \psi_2}{\partial x}\bigg|_{x, y = 0} &= 0. \label{eqn:TaylorBCSecond4b}
\end{align}
\end{subequations}
The  solution in this case is given by
\begin{align}
\psi_2 &= A x+ B y + ( C e^{-2y} + D y e^{-2y}) \cos 2(x-t). \label{eqn:TaylorBiharmonicHomo}
\end{align}
Eq.~(\ref{eqn:TaylorBCSecond2b}) gives $A =0$; Eq.~(\ref{eqn:TaylorBCSecond4b}) gives $C=0$; Eq.~(\ref{eqn:TaylorBCSecond3b}) gives $D = -1/2$ and $B= 1/2$; finally, Eq.~(\ref{eqn:TaylorBCSecond1b}) gives $B=U_2 = 1/2$. Therefore, the second-order solution reads
\begin{align}
\psi_2 = \frac{y}{2} - \frac{ye^{-2y}}{2}\cos 2(x-t),
\end{align}
and the swimming speed is given by
\begin{align}
U_2 = \frac{1}{2} \cdot \label{eqn:TaylorSpeed}
\end{align}
The leading-order swimming speed is hence given by $U = \epsilon^2/2 + o(\epsilon^2)$, with a dimensional form
\begin{align}
U \sim \frac{1}{2} a^2k^2 c . \label{eqn:TaylorSheetSpeed}
\end{align}
Since $U>0$, the swimming sheet propels in the  direction opposite to   the wave propagation. The propulsion speed scales quadratically as $\epsilon^2$, due to a $\epsilon \rightarrow -\epsilon$ symmetry: the swimming speed should be invariant upon reversing the sign of the amplitude, which is  equivalent to a phase shift of $\pi$ (the next term should therefore be of order  $\epsilon^4$). We will see in the next section that results using slender body theory  reproduce similar conclusions.

\subsection{Slender body theory}\label{sec:RFT}

\begin{figure}[t]
\begin{center}
\includegraphics[width=0.95\textwidth]{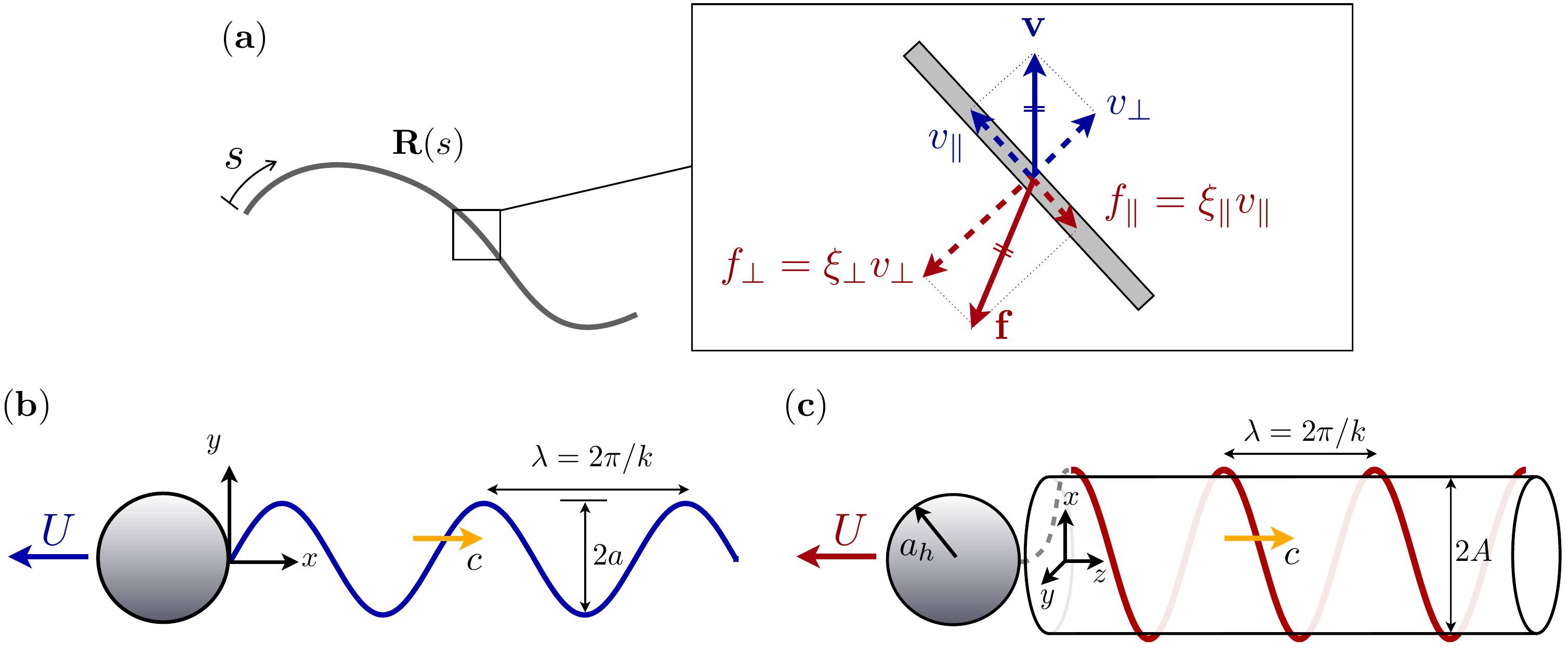}
\caption{Illustration of resistive force theory. $(\mathbf{a})$ Relating the local viscous force $\mathbf{f}$ to the local filament velocity $\mathbf{v}$ relative to the fluid in terms of the resistive coefficients $(\xi_\parallel, \xi_\perp)$. Geometrical setups for $(\mathbf{b})$ a planar sinusoidal flagellar wave and $(\mathbf{c})$ a helical flagellar wave. }
\label{fig:RFT}
\end{center}
\end{figure}

Taylor's infinite waving sheet analysis uncovers interesting features of flagellar swimming. However, flagella are  slender filaments and flagellar wave amplitudes are finite. These aspects may be handled by the use of slender body theory for  Stokes flows, which is the focus of this section. 

The main idea of slender body theory is to represent the flow induced by a deforming flagellum by a line of singular solutions to Stokes flow of appropriate strength (see Lighthill \cite{lighthill75}, Leal\cite{Leal2007}, or Lauga and Powers \cite{Lauga2009}   for an intuitive presentation).  This procedure is  accurate in the limit where  the flagellum is slender, which is the  case for real biological flagella (typical aspect ratio of one to a few hundreds).  Interested readers are referred to detailed theoretical analyses \cite{Cox1970, Batchelor1970, Lighthill1976, Keller1976b, Johnson1979, Johnson1980}. Here we take the results for granted and apply them to model flagellar swimming by planar and helical waves.

Slender body theory relates the force acting on the flagellum  to its distribution of velocity    relative to that of the fluid. The leading-order result of  slender body theory (in an asymptotic expansion in the filament aspect ratio \cite{Cox1970}) is a local theory stating that the viscous force on the body at a point scales linearly, in a tensorial fashion, with the  local velocity of the flagellum relative to the fluid. This local drag model,  called resistive force theory, ignores hydrodynamic interactions between distinct parts of the curved flagellum, and is expected to work well for simple geometries where different parts of the body are sufficiently well separated \cite{Gray1955, Wiggins1998, Yu2006, Kruse2007}.

The local velocity of the flagellum,  $\v$, relative to the background fluid can be decomposed along components  parallel ($v_{\parallel}$) and perpendicular ($v_{\perp}$) to its local tangent, as illustrated in Fig.~\ref{fig:RFT}$\mathbf{a}$. Resistive force theory states that the drag force is anisotropic with two distinct drag coefficients for motion parallel and perpendicular to the local flagellum orientation. Specifically,  the local force density (per unit length), $\f_{\text{vis}}$, acting in the  directions parallel to the local tangent is expressed as $f_\parallel = -\xi_\parallel v_\parallel$ while the perpendicular  one is given by $f_\perp = -\xi_\perp v_\perp$. The resistive coefficients, $\xi_\parallel$ and $\xi_\perp$, have units of viscosity, contain all the leading-order resistive hydrodynamics, and are approximately given by
\begin{align}
\xi_{\parallel} \approx \frac{2 \pi \mu}{\ln(L/r)-1/2}, \ \ \ \ \xi_{\perp} \approx \frac{4\pi \mu}{\ln(L/r)+1/2}, \label{eqn:RFTDragCoeff}
\end{align}
where $L$ and $r$ are, respectively, the length and radius of the filament, and $\mu$ is the dynamic  viscosity of the fluid. These resistive coefficients are valid when the filament is slender ($r \ll L$), and the ratio $\xi_{\perp}/\xi_{\parallel} \rightarrow 2$ as $L/r \rightarrow \infty$.  Clearly we have  $\xi_\perp \neq \xi_\parallel$: it is this property of drag anisotropy that allows the resultant drag force to be in a direction different from that of the deformation velocity,  inducing  net propulsion. This is schematically illustrated in Fig.~\ref{fig:RFT}$\mathbf{a}$, where we show how  drag anisotropy allows a horizontal thrust to be generated by a vertical deformation velocity. The importance of this property will be discussed further below. Note that the resistive coefficients may be refined to give more accurate results depending on the shape of the body as a whole  \cite{Hancock1953,Gray1955,Keller1976b,Cox1970, Lighthill1976}.

Within the context of resistive force theory, the viscous force density acting on the filament can be written in a mathematically compact form as
\begin{align}
\f_{\text{vis}} = - \left[ \xi_{\perp} \n\n + \xi_\parallel \t \t \right] \cdot \v = - \left[ \xi_{\perp} \left(\I - \t\t \right) + \xi_\parallel \t \t \right] \cdot \v = - \left[ \xi_{\perp} \v + \left(\xi_\parallel- \xi_\perp \right) \t \left( \t \cdot \v \right) \right], \label{eqn:RFT}
\end{align}
where $\t$ and $\n$ are, respectively, the local tangent and normal vectors along the filament. The local velocity distribution along the swimmer,  relative to any background flow $\v_0$, is given by $\v = \v_d + \U +\BOmega \times \r -\v_0$, where $\v_{\text{d}}$  is the deformation velocity of the filament,  $\U$ and $\BOmega$ the unknown swimming and rotational velocities of the  swimmer, and $\r(s)$  the instantaneous position vector describing the flagellum shape as a function of the  arclength $s$. Given the shape and deformation of the flagellum, the viscous force distribution along the flagellum can be computed according to Eq.~(\ref{eqn:RFT}), up to the unknown translational and rotational velocities, ($\U, \BOmega$), which will be determined by imposing the overall force-free and torque-free conditions in Stokes flows. To illustrate these steps, we will revisit below the classical use of resistive force theory in studying the propulsion of spermatozoa by Gray and Hancock \cite{Gray1955} and bacteria  by Chwang and Wu \cite{Chwang1971}.

\subsubsection{Planar flagellar waves}
Gray and Hancock \cite{Gray1955} applied  resistive force theory to model the propulsion of sea-urchin spermatozoa and obtained {remarkable agreement} with experimental data. A general analysis for any arbitrary flagellar waveform was given in their original work \cite{Gray1955}. Here we will consider a specific example with kinematics similar to that of Taylor's swimming sheet, as shown in Fig.~\ref{fig:RFT}$\mathbf{b}$. The filament is assumed to undulate  vertically and sinusoidally in a plane with a position vector $y(x,t) = a \sin(kx - \omega t)$, where $a$ and $k = 2\pi/\lambda$ denote the wave amplitude and the wavenumber ($\lambda$: wavelength) respectively. In the spirit of Gray and Hancock, we will further assume the swimmer  propels unidirectionally (in the $x$-direction) without any rotation. Such an assumption is valid when the swimmer is infinite (Taylor's swimming sheet, Sec.~\ref{sec:TaylorSheet}), but is still a good approximation when the number of wavelengths is large. A full three-dimensional study was offered in Keller and Rubinow \cite{Keller1976}. With the unidirectional swimming assumption, the local velocity distribution in a quiescent background flow is $\v= \v_{\text{d}}+\U$ since there is no rotation $\BOmega = \mathbf{0}$. We therefore have a combination of the deformation velocity of the filament, $\v_{\text{d}} =[0, \partial y/\partial t ]= [0,- a \omega \cos (kx-\omega t)]$, with  the unknown swimming velocity, $\U = \left[-U, 0 \right]$. The  minus sign in $\U$ arises because we expect  swimming to occur in the opposite direction of the wave propagation (see Fig.~\ref{fig:RFT}$\mathbf{b}$),  a lesson learned from Taylor's analysis (Sec.~\ref{sec:TaylorSheet}). The local velocity distribution is therefore given by $\v = [-U, \partial y/ \partial t]$. 

In order to further simplify the analysis, we again consider the small-amplitude limit $ak \ll 1$, where the wave amplitude $a$ is small compared with the wavelength $\lambda = 2\pi/k $. The local tangent vector is then given by
\begin{align}
\t = \begin{pmatrix}
1 \\
\frac{\partial y}{\partial x}
\end{pmatrix} \frac{1}{\sqrt{1+ (\partial y/\partial x)^2}} \sim  \begin{pmatrix}
1 \\
\frac{\partial y}{\partial x}
\end{pmatrix},
\end{align}
to leading order, because $1/\sqrt{1+(\partial y/ \partial x)^2} \sim 1 + O(a^2k^2)$. The leading-order  viscous force acting on the whole  filament is then given by
\begin{align}
\F_{\text{vis}} =\int_0^{L} \f_{\text{vis}} ds \sim
 -\int_0^{N \lambda} \left[ \xi_\perp \begin{pmatrix} -U \\ \frac{\partial y}{\partial t} \end{pmatrix} + \left( \xi_\parallel - \xi_\perp \right) \begin{pmatrix} 1 \\ \frac{\partial y}{\partial x} \end{pmatrix} \left(-U+\frac{\partial y}{\partial x} \frac{\partial y}{\partial t} \right) \right]dx, \label{eqn:Similar}
\end{align}
noting that $ds \sim dx$ for the small-amplitude wave assumption, and $N$ is the number of wavelengths. We next consider the total force balance in the $x$-direction in order to compute the swimming speed $U$. The $x$-component of the total viscous force acting on the filament, 
$\F_{\text{vis}} \cdot \e_x$, together with the viscous drag acting on the organism head, assumed to be characterized by  a resistive coefficient $R_h$,  $F_{\text{head}} = -(-R_h U)=R_h U$, should add up to be zero due to the overall force-free condition in Stokes flows
\begin{subequations}
\begin{gather}
\F_{\text{vis}} \cdot \e_x + F_{\text{head}} = 0, \\
\Rightarrow -\int_0^{N \lambda} \left[ -\xi_\parallel U + \left( \xi_\parallel - \xi_\perp \right) \frac{\partial y}{\partial x} \frac{\partial y}{\partial t}  \right] dx + R_h U = 0, \\
\Rightarrow \xi_\parallel U N \lambda + (\xi_\parallel - \xi_\perp) a^2 k \omega \int_0^{\frac{2N\pi}{k}} \cos^2 (kx-\omega t) dx + R_h U = 0, \\
\Rightarrow U= \frac{a^2 k^2 c}{2} \left(\frac{\xi_\perp}{\xi_\parallel} -1 \right)\left(\frac{1}{1+ \frac{R_h}{N \lambda \xi_\parallel}} \right) \cdot \label{eqn:PlanarHead}
\end{gather}
\end{subequations}
Since ${\xi_\perp}>{\xi_\parallel}$, and as expected  from Taylor's analysis, the propulsion speed occurs in the direction opposite to the wave propagation  ($U>0$). We see that the speed decreases monotonically with the size of the organism head -- this it to be contrasted with the behaviour obtained in the case of  helical swimming, see Sec.~\ref{sec:RFThelical}. When a sperm head is absent ($R_h = 0$), Eq.~(\ref{eqn:PlanarHead}) reduces to $U = a^2k^2c (\xi_\perp/\xi_\parallel -1)/2$. If we assume an infinitely slender filament, we have $\xi_\perp /\xi_\parallel \rightarrow 2$, and the propulsion speed becomes $U = a^2k^2c/2$, a result identical to that derived by Taylor for a waving sheet (Eq.~\ref{eqn:TaylorSheetSpeed}) and a waving cylindrical tail \cite{Taylor1951, Taylor1952}. Furthermore, we see the importance of drag anisotropy $\xi_\perp \neq \xi_\parallel$: under isotropic drag, $\xi_\perp = \xi_\parallel$, the propulsion speed vanishes in all cases. Note that beyond this small-amplitude approach,  finite-amplitude calculations can also be computed  \cite{Gray1955, Keller1976}. We now present such a calculation in the case of helical kinematics.

\subsubsection{Helical flagellar waves}\label{sec:RFThelical}

In this section we consider another common flagellar geometry -- a helical structure. Different from planar flagellar waves discussed in the previous section, {helical flagellar waves} are spatially three-dimensional structure (see notation in Fig.~\ref{fig:RFT}$\mathbf{c}$). {Helical flagellar waves are observed in both prokaryotic and eukaryotic cells (see Sec.~\ref{sec:Microorganisms}). Prokaryotic cells propagate these waves by rotating rigid helical flagella, and eukaryotic cells generate them by bending. While their actuation mechanisms are different, the kinematics of the centerline of the flagellum in both cases are exactly the same: a helical wave (Eq.~\ref{HelicalWave} below). A subtle difference lies in the contribution of torques due to spinning about the local centerline of the flagellum, which we ignore  in the analysis below but comment on at the end of the section.} 

Due to the lesson learned in previous sections, we  assume the swimming occurs along  the $z$-direction. 
With the coordinate system shown in Fig.~\ref{fig:RFT}$\mathbf{c}$, a regular helix can be parametrized in terms of the $z$-coordinate with the position vector $\h$
\begin{align}
\h(z) = [A \cos(kz), A \sin(kz), z] .
\end{align}
A regular, right-handed helical wave with angular frequency 
$\omega$ is then given by 
\begin{align}
\r(z, t) = [A \cos(kz - \omega t), A \sin(kz - \omega t), z], \label{HelicalWave}
\end{align}
which propagates in the positive $z$-direction.
It is in general more convenient to parametrize a helix in terms of the arclength $s$ along the helix, which is linearly proportional to the $z$-coordinate as $z = \alpha s$. The constant $\alpha$ is such that the local tangent $\partial \r/\partial s$ is of unit length
\begin{align}
\bigg| \frac{\partial \r}{\partial s} \bigg|^2 = 1 \Rightarrow \alpha = \frac{1}{\sqrt{1+A^2 k^2}} \cdot
\end{align}
Note that we do not assume we have small amplitudes in this section in order to illustrate the steps involved in calculations for finite-amplitude shapes. Under the arclength parametrization, the local unit tangent vector $\t$ is simply given by a derivative with respect to the arclength parameter $s$
\begin{align}
\t = \frac{\partial \r}{\partial s}=  [-Ak \alpha \sin(k \alpha s - \omega t), Ak \alpha \cos(k \alpha s - \omega t), \alpha] .
\end{align}
{Similar to} the analysis in the case of planar undulating waves, the velocity distribution along the swimmer is given by $\v = \v_d + \U +\BOmega \times \r$, where the deformation velocity has the form
\begin{align}
\v_d = \frac{\partial \r}{\partial t}=[ A \omega \sin (k \alpha s -\omega t), - A\omega \cos (k \alpha s - \omega t), 0],
\end{align}
with $\U$ and $\BOmega$ are the unknown translational and rotational swimming velocities respectively. For simplicity, here we follow Chwang and Wu's analysis \cite{Chwang1971} and assume unidirectional swimming and rotation in the $z$-direction. A full three-dimensional analysis can be found in Keller and Rubinow \cite{Keller1976}. With this assumption, we therefore have  $\U = [0 ,0, -U]$ and $\BOmega = [0, 0, \Omega]$ where again we have assumed that swimming occurs in the opposite direction as wave propagation. Note that the propagation of the helical wave above can be seen as a rotation about the negative $z$-direction (Fig.~\ref{fig:RFT}$\mathbf{c}$), and we expect the rotational velocity induced by hydrodynamics to be in the opposite direction (the positive $z$-direction) to satisfy the torque-free condition. The total velocity distribution hence reads
\begin{align}
\v = [A (\omega - \Omega) \sin (k \alpha s - \omega t), -A(\omega - \Omega) \cos (k \alpha s - \omega t), -U] .
\end{align}
The overall viscous force acting along the helical flagellum is given by
\begin{align}
\F_{\text{vis}} = \int_0^L \f_{\text{vis}} ds &= -\int_0^{\frac{2N\pi}{k \alpha}} \left[ \xi_{\perp} \v + \left(\xi_\parallel- \xi_\perp \right) \t \left( \t \cdot \v \right) \right] ds,
\end{align}
with its $z$-component equal to
\begin{align}
\F_{\text{vis}} \cdot \e_z= \frac{2N \pi (\xi_\parallel+\xi_\perp A^2K^2) }{k \sqrt{1+ A^2 k^2}} U + \frac{2 N \pi (\xi_\perp -\xi_\parallel)A^2}{\sqrt{1+A^2 k^2}} \Omega - \frac{2 N \pi (\xi_\perp-\xi_\parallel)A^2}{\sqrt{1+A^2k^2}} \omega .
\end{align}
Similarly, the overall viscous torque about the origin reads
\begin{align}
\M_{\text{vis}} &= \int_0^L \r \times \f_{\text{vis}} ds,
\end{align}
and has a $z$-component given by
\begin{align}
\M_{\text{vis}}  \cdot \e_z = - \frac{2 N \pi (\xi_\perp - \xi_\parallel) A^2}{\sqrt{1+A^2 k^2}} U - \frac{2 N\pi (\xi_\perp + \xi_\parallel A^2 k^2) A^2}{k \sqrt{1+A^2 k^2}} \Omega + \frac{2 N\pi (\xi_\perp + \xi_\parallel A^2 k^2) A^2}{k \sqrt{1+A^2 k^2}} \omega .
\end{align}

We now consider the overall force and torque balances in the $z$-direction. The $z$-component of the overall viscous force together with the viscous drag on a spherical sperm head of radius $a_h$, $F_{\text{head}} = - (-R_h U)=6 \pi \eta a_h U$,  should sum up to be zero due to the overall force-free condition. Similarly, the $z$-component of the overall viscous torque together with the viscous torque on the head,  $M_{\text{head}} = - (R^T_h \Omega)= -8 \pi \eta a^3_h \Omega$, should sum up to be zero due to the  torque-free condition. Mathematically, we thus have
\begin{subequations}
\begin{gather}
\F_{\text{vis}} \cdot \e_z + F_{\text{head}} = 0, \\
\M_{\text{vis}} \cdot \e_z + M_{\text{head}} = 0,
\end{gather}
\end{subequations}
which leads to the system
\begin{subequations}
\begin{gather}
\frac{2N \pi (\xi_\parallel+\xi_\perp A^2K^2) }{k \sqrt{1+ A^2 k^2}} U + \frac{2 N \pi (\xi_\perp -\xi_\parallel)A^2}{\sqrt{1+A^2 k^2}} \Omega - \frac{2 N \pi (\xi_\perp-\xi_\parallel)A^2}{\sqrt{1+A^2k^2}} \omega + 6\pi \eta a_h U = 0,\\
- \frac{2 N \pi (\xi_\perp - \xi_\parallel) A^2}{\sqrt{1+A^2 k^2}} U - \frac{2 N\pi (\xi_\perp + \xi_\parallel A^2 k^2) A^2}{k \sqrt{1+A^2 k^2}} \Omega + \frac{2 N\pi (\xi_\perp + \xi_\parallel A^2 k^2) A^2}{k \sqrt{1+A^2 k^2}} \omega - 8\pi \eta a^3_h \Omega = 0.
\end{gather}
\end{subequations}
 Solving these equations  for  $U$ and $\Omega$ yields the solution
 \begin{subequations}
\begin{align}
\frac{U}{c} &= \frac{4 N a_h^{*3} (\gamma-1) A^{*2} \sqrt{1+A^{*2}}}{\mathcal{C}} \label{eqn:RFTHelicalU}, \\
\frac{\Omega}{\omega} &= \frac{NA^{*2} \left[ N \xi^*_\perp (1+A^{*2})^2 + 3 a_h^* \sqrt{1+A^{*2}} \left( \gamma + A^{*2} \right)  \right] }{\mathcal{C}}\label{eqn:RFTHelicalO},
\end{align}
\end{subequations}
where
\begin{subequations}
\begin{gather}
A^* = A k, \ \ a_h^* = a_h k, \ \ \gamma = \xi_\perp/ \xi_\parallel, \ \  \xi_\perp^*= \xi_\perp/\eta, \ \ \xi_\parallel^*= \xi_\parallel/\eta, \\
\mathcal{C} = N^2 \xi_\perp^* A^{*2} (1+A^{*2})^2 +Na_h^* \sqrt{1+A^{*2}} \left[ 3 A^{*2} (\gamma + A^{*2}) +4 a_h^{*2} (1+\gamma A^{*2}) \right] \notag \\
+ 12 a_h^{*4} (1+A^{*2})/\xi_\parallel^* .
\end{gather}
\end{subequations}

\begin{figure}[t]
\begin{center}
\includegraphics[width=0.45\textwidth]{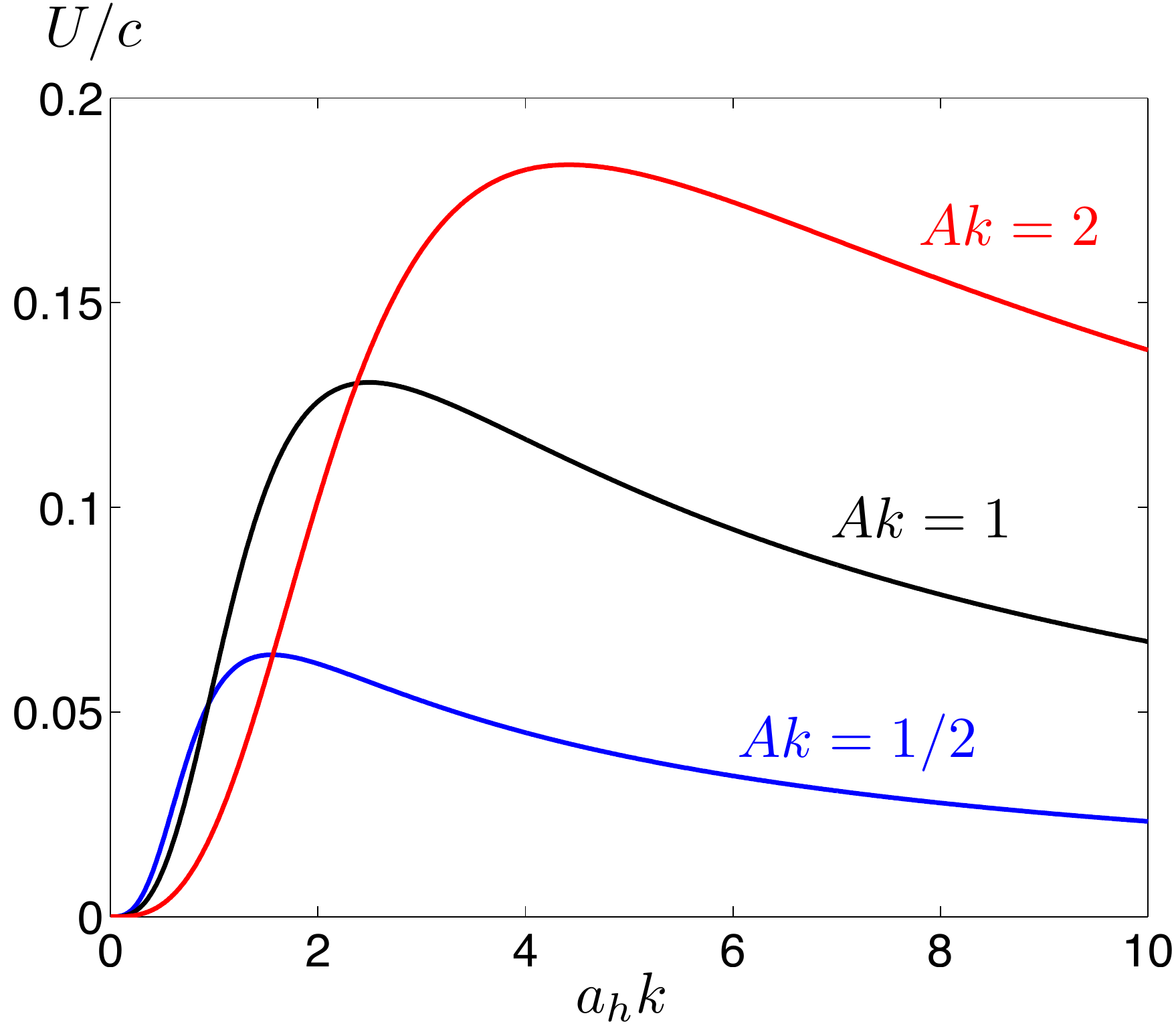}
\caption{Helical flagellar swimming. Non-monotonic variation of the dimensionless swimming speed, $U/c$, as a function of the dimensionless head radius, $a_h k$, with $N=5$ and an aspect ratio of the filament $L/r =500$.}
\label{fig:HelicalHead}
\end{center}
\end{figure}

A few remarks should be made about these results. First, and similarly to the case of planar flagellar waves,  we see in Eq.~(\ref{eqn:RFTHelicalU}) that propulsion by a helical flagellum also relies on drag anisotropy ($U/c = 0$ when $\gamma = \xi_\perp/\xi_\parallel =1$). In contrast, while increasing the size of the cell body monotonically decreases the swimming speed for planar waves (Eq.~\ref{eqn:PlanarHead}), it is interesting to notice from Eq.~(\ref{eqn:RFTHelicalU}) that, in the case of helical propulsion, the swimming speed vanishes if a cell body is absent ($a_h^* = 0$). A cell body is therefore necessary for helical swimming \cite{Keller1976}. 
This surprising result arises because of the balance of moments. Without a cell body, the wave propagation is equivalent to a rotating rigid helix, which exerts a net torque on the fluid. In order to satisfy the zero net-torque condition, the fluid forces cause the rotating helix to counter-rotate at exactly the same rate, resulting in no apparent rotation and hence zero propulsion velocity. Swimming can only occur when a cell body is present so that the fluid forces induce a counter-rotation of the  helix   at a smaller  rate due to the additional contribution of the cell body to the torque balance. On the other hand, the viscous drag acting on the cell body  increases with its size, hampering the swimming performance. We therefore expect a non-monotonic variation of swimming speed with the cell head size, and hence an optimal, intermediate size of sperm head for the greatest swimming speed. This is illustrated in Fig.~\ref{fig:HelicalHead} where we plot the dependence of the dimensionless swimming speed, $U/c$, with the dimensionless sperm head radius, $a_hk$, for a helical flagellum with $N=5$ and an aspect ratio of the filament $L/r =500$.

Note that for a right-handed helix, a helical wave propagation in the positive $z$-direction (considered in this example) can be seen as a rotation of the helix about the negative $z$-direction (Fig.~\ref{fig:HelicalHead}$\mathbf{c}$).
We further find, by inspecting Eq.~(\ref{eqn:RFTHelicalO}),  that the hydrodynamically induced rotational rate of the cell body $\Omega$ is positive, meaning that the induced rotation occurs about the positive $z$-direction, which is opposite to  that of the helical wave. The apparent rotation of the flagellum is a competition between the two. As the ratio $\Omega / \omega $ can be shown to be smaller than unity \cite{Chwang1971}, the apparent rotational rate of the helical flagellum,  reduced to $\omega - \Omega$, still occurs in the negative $z$-direction.  As a result, the head and flagellum of a helical swimmer such as \textit{E. coli} would be experimentally observed to rotate in opposite directions.

{At this point the  subtle differences between a rotating prokaryotic helical flagellum and an eukaryotic flagellum propagating a bending helical wave should be noted. Bacteria propagate apparent helical waves due to the rotation of  their  rigid helical flagella. In this case, in the absence of a head, the fluid forces induce a (passive) rigid body counter-rotation of the flagellum at exactly the same magnitude but in opposite direction to satisfy the overall torque-free condition. This results in zero apparent rotation and leads to strictly zero propulsion. On the other hand, for an eukaryotic flagellum propagating a bending helical wave, the torque-free condition cannot be satisfied by simply counter-rotating the flagellum at the same rotational rate, because the torque due to the rotation about the local centerline of the flagellum is absent in the active propagation of the bending helical wave. The overall torque-free condition in this case is satisfied with a non-zero apparent rotational rate. Therefore, theoretically an eukaryotic cell can swim without a sperm head by propagating a bending helical wave. However, because the flagellum is very thin compared to the helical pitch or radius, the rotation and resulting swimming speed are very small and  typically always neglected.  The contribution from this spinning torque  was discussed by Chwang and Wu \cite{Chwang1971} . }

\section{Ciliary Propulsion}\label{sec:squirmer}

In this section, we move to  another mode of locomotion by microorganisms, namely ciliary propulsion. Certain ciliates (\textit{e.g.~Opalina}) and colonies of flagellates (\textit{e.g.~Volvox})  swim by beating arrays of cilia (short flagella) covering their surfaces. The tips of cilia are closely packed during beating and  form a continuously deforming surface refereed to as an ``envelope''. Assuming for simplicity a spherical geometry, Lighthill \cite{Lighthill1952} first considered this envelope model, an analysis which was later completed by Blake \cite{Blake1971b}. To leading order, the surface distortion may be approximated by small-amplitude radial and tangential motion on the spherical surface -- squirming motion.  In recent years, such a squirmer model has been adopted widely to study hydrodynamic interactions of swimmers \cite{Ishikawa2006, Drescher2009}, suspension dynamics \cite{Ishikawa2007, Ishikawa2007b}, nutrient transport and uptake by microorganisms \cite{Magar2003, Magar2005, Michelin2011}, and optimal locomotion \cite{Michelin2010}. 

Several common assumptions have been made in the literature in order to simplify the mathematical analysis. First, the radial motion of the envelope is usually neglected and the squirmer propels only by tangential motion on the surface. Second, the tangential squirming motion is assumed to be axisymmetric. Finally, the tangential velocity profile prescribed on the sphere is assumed steady in time. While the squirming motion of ciliates is clearly time-dependent, it is common to consider an average motion over many beat cycles, so that a time-independent tangential squirming motion can be prescribed on the spherical surface. We will follow these assumptions below to present the derivation for  squirmer dynamics.

\subsection{Lamb's general solution}\label{sec:Lamb}

There are different manners to derive the propulsion speed and velocity field of a swimming squirmer given a prescribed tangential velocity  on the squirmer's surface. We present here a formulation  taking advantage of the general solution for Stokes flows outlined by Lamb \cite{Lamb1932} ideally suited for problems with spherical or nearly spherical \cite{Brenner1964} geometries. This is different from the original analysis by Lighthill \cite{Lighthill1952} and Blake \cite{Blake1971b}, and the interested reader  is referred to these studies for an alternative method (see also below for the link between both approaches).  A detailed description of Lamb's general solution and its applications can be found in Happel and Brenner \cite{Happel1973} and Kim and Karrila \cite{Kim1991}. 

\begin{figure}[t]
\begin{center}
\includegraphics[width=0.30\textwidth]{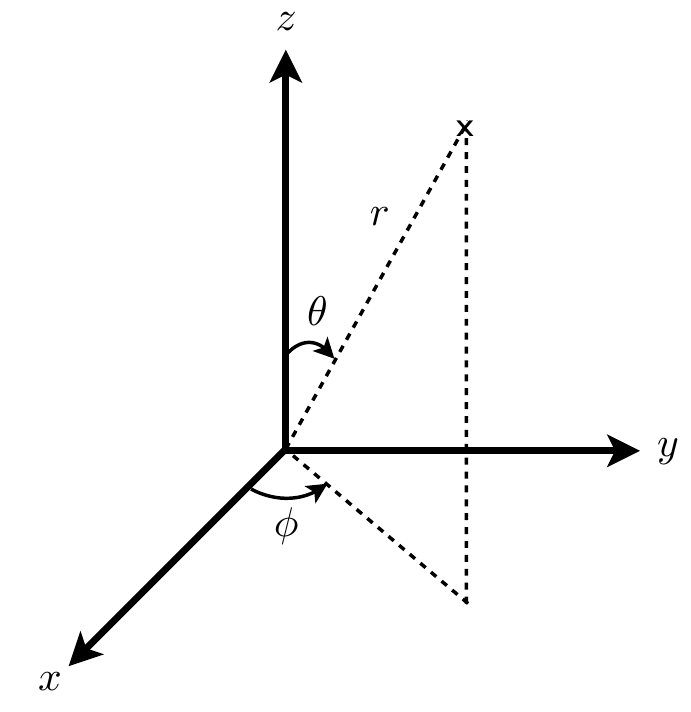}
\caption{Spherical coordinate system for the study of a spherical squirmer of radius $r=a$.}
\label{fig:Squirmer}
\end{center}
\end{figure}

Assuming that the problem is axisymmetric and the flow field decays at infinity, Lamb's general solution in spherical coordinates (Fig.~\ref{fig:Squirmer}) reads  
\begin{align}
\v (r, \theta) = \sum_{n=1}^{\infty} \left[- \frac{(n-2) r^2 \nabla p_{-n-1}}{2 \mu n (2n-1)}+ \frac{(n+1) \r p_{-n-1}}{\mu n(2n-1)}\right] + \sum_{n=1}^{\infty}  \nabla \Phi_{-n-1},
\end{align}
where the pressure field $p$ and the function $\Phi$ are both harmonic functions with
\begin{subequations}
\begin{align}
p_{-n-1} &= r^{-n-1}  P_n(\eta) A_n , \label{eqn:pn} \\
\Phi_{-n-1} &= r^{-n-1} P_n(\eta) B_n, \label{eqn:phin}
\end{align}
\end{subequations}
 $P_n(\eta \equiv  \cos\theta)$ is the $n$-th degree Legendre polynomial, and $A_n$ and $B_n$ arbitrary constants. The total  pressure field is given by $p = \sum_{n=1}^\infty p_{-n-1}$. After performing the differential operations in spherical coordinates, Lamb's general solution  for axisymmetric Stokes flows, $\v = v_r \e_r + v_\theta \e_\theta$,  has the explicit form
\begin{align}
\v(r,\theta)= \sum_{n=1}^{\infty} \frac{(n+1) P_n}{2 (2n-1) r^{n+2}}  \left[ \frac{A_{n} r^2}{\mu} - 2(2n-1)B_{n}  \right] \e_r+ \sum_{n=1}^{\infty}  \frac{\sin \theta P_n^{'}}{2 r^n} \left[ \frac{(n-2)A_n}{n(2n-1)\mu} - \frac{2 B_n}{r^2} \right] \e_\theta, \label{eqn:vgeneral}
\end{align}
where the prime denotes differentiation with respect to $\eta$. The values of  $A_n$ and $B_n$ are to be  determined using  the boundary conditions. 

We now make the assumption that a squirmer swims by a purely tangential velocity profile on the surface. The value of the radial velocity on the surface of a spherical squirmer of radius $a$ is given by 
\begin{equation}
v_r(r=a,\theta) = \sum_{n=1}^{\infty} \frac{(n+1) P_n}{2 (2n-1)  a^{n+2}}  \left[ \frac{A_{n} a^2}{\mu} - 2  (2n-1)B_{n} \right] .
\end{equation}
The condition of purely tangential squirming motion is $v_r(r=a, \theta)=0$, leading therefore  to
\begin{align}
A_{n} &= \frac{2 (2n-1) \mu}{a^2} B_{n} . \label{eqn:PurelyTangential}
\end{align}
Substituting this condition into Eq.~(\ref{eqn:vgeneral}), the velocity field due to purely tangential squirming motion becomes
\begin{align}
\v (r, \theta)  = \sum_{n=1}^\infty  \frac{(n+1) P_n}{  r^{n+2}} \left( \frac{r^2}{a^2}-1 \right) B_n \e_r + \sum_{n=1}^\infty  \frac{\sin \theta P^{'}_n}{r^n}    \left( \frac{n-2}{n a^2}-\frac{1}{r^2} \right) B_n \e_\theta, \label{eqn:SquirmerPumping}
\end{align}
with the boundary values 
\begin{align}
\v (r=a, \theta) =- \sum_{n=1}^\infty  \frac{2\sin \theta P^{'}_n}{n a^{n+2}}B_n \e_\theta . \label{eq:SqBCPak}
\end{align}

Given the assumed axisymmetry, the swimming velocity, $\mathbf{U}=[0, 0, U]$, will be directed along the $z$-direction. When studying the swimming of a squirmer, is it then convenient to  consider the problem in two separate steps. In the first step, we consider the above solution (Eq.~\ref{eqn:SquirmerPumping}) and boundary conditions (Eq.~\ref{eq:SqBCPak}) as the case when the squirmer is fixed in space, held by an external force, and not allowed to move -- this is usually referred to as the pumping problem. In the second step, we allow the squirmer to move freely and compute the induced swimming velocity ($U$) given the boundary actuation, Eq.~(\ref{eq:SqBCPak}), in the pumping problem. This allows the separation of the boundary values due to the squirming actuation  from that due to  the induced swimming. 

To obtain the overall flow field  of a swimming squirmer, $\bar{\v}$, we thus superimpose the solution of the pumping problem, $\v$, with  the flow field, $\v_T$,  due to a rigid sphere translating  at the induced swimming speed $U$ and given by 
\begin{align}
\v_T = U \cos \theta \left( \frac{3a}{2r} - \frac{a^3}{2 r^3} \right) \e_r -U \sin \theta \left( \frac{3a}{4r} + \frac{a^3}{4r^3}  \right) \e_\theta .
\end{align}
The overall flow field due to a swimming squirmer, $\bar{\v} = \bar{v}_{r} \e_r + \bar{v}_\theta \e_\theta$, is finally given by
\begin{subequations}\label{eqn:SquirmerSwimmingF}
\begin{align}
\bar{v}_{r} (r, \theta)  &= U \cos \theta \left( \frac{3a}{2r} - \frac{a^3}{2 r^3} \right)+ \sum_{n=1}^\infty  \frac{(n+1) P_n}{  r^{n+2}} \left( \frac{r^2}{a^2}-1 \right) B_n, \label{eqn:SquirmerSwimmingR}\\
\bar{v}_{\theta} (r, \theta) &= -U \sin \theta \left( \frac{3a}{4r} + \frac{a^3}{4r^3}  \right)+\sum_{n=1}^\infty  \frac{\sin \theta P^{'}_n}{r^n}    \left( \frac{n-2}{n a^2}-\frac{1}{r^2} \right) B_n , \label{eqn:SquirmerSwimmingTheta}
\end{align}
\end{subequations}
with the value of $U$ still to be determined.

\subsubsection{Swimming velocity}

We now compute  the   swimming speed, $U$, as a function of the imposed coefficients $B_n$ from the surface squirming motion  (Eq.~\ref{eq:SqBCPak}). We calculate the total hydrodynamic force acting on the squirmer and solve for  the value of $U$ which  enforces the overall force-free condition. The  hydrodynamic force on the squirmer has two components: the net force acting on the sphere due to the surface squirming motion  in the pumping problem ($\F_{\text{squirm}}$) and the drag force acting on the squirmer due to the induced swimming motion ($\F_{\text{swim}}$). By axisymmetry, both forces only act in the $z$-direction. Using Lamb's general solution, the net force in the pumping problem can be computed easily according to the formula\cite{Happel1973, Kim1991} $\F_{\text{squirm}} = - 4\pi \nabla \left( r^3 p_{-2} \right)$. The force due to the swimming motion is simply the Stokes drag $\F_{\text{swim}}= -6\pi\eta a U$. The overall force-free condition reads therefore
\begin{subequations}
\begin{align}
\F_{\text{squirm}} + \F_{\text{swim}} &= 0, \\
\Rightarrow -6 \pi \eta a U \e_z - 4\pi \nabla \left[ r P_1(\mu) A_1 \right] &= 0, \\
\Rightarrow U = -\frac{2 A_1}{3 \eta a} &= - \frac{4 B_1}{3 a^3} , \label{eq:SquirmU}
\end{align}
\end{subequations}
in which we have employed the no-radial surface velocity condition, Eq.~(\ref{eqn:PurelyTangential}), to relate $A_1$ to $B_1$. Substituting the calculated swimming velocity, Eq.~(\ref{eq:SquirmU}), into Eq.~(\ref{eqn:SquirmerSwimmingF}), we find the  flow field due to a swimming squirmer in the laboratory frame is given by
\begin{subequations}\label{ss}
\begin{align}
\bar{v}_{r}(r, \theta) &= - \frac{4 \cos \theta}{3 r^3} B_1 +  \sum_{n=2}^\infty  \frac{(n+1) P_n}{  r^{n+2}} \left( \frac{r^2}{a^2}-1 \right) B_n,\label{s1} \\
\bar{v}_{\theta} (r, \theta)&= - \frac{2 \sin \theta}{3 r^3} B_1 + \sum_{n=2}^\infty  \frac{\sin \theta P^{'}_n}{r^n}    \left( \frac{n-2}{n a^2}-\frac{1}{r^2} \right) B_n .\label{s2}
\end{align}
\end{subequations}
Note that the result in Eq.~(\ref{eq:SquirmU}) could alternatively been found by requiring the value of $U$ to cancel the $1/r$ terms in Eq.~(\ref{eqn:SquirmerSwimmingR}) or (\ref{eqn:SquirmerSwimmingTheta}) as they are  the signature of a net force on the sphere.

\subsubsection{Structure of the flow field}
\begin{figure}[t]
\begin{center}
\includegraphics[width=1\textwidth]{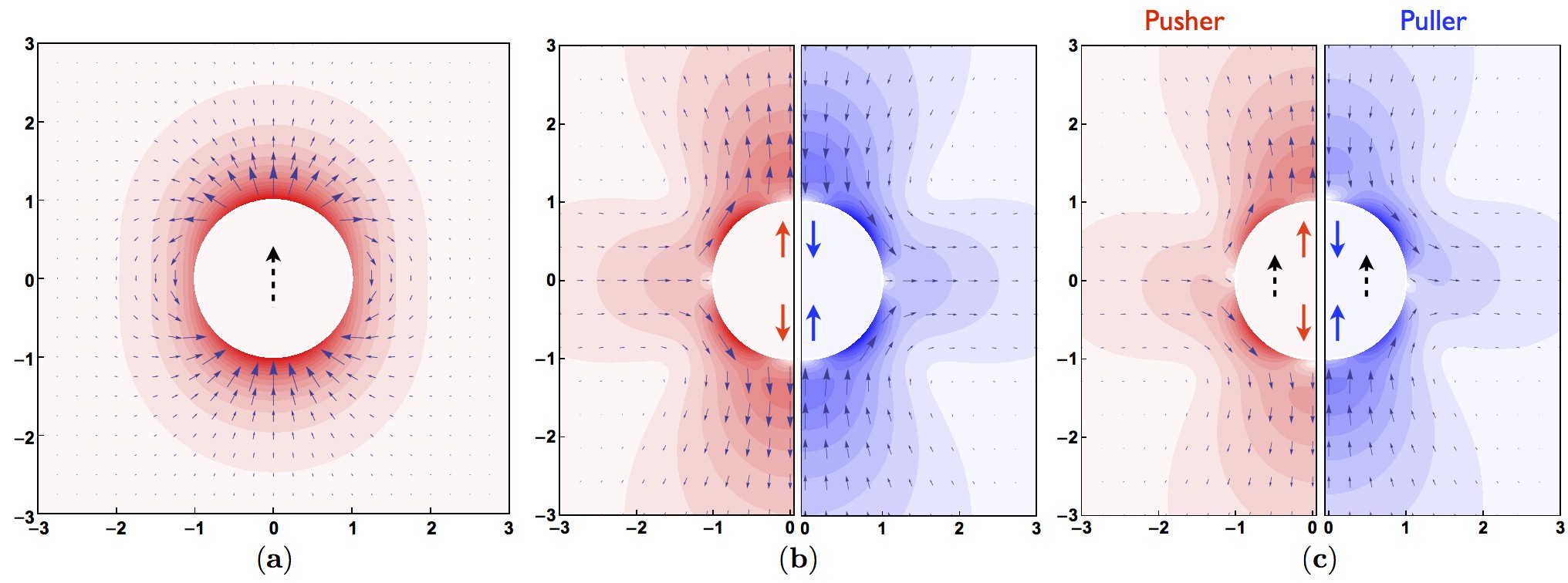}
\caption{Flow fields in squirming motion shown in the laboratory frame. $(\mathbf{a})$ Velocity field due solely to the  swimming mode with $B_1=-1$ (swimming upward), corresponding to a (potential) source dipole. $(\mathbf{b})$ Velocity field due solely to the $B_2$ mode with $B_2=1$ (left panel) and $B_2=-1$ (right panel). Both correspond to Stokes dipoles (stresslet) in the far field and show no swimming; only half of the domain is shown in each case due to axisymmetry. $(\mathbf{c})$ Total velocity field around two upward swimming squirmers (both with $B_1=-1$). Left panel: Pusher with $B_2=4$; Right panel:  Puller with $B_2=-4$. The black dotted arrow indicates the swimming direction of the squirmer, and the two pairs of solid arrows indicate the configuration of the Stokes dipoles.}
\label{fig:squirmer}
\end{center}
\end{figure}

It is interesting to notice that among all modes of boundary actuation, $B_n$'s, in Eq.~(\ref{eq:SqBCPak}), only the $B_1$ mode contributes to swimming (Eq.~\ref{eq:SquirmU}). This swimming mode generates a flow field decaying as $1/r^3$
\begin{align}
\bar{\v}_{B_{1}} = -\frac{2}{3r^3} \left(2\cos\theta \e_r + \sin\theta \e_\theta \right) B_1, \label{eq:B1}
\end{align}
which physically corresponds to a (potential) source dipole. The flow field due to that mode, for $B_1=-1$ , is illustrated in  Fig.~\ref{fig:squirmer}$\mathbf{a}$.

From Eq.~(\ref{ss}), we see that the slowest spatially decaying flow field however is due to the $B_2$ mode, and is given by
\begin{align}
\bar{\v}_{B_{2}} = \frac{3 B_{2}}{4 a^2r^2} (1+3\cos2\theta) \e_r - \frac{3 B_{2}}{4r^4} \left[ (1+3\cos2\theta)\e_r + 2\sin2\theta \e_\theta   \right] . \label{eq:B2}
\end{align}
We plot in Fig.~\ref{fig:squirmer}$\mathbf{b}$ the velocity fields due to positive (left panel) and negative (right panel) $B_2$ modes. {Note that in Fig.~\ref{fig:squirmer}$\mathbf{b}$, we plot only the flow induced by the $B_2$ mode in order to illustrate the features of this particular mode.} This $B_2$ mode leads to a flow field decaying as $1/r^2$, which is purely radial and physically corresponds to a Stokes (force) dipole. A positive (resp.~negative) Stokes dipole represents two equal and opposite  forces acting away from (resp.~towards) each other (Fig.~\ref{fig:squirmer}$\b$). These force dipoles exert zero net force on the surrounding  fluid and can represent two different propulsion mechanisms of swimmers: so-called ``pushers'' ($B_2 > 0$, Fig~\ref{fig:squirmer}$\b$ left panel) and ``pullers'' ($B_2 <0$, Fig~\ref{fig:squirmer}$\b$ right panel). A pusher  obtains  thrust from the rear part of the body, such as all peritrichous bacteria (including \textit{E. coli}) or flagellated spermatozoa. As a result, a pusher repels fluid along, and behind, its swimming direction and draws fluid in from the sides. In contrast, for a puller, the thrust comes from its front, such as for the breaststroke swimming of algae \textit{Chlamydomonas}. Thus, a puller draws fluid in along its swimming direction, and repels fluid from the sides.

Finally, the other component decaying as $1/r^4$ in {the $B_2$ mode (Eq.~\ref{eq:B2})} corresponds to a source quadrupole, which decays  faster than the Stokes dipole in the far field but is more noticeable close to the squirmer, as can be appreciated visually by the tangential velocity component in close proximity to the squirmer (Fig.~\ref{fig:squirmer}$\b$).  Further discussion on the far-field hydrodynamic description of swimming organisms will be presented in Sec.~\ref{sec:FarField}.

Since the $B_1$ and $B_2$ modes capture the essential and dominant features of free swimming microorganisms, it is common in many studies to retain only these two modes and formally take $B_{n} = 0$ for $n\ge3$ as a simplified  swimmer. The squirming profile on the swimmer surface, Eq.~(\ref{eq:SqBCPak}), then reduces  to
\begin{align}
\v(r=a,\theta) = \left[- \frac{2}{a^3}\sin\theta B_1 -\frac{3}{2 a^4}\sin 2\theta B_2 \right] \e_\theta,
\end{align}
which generates the flow field found by superimposing  Eqs.~(\ref{eq:B1}) and (\ref{eq:B2}). The sign of $B_1$ determines the swimming direction, Eq.~(\ref{eq:SquirmU}), as illustrated with $B_1=-1$ in Fig.~\ref{fig:squirmer}$\a$ with upward swimming. The  sign of $B_2$ determines the configuration of the Stokes dipole and hence the basic propulsion mechanism of the microorganism and its far-field signature. Depending on the chosen parameters in the squirming profile, a squirmer can either be a pusher or puller, making it a useful idealized model for studying general features of motility for different cells. Superimposing the $B_1$ and $B_2$ modes, we plot in Fig.~\ref{fig:squirmer}$\c$ the flow field around a squirmer swimming upward ($B_1=-1$)  with a pusher ($B_2=4$) and puller ($B_2=-4$) on the left and right panels respectively.

As a final remark, in this alternative formulation, the squirming profile on the boundary (Eq.~\ref{eq:SqBCPak}) is expressed in terms of the natural basis employed in Lamb's general solution. It has a form different from, but equivalent  to, that adopted by Lighthill \cite{Lighthill1952} and Blake \cite{Blake1971b}, which is given by
\begin{align}
\v(r=a,\theta) = \sum_{n=1}^\infty \frac{2\sin\theta P_n^{'}}{n(n+1)} \tilde{B}_n \e_\theta, \label{eq:LighthillBlake}
\end{align}
where $\tilde{B}_n$'s are the coefficients used in their studies. Comparing Eq.~(\ref{eq:SqBCPak}) with Eq.~(\ref{eq:LighthillBlake}), the relation between the two sets of coefficients is given simply by $B_n = -a^{n+2}\tilde{B}_n/ (n+1)$. Transforming our results in terms of Lighthill and Blake's notation \cite{Lighthill1952, Blake1971b}, the swimming speed of a squirmer, Eq.~(\ref{eq:SquirmU}), is given by $U = -4 B_1/(3 a^3) =  2\tilde{B}_1/3$.

\subsection{Reciprocal theorem}\label{sec:reciprocal}
Stone and Samuel \cite{Stone1996} exploited the reciprocal theorem of low-Reynolds-number hydrodynamics  \cite{Happel1973} to analyze the motion of a squirmer. They were able to derive analytical expressions relating the translational and rotational velocities of the swimmer to its  arbitrary surface squirming profile without having to solve for the entire flow field. The use of  the reciprocal theorem in this fashion is handy in scenarios where only the swimming kinematics, but not the detailed flow field, is of interest. It is also shown useful in the study of non-Newtonian and inertial effects \cite{Leal1980}. Here, we introduce this technique by following Stone and Samuel's calculation \cite{Stone1996}.

Let $(\v, \Bsigma)$ be the velocity and stress fields of the original squirming problem discussed in the previous section (Sec.~\ref{sec:Lamb}), subject to an arbitrary, prescribed,  squirming profile $\v(r=a)= \v'$ on the surface. Let us then   consider an appropriate auxiliary problem with the same geometry in a Stokes flow $(\hat{\v}, \hat{\Bsigma})$. In this case, the auxiliary problem is the translation of a rigid sphere at a velocity $\hat{\U}$ due to an external force $\hat{\F}$ for reasons explained below. We have the original and auxiliary problems both satisfying the incompressible Stokes equations:
\begin{subequations}
\begin{align}
\nabla \cdot \Bsigma &= \mathbf{0}, \label{eq:DStressO} \\ 
\nabla \cdot \v &= 0, \label{eq:DUO}
\end{align}
\end{subequations}
and 
\begin{subequations}
\begin{align}
\nabla \cdot \hat{\Bsigma} &= \mathbf{0}, \label{eq:DStressA}\\
\nabla \cdot \hat{\v} &= 0. \label{eq:DUA}
\end{align}
\end{subequations}
We take the inner product of Eq.~(\ref{eq:DStressO}) with the velocity field of the auxiliary problem $\hat{\v}$ minus, reciprocally, the inner product of Eq.~(\ref{eq:DStressA}) with the velocity field of the original problem $\v$ to obtain the relation
\begin{align}
\hat{\v} \cdot (\nabla \cdot \Bsigma) - \v \cdot (\nabla \cdot \hat{\Bsigma}) = 0. \label{eq:Reciprocal}
\end{align}
By a general vector identify
\begin{align}
\hat{\v} \cdot \nabla \cdot \Bsigma - \v \cdot \nabla \cdot \hat{\Bsigma} = \nabla \cdot (\hat{\v} \cdot \Bsigma - \v \cdot \hat{\Bsigma}) + (\nabla \v : \hat{\Bsigma}- \nabla \hat{\v}: \Bsigma),
\end{align}
we rewrite Eq.~(\ref{eq:Reciprocal}) as
\begin{align}
\nabla \cdot (\hat{\v} \cdot \Bsigma - \v \cdot \hat{\Bsigma}) + (\nabla \v : \hat{\Bsigma}- \nabla \hat{\v}: \Bsigma) = 0. \label{eq:Reciprocal2}
\end{align}
The advantage of such a construction  is that it renders the second bracket in Eq.~(\ref{eq:Reciprocal2}) identically zero since
\begin{subequations}
\begin{align}
&\nabla \v : \hat{\Bsigma}- \nabla \hat{\v}: \Bsigma \notag \\
&= \nabla \v : \left[ -\hat{p} \I + \mu\left(\nabla \hat{\v} + \nabla \hat{\v}^T \right) \right]- \nabla \hat{\v}: \left[ -p \I + \mu \left(\nabla \v + \nabla \v^T\right) \right] \\
&= -\hat{p} \nabla \cdot \v + \mu \left( \nabla \v : \nabla \hat{\v} + \nabla \v : \nabla \hat{\v}^T \right) + p \nabla \cdot \hat{\v} - \mu \left( \nabla \hat{\v}: \nabla \v + \nabla \hat{\v}: \nabla \v^T \right) = 0,
\end{align}
\end{subequations}
due to the continuity equation ($\nabla \cdot \v = \nabla \cdot \hat{\v}=0$), and the identities $\mathbf{A} : \mathbf{B} = \mathbf{B} : \mathbf{A}$ and  $\mathbf{A}: \mathbf{B}^T = \mathbf{A}^T: \mathbf{B}$, which are true for any tensors $\mathbf{A}$ and $\mathbf{B}$ and follow trivially from the definition of the double-dot product.
Integrating Eq.~(\ref{eq:Reciprocal2}) over the entire fluid domain $V$ external to the sphere, we then   obtain
\begin{align}
\int_V \nabla \cdot (\hat{\v} \cdot \Bsigma - \v \cdot \hat{\Bsigma}) dV = 0 .
\end{align}
Using the divergence theorem, we convert the volume integral to the following surface integrals
\begin{align}
\int_{S_\infty} \left( \n \cdot \hat{\Bsigma} \cdot \v - \n \cdot \Bsigma \cdot \hat{\v} \right) \ dS -\int_S \left( \n \cdot \hat{\Bsigma} \cdot \v - \n \cdot \Bsigma \cdot \hat{\v} \right) \ dS = 0, \label{eq:decay}
\end{align} 
where $\n$ is the outer normal from the body into the fluid, $S$ is the spherical surface, and $S_\infty$ is the surface enclosing the sphere at infinity. Denoting $r$ as the distance from the origin and assuming the velocity fields ($\v$ and $\hat{\v}$) decay as $r^{-1}$ or faster, and the pressure fields ($p$ and $\hat{p}$) decay as $r^{-2}$ or faster, we see that the integrand of the integral over $S_\infty$ decays at least as $r^{-3}$ as $r \rightarrow \infty$. Since the surface area grows as $r^2$, the integral over $S_\infty$ decays at least as $r^{-1}$ and therefore vanishes at infinity, leaving us with
\begin{align}
\int_S \n \cdot \hat{\Bsigma} \cdot \v \ dS  = \int_S \n \cdot \Bsigma \cdot \hat{\v}  \ dS \label{eq:Reciprocal3} .
\end{align}
Because of our choice of the auxiliary problem -- a translating sphere --  we have a constant boundary condition $\hat{\v} = \hat{\U}$ on the spherical surface $S$. Moving the constant $\hat{\U}$ out of the integral we get
\begin{align}
\int_S \n \cdot \hat{\Bsigma} \cdot \v \ dS = \left(\int_S \n \cdot \Bsigma dS \right) \cdot \hat{\U} .  \label{eq:Reciprocal4}
\end{align}
Since free swimming occurs with no net force, the  right hand side of that equation should vanish, $\int_S \n \cdot \Bsigma dS = \mathbf{0}$. Under this choice of auxiliary problem, Eq.~(\ref{eq:Reciprocal4}) then becomes simply 
\begin{align}
\int_S \n \cdot \hat{\Bsigma} \cdot \v \ dS = 0 \label{eq:Reciprocal5} .
\end{align}
Next, one  decomposes the surface velocity of the original problem into the unknown translational swimming velocity, $\U$, and the arbitrary surface squirming motion, $\v'$, \textit{i.e.}~$\v(S) = \U + \v'$. With these boundary conditions, Eq.~(\ref{eq:Reciprocal5}) can be split in two integrals to become
\begin{align}
\left( \int_S \n \cdot \hat{\Bsigma} dS \right) \cdot \U = - \int_S \n \cdot \hat{\Bsigma} \cdot \v' dS . \label{eq:ReciporcalIntegrals}
\end{align}
The unknown swimming velocity $\U$ can be determined if all the integrals in Eq.~(\ref{eq:ReciporcalIntegrals}) are evaluated. This requires knowledge of the stress field of the auxiliary problem. For the translation of a rigid sphere, we have the Stokes' law, $\int_S \n \cdot \hat{\Bsigma} dS = -6 \pi \mu a \hat{\U}$, and a useful fact that the surface traction is  constant \cite{Stone1996}, $\n \cdot \hat{\Bsigma} = -3 \mu/2a \hat{\U}$. As a result, Eq.~(\ref{eq:ReciporcalIntegrals}) becomes
\begin{align}
- 6 \pi \mu a \hat{\U} \cdot \U &= \frac{3 \mu}{2a} \hat{\U} \cdot \int_S \v' dS  \ \ \ \Rightarrow \ \ \ \U = - \frac{1}{4 \pi a^2} \int_S \v' dS \label{eq:ReciprocalResult} .
\end{align}
We have now obtained the swimming speed of a squirmer, $\U$, as a simple surface integral of its  surface motion, $\v'$, without actually solving for the flow field around the swimmer. We do however require the stress field of the auxiliary problem, which means that at some point a flow calculation had to be carried out. {Note that the Stokes equations being steady, the analysis above also holds for the time-dependent case with Eq.~(\ref{eq:ReciprocalResult}) being understood as  an instantaneous identity.  

{Furthermore, similarly to the calculations above, the angular velocity, $\BOmega$, of a spherical squirmer can be related to its surface deformation using the reciprocal theorem as well and one gets\cite{Stone1996}
\begin{align}
\BOmega = -\frac{3}{8 \pi a^3} \int_S \n \times \v' ds,
\end{align}
with details  left as an exercise for the readers.}

As a verification of the final result, we use Eq~\ref{eq:ReciprocalResult} to compute the swimming speed of a squirmer subject to the general squirming profile expressed in terms of the basis given by Eq.~(\ref{eq:SqBCPak}):
\begin{align}
\U &= - \frac{1}{4 \pi a^2} \int_S \v' dS = -\frac{1}{4 \pi a^2} \int_0^{2\pi} \int_0^\pi \left( \sum_{n=1}^\infty  -\frac{2\sin \theta P^{'}_n}{n a^{n+2}}   B_n \e_\theta \right) a^2 \sin \theta d \theta d \phi .
\end{align}
Expressing the unit vector $\e_\theta$ in terms of the basis vectors in Cartesian coordinates we have
\begin{align}
\U &= \frac{1}{2 \pi a^2} \int_0^{2\pi} \int_0^{\pi} \sum_{n=1}^\infty  \frac{\sin^2\theta P^{'}_n}{n a^{n}} B_n \left( \cos \theta \cos \phi \e_x + \cos \theta \sin \phi \e_y - \sin\theta \e_z  \right) d\theta d\phi, 
\end{align}
only the $z$-component survives due to axisymmetry, leaving the integrals
\begin{align}
\U&=- \frac{1}{a^2} \int_0^\pi \sum_{n=1}^\infty   \frac{\sin^3\theta P^{'}_n}{n a^{n}} B_n d\theta \ \e_z . 
\end{align}
By a change of variable from $\theta$ to $\eta = \cos\theta$, the evaluation of the integral can be computed using properties of Legendre polynomials as
\begin{align}
\U&= - \frac{1}{a^2} \int_{-1}^1 \sum_{n=1}^\infty \frac{\sin^2 \theta P_n^{'}}{n a^n} B_n d\eta \ \e_z = -\frac{1}{a^2} \sum_{n=1}^{\infty} \int_{-1}^1 \frac{(1-\mu^2) P_n^{'} P_1^{'}}{n a^n} B_n d\eta \ \e_z = -\frac{4}{3a^3}  B_1, 
\end{align}
verifying the result obtained analytically in the previous section (Eq.~\ref{eq:SquirmU}).  The reciprocal theorem is therefore a useful tool for determining the swimming kinematics, bypassing detailed calculation of the flow field provided the swimmer geometry is one for which the stress profile in the auxiliary problem has been determined. It  provides however (obviously) no information on the flow around the squirmer, which is required for problems such as  nutrient transport and uptake by microorganisms \cite{Magar2003, Magar2005, Michelin2011}.

\section{Far-field Description of a Swimmer}\label{sec:FarField}
In this section, we introduce the mathematical framework necessary to  quantify the   swimming hydrodynamics in the far-field  \cite{Pedley1992}. This concept is useful for cases in which the flow field in close proximity of a swimmer is not of interest but the far-field behaviour is, for example to determine the influence of a nearby boundary, or of another swimmer nearby. Physically, it is equivalent to zooming-out and observing the swimmer over length scales much larger than its intrinsic length. Under this far-field approximation, the geometrical details of the swimmer are therefore unimportant and some generic features of low-Reynolds-number swimming may be obtained.

\subsubsection{Stokeslet}
To formulate such a perspective, it is useful to first introduce the Green's function for the Stokes equations. This is obtained by placing a point force $ f\e \delta(\x)$ at the origin in an otherwise quiescent infinite fluid, where $\delta(\x)$ is the Dirac delta function centered at $\x = \mathbf{0}$\footnote{Without loss of generality, in this section we present the results for a point force located at the origin for convenience. Results for a point force at other locations $\x_0$ can be readily obtained by a simple translation of coordinates by replacing $\x$ as $\x -\x_0$.}, $\e$ a unit vector  represents the direction of the point force, and $f$ the magnitude of the force. The forced Stokes equations are given by 
\begin{subequations}
\begin{align}
\nabla p &= \mu \nabla^2 \v +  f \e  \delta, \\
\nabla \cdot \v &= 0 .
\end{align}
\end{subequations}
The solution $\v(\x)$ can be obtained by a variety of methods, such as Fourier transformation and superposition of vector harmonic functions \cite{Leal2007, Pozrikidis1992}, and is given by
\begin{align}
\v(\x) =  f\G(\x; \e)=\frac{f}{8\pi\mu} \left[ \frac{\e}{r} + \frac{(\e \cdot \x)\x}{r^3} \right]\label{eq:Stokeslet},
\end{align}
where $r=|\x|$ is the distance from the singularity. This fundamental singular solution  in viscous flows is called a Stokeslet \cite{Hancock1953}. It decays as $1/r$ and is therefore long-ranged \cite{Chwang1975}. The flow field of a Stokeslet is shown in Fig.~\ref{fig:FarField}$\a$  {in the laboratory frame}, 
and is the one given by  a translating sphere in the far field as in both cases a net force is exerted  on the fluid. Physically, as we increasingly zoom-out from a translating sphere, it becomes sufficiently small that it can be regarded as a point acted on by a force, and thus a Stokeslet may be physically understood as the far-field approximation of a translating sphere. Since a swimmer does not exert a net force on the surrounding fluid, only force dipoles and above will be allowed, as we now detail.

\begin{figure}[t]
\begin{center}
\includegraphics[width=0.8\textwidth]{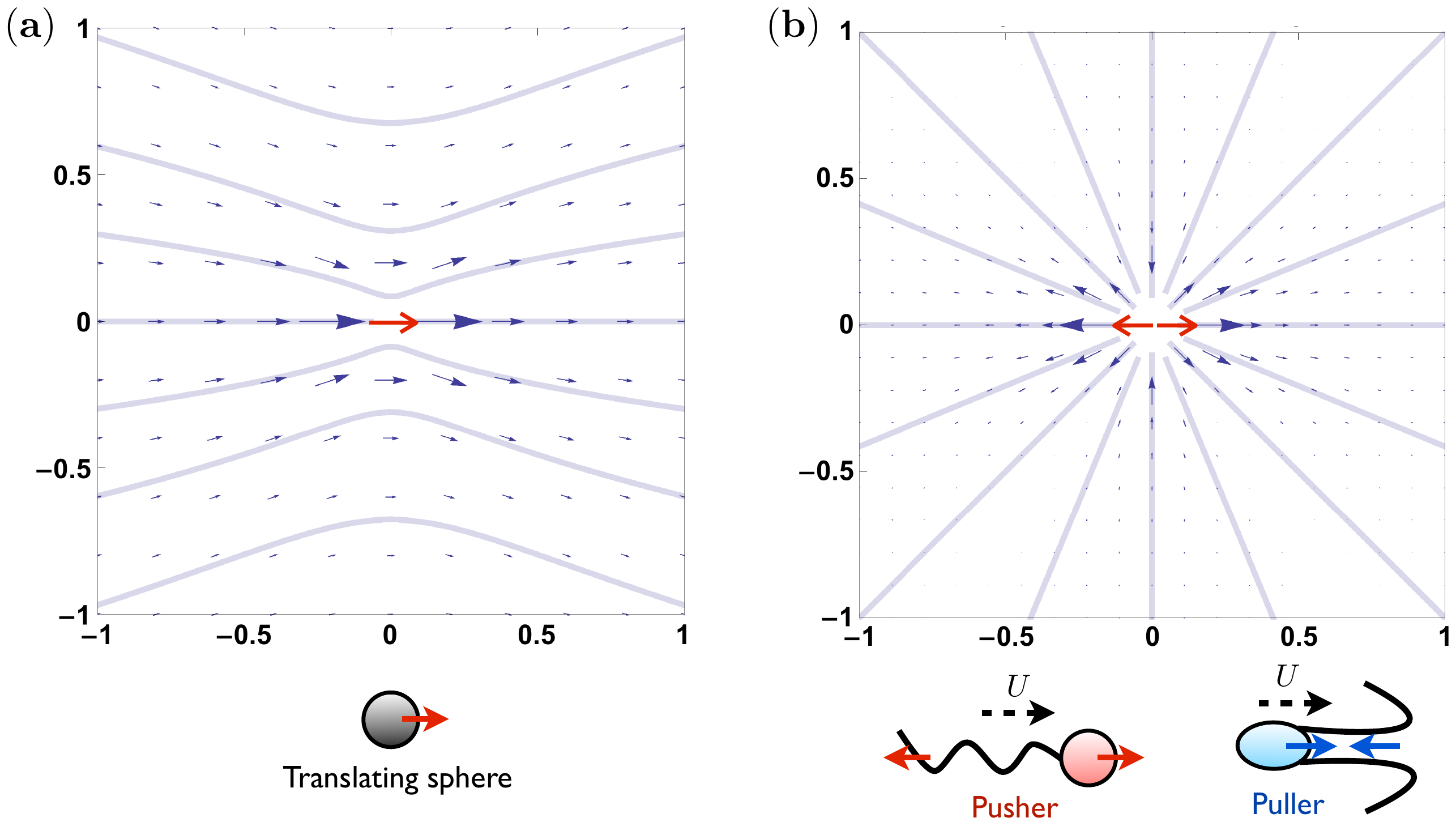}
\caption{Flow singularities. $(\mathbf{a})$ Velocity field due to a Stokeslet at the origin, the fundamental singular solution in Stokes flows due to a point force, and a model for the  far-field approximation of a translating sphere. $(\mathbf{b})$ Velocity field due to a positive Stokes dipole at the origin. The vectors represent local fluid velocity and the lines represent streamlines.  The flow field of a negative Stokes dipole  (two point forces acting towards each other) has the same streamline pattern with the sign of the velocity field reversed. A positive (resp.~negative) Stokes dipole is the far-field approximation of for a pusher (resp.~puller) swimmer. The arrows on the swimmers indicate local forces exerted on the fluid.}
\label{fig:FarField}
\end{center}
\end{figure}

\subsubsection{Stokes dipole}

Since the Stokes equations are linear, a derivative of any order of a Stokeslet is also a solution, forming higher-order singular solutions. By taking one derivative of an  $\e$-directed Stokeslet along the $\d$ direction, one obtains a Stokes dipole
\begin{align}
\G_{D}(\x; \d, \e) = -\d \cdot \nabla \G(\x; \e)  &= \frac{1}{8\pi \mu} \left\{ \frac{(\e \d - \d \e) \cdot \x}{r^3} + \left[  \frac{3}{2}(\e \d + \d \e) -(\e\cdot \d) \I \right] : \frac{\x\x\x}{r^5}  \right\}, \\
&=  \frac{1}{8\pi \mu} \left\{ \frac{(\d \times \e) \times \x}{r^3} + \left[ -\frac{(\e \cdot \d) \x}{r^3} + 3 \frac{(\e \cdot \x)(\d \cdot \x) \x}{r^5}  \right] \right\},
\end{align}
which is the most useful singular solution in the study of swimming microorganisms (see also discussion in Sec.~\ref{sec:squirmer} for the flow field around a squirmer). The flow due to the symmetric part of the Stokes dipole tensor is termed a stresslet \cite{Batchelor1970S}
\begin{align}
\S(\x; \d, \e) =\frac{1}{8\pi \mu} \left[- \frac{(\e \cdot \d) \x}{r^3}+3 \frac{(\e \cdot \x)(\d \cdot \x) \x}{r^5}  \right],
\end{align}
physically representing straining motion of the fluid, whereas the flow due to the antisymmetric part is termed a rotlet
\begin{align}
\R(\x; \d, \e)   =\frac{(\d \times \e) \times \x}{8 \pi \mu r^3}, \label{eqn:rotlet}
\end{align}
physically representing the flow due to a point torque. Other higher-order singularities such as Stokes quadrupole, potential source dipole, and source quadrupole can be obtained by taking derivatives of the corresponding lower-order singularities. These singularities will be useful later when we investigate the effects of a boundary on swimming cells (Sec.~\ref{sec:WallSingularities}). 

A Stokes dipole decays as $1/r^2$, one order of magnitude faster than a Stokeslet. Physically, a  Stokes dipole can be understood  as the limit when the distance between two Stokeslets of equal magnitudes but opposite directions becomes vanishingly small and the strength is adjusted to give a finite flow field. Consider a Stokeslet, 
$\v^{+}(\x)$, of strength $f \e$ acting at a small  distance $d/2$ from the origin along the   direction  $\d$. By Taylor's expansion about the origin, we can express the Stokeslet as
\begin{align}
\v^{+}(\x) = f\G(\x-d \d/2; \e) = f \left[ \G(\x; \e)-\frac{d}{2} \d \cdot \nabla \G(\x, \e)  + ... \right].
\end{align}
Consider another Stokeslet, $\v^{-}(\x)$, of opposite strength $-f \e$ acting at an opposite position of $-d \d/2$ from the origin. Again by Taylor's expansion, we expand this Stokeslet about the origin to obtain
\begin{subequations}
\begin{align}
\v^{-}(\x) = f\G(\x+d \d/2; -\e) &= f \left[ \G(\x; -\e) + \frac{d}{2} \d \cdot \nabla \G(\x, -\e)  +  ... \right] \\
&= f \left[ -\G(\x; \e)- \frac{d}{2} \d \cdot \nabla \G(\x, \e)  + ... \right] \cdot
\end{align}
\end{subequations}
Superposing the two Stokeslets gives an overall flow field 
\begin{align}
\v(\x) = \v^{+}(\x) + \v^{-}(\x) = -df \d \cdot \nabla \G(\x; \e)+ ... \ ,
\end{align}
where the leading-order contribution $\d \cdot \nabla \G(\x; \e)$ is a Stokes dipole. The distance $d$ and strength $f$ of the Stokeslets can be adjusted so that higher-order terms vanish upon taking the limit, leaving only the flow field due to a Stokes dipole. We can thus understand a Stokes dipole as the leading-order contribution from two point forces acting at a fixed and sufficiently small separation distance $d$. 

In the case relevant to axisymmetric swimmers, the two point forces in the dipole are aligned in the same direction as the one along which derivatives are taken  ($\e \times \d = \mathbf{0}$, or $\d = \pm \e$). In other words, it is the case where the rotlet component (Eq.~\ref{eqn:rotlet}) of the Stokes dipole vanishes as there is no net torque on a swimmer, leaving only the symmetric stresslet component. The stresslet resulting in this case has the general expression
\begin{align}
\S(\x; \d, \e) &= \frac{1}{8\pi \mu} \left[- \frac{(\e \cdot \d) \x}{r^3} + 3 \frac{(\e \cdot \x)(\d \cdot \x)\x}{r^5} \right]  = \frac{\alpha}{8\pi \mu} \left[ -\frac{\x}{r^3}+ 3 \frac{(\e \cdot \x)^2\x}{r^5} \right], \label{eq:stresslet}
\end{align}
where $\alpha = \e \cdot \d = \pm1$ represents the two different opposite configurations. 
When $\alpha=1$, the two points forces act away from each other (see Fig.~\ref{fig:FarField}$\b$ for a Stokes dipole with $\e=\e_z$ and $\alpha=1$),  pushing fluid away along the direction of the dipole and drawing fluid towards the dipole from the side. Notice that the streamline pattern remains exactly the same for the other case $\alpha=-1$ but the velocity field changes by a sign. Therefore, a Stokes dipole with $\alpha=-1$ draws fluid along the direction of the dipole and repels fluid to the side.

With this concept in mind, we can now analyze the motion of self-propelled microorganisms from a sufficiently large distance  that the geometrical details of the swimmer may be ignored. For many self-propelled microorganisms such as a spermatozoon or an \textit{E. coli} cell (Fig.~\ref{fig:FarField}$\b$, pusher), one can identify two parts, namely the cell body and the flagellum. As the cell moves through a viscous fluid (to the right in Fig.~\ref{fig:FarField}$\b$), the cell body experiences a viscous drag acting to the left. Since a self-propelled swimmer is force-free (for neutrally buoyant cells), the fluid has to exert (due to the action of the flagellum) a force of equal magnitude acting to the right to balance the drag force on the cell body, forming a pair of force acting towards each other on the cell. By Newton's third law, the swimmer exerts therefore on the fluid a pair of forces acting away from each other  (indicated by red arrows in Fig.~\ref{fig:FarField}$\b$). Essentially, the force to the right is the drag while that on the left is the propulsive thrust.  
Observing the motion of these swimmers in the   far field,  to leading order they generate a positive Stokes dipole with $\alpha =1$, and are called pushers. In contrast,  the type of swimmers called pullers obtain their thrust from the front part of the body and hence exerts a pair of force towards each other on the fluid (Fig.~\ref{fig:FarField}$\b$, puller), generating a negative Stokes dipole ($\alpha =-1$). This is, for example, the case for the algae  \textit{Chlamydomonas} which uses two flagella.  The squirmer model introduced in Sec.~\ref{sec:squirmer} can model both pushers and pullers -- depending on the sign of $B_2$ -- since the flow field generated by the squirming motion are asymptotically Stokes dipoles in the far field. The fact that the  flow fields around a pusher and puller differ only by a sign  can lead to qualitatively different types of hydrodynamic interactions, as we will see in Sec.~\ref{sec:HI}.

\section{Hydrodynamic Interactions}\label{sec:HI}

In the previous sections we introduced theoretical models for an isolated swimming microorganism in an unbound fluid. Actual biological environments are however more complicated in a number of ways. Microorganisms do not usually swim alone and a swimming cell experiences physical effects due to the presence of other co-swimming organisms. Instead  of an infinite fluid, microorganisms encounter  surfaces, for example during the  locomotion of spermatozoa in mammalian cervical mucus \cite{Fauci2006, Lauga2009}. Furthermore, during most laboratory experiments, coverslips impose solid boundaries near the microorganisms. These boundaries and the presence of other swimmers modify the fluid flow around a given microorganism and has important consequences on its dynamics. We review in this section classical ideas on hydrodynamic interactions between cells and boundaries. 

\subsection{Swimming near a boundary: Lubrication theory}\label{sec:lubrication}

Reynolds \cite{Reynolds1965}  first adopted Taylor's swimming sheet model to consider locomotion near solid walls for a prescribed small-amplitude waving motion. He found that the effect of a solid wall is to increase the swimming speed when the sheet is swimming closer to a wall.  Katz \cite{Katz1974} subsequently performed a lubrication analysis for a sheet swimming in close proximity to the wall --  another useful tool allowing analytical progress  in certain asymptotic regimes. Recently, this lubrication calculation was reviewed and extended to consider swimming near a wall in complex fluids \cite{Pachmann2008, Balmforth2010}. Here we follow the review \cite{Pachmann2008} to reproduce Katz's results in the Newtonian case and illustrate the use of lubrication theory for analyzing swimming near a wall.

\subsubsection{Formulation}

Adopting Taylor's swimming sheet with a wave propagating in the positive $x$-direction (Sec.~\ref{sec:TaylorSheet}), the dimensional vertical displacement is given by $Y(x,t)= a \sin (kx-\omega t)$, where $a$, $\omega$, and $k$ are the amplitude, angular frequency, and wave number respectively. The sheet swims at an average distance $h$ from the wall (see Fig.~\ref{fig:Wall}$\a$). We consider the lubrication limit, \textit{i.e.}~assume that $hk  \ll 1$ and thus that the distance from the wall, $h$, is small compared with the wavelength, $\lambda = 2\pi/k$, of the sheet. From the results in Sec.~\ref{sec:TaylorSheet}, we assume that the sheet swims in the negative $x$-direction (the direction opposite to the wave propagation, see Fig.~\ref{fig:Wall}$\a$). We denote the swimming velocity as $-U \e_x$, where $U$ is the unknown swimming speed. We approach this problem by observing the motion in a frame moving with the sheet (at the velocity $-U \e_x$). In this moving frame (Fig.~\ref{fig:Wall}$\b$), the wall moves at a velocity $U \e_x$, and the sheet displaces only vertically with velocity $\partial Y(x,t)/\partial t = Y_t$. The dimensional boundary conditions in this case are therefore given by
\begin{subequations}
\begin{align}
u(x,y=h) &= U, \\
u(x,y=Y)&=0,\\
v(x,y=h) &= 0, \\
v(x,y=Y)&= Y_t = - a \omega \cos(kx-\omega t).
\end{align}
\end{subequations}

\begin{figure}[t]
\begin{center}
\includegraphics[width=0.9\textwidth]{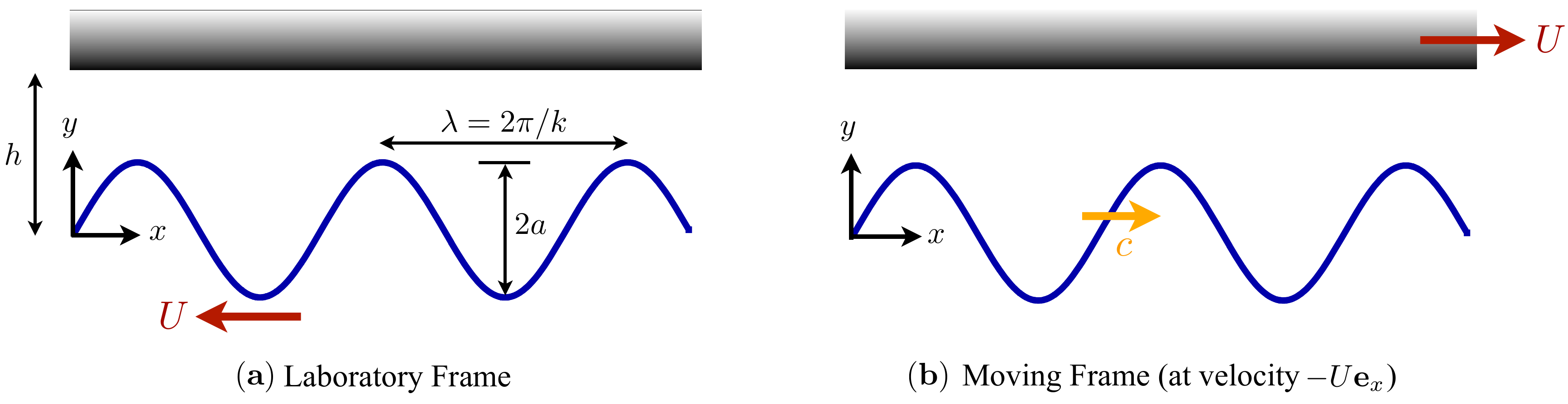}
\caption{Geometrical setup and notations for a Taylor's two-dimensional swimming sheet model near a rigid wall. $(\mathbf{a})$ Motion observed under the laboratory frame, where the unknown swimming velocity, $-U \e_x$, is assumed to occur in the negative $x$-direction. $(\mathbf{b})$ Motion observed in a frame moving with the sheet at the swimming velocity $-U \e_x$. In this frame, the sheet  undergoes vertical displacements propagating a travelling wave to the right at phase speed $c$ and the solid wall  moves with velocity $U \e_x$ to the right.}
\label{fig:Wall}
\end{center}
\end{figure}

\subsubsection{Non-dimensionalization}
Similar to Taylor's original calculations (Sec.~\ref{sec:TaylorSheet}), we non-dimensionalize time by $1/\omega$, the horizontal length scale by $1/k$, and hence horizontal velocity by $c = \omega/k$. In lubrication problems, lengths and velocities in the vertical direction are scaled differently to capture the correct physics of the problem. We scale the vertical length scale by $h$, and obtain the scaling for the vertical velocity $v_c$ by considering the continuity equation 
\begin{align}
\frac{c}{1/k} \frac{\partial u^*}{\partial x^*} + \frac{v_c}{h}\frac{\partial v^*}{\partial y^*} = 0,
\end{align}
which gives $v_c= c \delta$, where we have introduced the dimensionless parameter $\delta = hk$. Recall that the  lubrication limit is the one for which   $\delta \ll 1$. The scaling for the pressure is obtained by considering the two-dimensional Stokes equations as
\begin{subequations} \label{eqn:lub}
\begin{align}
\frac{p_c}{\mu c/ h^2 k}  \frac{\partial p^*}{\partial x^*} &= \delta^2 \frac{\partial^2 u^*}{\partial x^{*2}} + \frac{\partial^2 u^*}{\partial y^{*2}} \label{eqn:lubx}, \\
\frac{p_c}{\mu c k} \frac{\partial p^*}{\partial y^*}&= \delta^2 \frac{\partial^2 v^*}{\partial x^{*2}} + \frac{\partial^2 v^*}{\partial y^{*2}} \label{eqn:luby},
\end{align}
\end{subequations}
where $p_c$ represents the characteristic pressure. We are interested in the limit $\delta = hk \ll 1$ and by considering the dominant balance in Eqs.~(\ref{eqn:lubx}) and (\ref{eqn:luby}), the pressure may be scaled as $p_c \sim \mu c/h^2k$ or $p_c \sim \mu c k$ respectively. Given the boundary conditions, it can be shown that the consistent balance is given by $p_c \sim \mu c/h^2k$, which is a general feature of lubrication problems\cite{Leal2007}. Adopting this pressure scale, we see that the pressure gradient in the $y$-momentum equation, Eq.~(\ref{eqn:luby}), scales as $1/\delta^2$, much greater than the viscous terms on the right-hand side in the lubrication limit ($\delta \ll 1$), and hence  vanishes to leading order. Therefore, the leading-order governing equations in lubrication theory become
\begin{subequations} \label{eqn:lubGov}
\begin{align}
\frac{\partial u^*}{\partial x^*} + \frac{\partial v^*}{\partial y^*} &= 0, \label{eqn:lubContinuity}\\
\frac{\partial p^*}{\partial x^*} &= \frac{\partial^2 u^*}{\partial y^{*2}}, \label{eqn:lubMomX} \\
\frac{\partial p^*}{\partial y^*} &= 0, \label{eqn:lubMomY}
\end{align}
\end{subequations}
with boundary conditions
\begin{subequations}
\begin{align}
u^*(x^*, y^* = 1) &= U/c =U^*, \label{eqn:lubBC1}\\
u^*(x^*, y^* = Y^*) &= 0, \label{eqn:lubBC2}\\
v^*(x^*, y^*=1) &= 0,\label{eqn:lubBC3}\\
v^* (x^*, y^* =Y^*) &= - a^* \cos(x^*-t^*) \label{eqn:lubBC4},
\end{align}
\end{subequations}
where the displacement of the sheet is given by
\begin{align}
Y^* &= \frac{a}{h} \sin (x^* - t^*)=a^* \sin (x^* - t^*) .
\end{align}
All variables hereafter are dimensionless and we drop the stars for convenience. 

\subsubsection{Lubrication analysis}
From the $y$-momentum equation, Eq.~(\ref{eqn:lubMomY}), we get that the pressure,  $p$, and hence the pressure gradient, $\partial p/\partial x$, are independent of $y$. We exploit this to integrate Eq.~(\ref{eqn:lubMomX}) twice with respect to $y$ and obtain an explicit expression for the horizontal velocity component as
\begin{align}
u(x, y) = \frac{1}{2}\frac{\partial p}{\partial x} (y-Y)(y-1) + U \frac{y-Y}{1-Y} \label{eqn:lubU}, 
\end{align}
where the boundary conditions for $u$, Eqs.~(\ref{eqn:lubBC1}) and (\ref{eqn:lubBC2}),  have been implemented. Next, we integrate the continuity equation, Eq.~(\ref{eqn:lubContinuity}), with respect to $y$ from the sheet ($y=Y$) to the wall ($y=1$) and apply the boundary conditions for $v$, Eqs.~(\ref{eqn:lubBC3}) and (\ref{eqn:lubBC4}), in order to obtain
\begin{align}
\int_{Y}^1 \frac{\partial u}{\partial x} dy + a \cos(x -t) = 0 . \label{eqn:lubReynoldsEq}
\end{align}
In many lubrication problems, after differentiating Eq.~(\ref{eqn:lubU}) and carrying out the integral in Eq.~(\ref{eqn:lubReynoldsEq}), one obtains the famed ``Reynolds equation" in lubrication theory allowing to determine the unknown pressure  \cite{Katz1974}. Here, since our primary interest is to compute the swimming speed, $U$, we take a slightly different route bypassing the computation of the pressure. Instead, we use  Leibniz's rule to interchange the differential and integral operations in Eq.~(\ref{eqn:lubReynoldsEq}) as
\begin{align}
\frac{\partial}{\partial x}\int_{Y}^1 u dy + a \cos(x -t) = 0, \label{eqn:lubIntegral}
\end{align}
with the boundary condition $u(y=Y)=0$. We then integrate Eq.~(\ref{eqn:lubIntegral})  in $x$ to obtain
\begin{align}
\int_{Y}^1 u dy + a \sin(x -t) = q(t), \label{eqn:lubIntegral2} 
\end{align}
where $q(t)$ represents the mass flux. We substitute the expression for $u$, Eq.~(\ref{eqn:lubU}), into Eq.~(\ref{eqn:lubIntegral2}) and evaluate the integral to obtain an explicit expression for the pressure gradient
\begin{align}
\frac{\partial p}{\partial x} = \frac{12(1-q)}{(1-Y)^3} +\frac{6(U-2)}{(1-Y)^2}, \label{eqn:lubDpDx}
\end{align}
where the two unknowns are $q$ and $U$. They are determined   by enforcing first the periodicity of the problem 
\begin{align}
\int_0^{2\pi}  \frac{\partial p}{\partial x} d x = 0 \label{eqn:lubPeriodicity},
\end{align}
and second  the dynamic condition requiring the swimmer to be overall force-free. To derive the latter condition, we compute the total dimensionless force acting on the swimming sheet over one complete wavelength 
\begin{align}
\F &= \int_{S} \Bsigma\cdot \n ds \sim \int_{0}^{2\pi} 
\begin{pmatrix}
-p+2 \delta^2 \frac{\partial u}{\partial x} & \delta \frac{\partial u}{\partial y} + \delta^3 \frac{\partial v}{\partial x} \\
\delta \frac{\partial u}{\partial y} + \delta^3 \frac{\partial v}{\partial x} & -p + 2\delta \frac{\partial v}{\partial y}
\end{pmatrix}
\begin{pmatrix}
\delta Y_{x} \\
-1
\end{pmatrix}
dx,
\end{align}
where $\Bsigma$ denotes the dimensionless stress (scaled similarly to the  pressure with $\mu c/h^2k = \mu \omega/\delta^2$), $\n$ denotes the unit normal vector, $\n = (\delta Y_{x} \ \ -1)^T/\sqrt{1+\left( \delta Y_{x} \right)^2}$ , and $s$ the arc-length along the sheet, $s=\sqrt{1+(\delta Y_x)^2}$. Note  that the $x$-component of the normal $\n$ is of order $O(\delta)$ due to the different scalings in  the horizontal and vertical directions. The leading-order force-free condition in the horizontal ($x$) direction is then given by
\begin{align}
\int_0^{2\pi} \left(- p Y_x -\frac{\partial u}{\partial y}\right) \bigg|_{y= Y} dx =0 .
\end{align}
To facilitate the use of this condition for determining the constant in Eq.~(\ref{eqn:lubDpDx}), we integrate by parts (employing the periodicity of pressure and $Y$) to rewrite the above force-free condition as
\begin{align}
\int_0^{2\pi} \left(Y \frac{\partial p}{\partial x} - \frac{\partial u}{\partial y}\right) \bigg|_{y= Y} dx &= 0 \label{eqn:lubForceFree} .
\end{align}
We finally have two equations, Eqs.~(\ref{eqn:lubPeriodicity}) and (\ref{eqn:lubForceFree}), allowing to determine the two unknowns $q$ and $U$ in Eq.~(\ref{eqn:lubDpDx}),  leading to the dimensionless swimming speed
\begin{align}
U = q = \frac{3 a^{2}}{2 a^{2}+1} \cdot
\end{align}
In dimensional form, the swimming speed reads
\begin{align}
U = \frac{3c}{2 + (h/a)^2} \cdot
\end{align}
This lubrication analysis, due to  Katz \cite{Katz1974}, obtained the same conclusion as the small-amplitude analysis by Reynolds \cite{Reynolds1965} in that the propulsion speed increases as the swimmer comes closer to the wall. In addition, since $h \ge a$, the propulsion speed is bounded above by the wave propagation speed. Importantly, the analysis assumes that the prescribed swimming waveform remains  the same and does not depend on the value of $h$. Actual organisms however may very well modify their flagellar waveforms as they approach walls, potentially leading to a decrease of the swimming speed  close to a  wall \cite{Reynolds1965}.

\subsection{Swimming near a boundary: Far-field approximation}\label{sec:WallSingularities}

In Sec.~\ref{sec:FarField}, we have introduced the far-field approximations of swimming cells and shown that they can be described as Stokes dipoles in an unbounded fluid. The presence of rigid boundaries requiring that the no-slip and no-penetration boundary conditions be enforced modifies the flow around the singularity, and has an impact on the swimmer motion. Similarly to the method of images in electrostatics -- albeit somewhat more involved due to the requirement of enforcing three scalar boundary conditions -- Blake \cite{Blake1971} showed how to derive the Green's function for Stokes flows near a rigid surface by placing a system of image singularities on the other side of the surface (\textit{i.e.}~inside the wall). Blake's results are equivalent to those obtained earlier using an alternative method (a reciprocal theorem approach) by Lorentz \cite{Lorentz1896}. These results have subsequently been applied to describe the far-field dynamics of a swimming cell in the presence of a wall, providing a hydrodynamic explanation for cell concentration at the boundaries observed in experiments \cite{Berke2008}. The accuracy of such far-field description of low-Reynolds-number swimming near a surface has been thoroughly discussed by Spagnolie and Lauga \cite{Spagnolie2012}.

\begin{figure}[t]
\begin{center}
\includegraphics[width=1\textwidth]{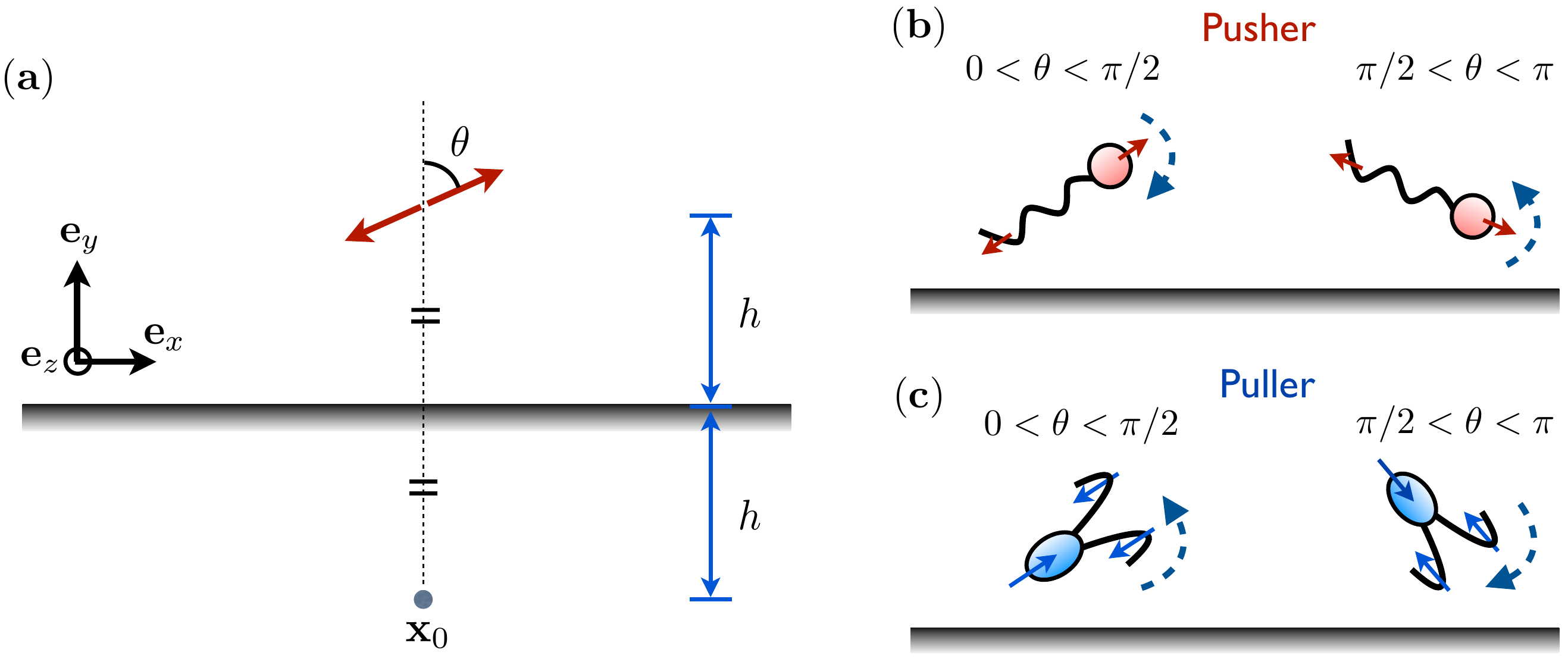}
\caption{($\mathbf{a}$) Geometrical setup for the image system of a singularity (here a force-dipole as a model swimmer). Hydrodynamic reorientation near a surface of a ($\mathbf{b}$) pusher and a ($\mathbf{c}$) puller. }
\label{fig:WallFarField}
\end{center}
\end{figure}

\subsubsection{Image system of a Stokes dipole}
The image singularity system of a Stokeslet parallel or perpendicular to a no-slip boundary can be linearly combined to give the image system of a Stokeslet at an arbitrary angle relative to the wall \cite{Blake1971}. Care has then to be taken in order to derive the image system for higher-order singularities (dipoles and higher order) \cite{Blake1974}. Full vector expressions of these singularities and their image systems can be found in the literature \cite{Spagnolie2012}. Let us consider a Stokes dipole of strength $\alpha$, Eq.~(\ref{eq:stresslet}), oriented at an arbitrary angle $\theta$ and located at an arbitrary distance $h$ from the wall, as illustrated in Fig.~\ref{fig:WallFarField}$\a$; note that we assume without loss of generality that the singularity lies on the $x-y$ plane. Its image system can be obtained by taking appropriate derivatives of the Stokeslet image system, leading to
\begin{align}
\G^*_{D}(\x-\x_0^*; \alpha) = \alpha &\big\{ \sin^2 \theta \left[ -\G_D(\e_x, \e_x) + 2h \G_Q (\e_x, \e_x, \e_y) -2h^2 \Q(\e_x,\e_x) \right] \notag\\
&+\cos^2 \theta \left[ -\G_D(\e_y,\e_y)+4h \D (\e_y)+2h \G_Q(\e_y,\e_y,\e_y)-2h^2 \Q(\e_y,\e_y)  \right] \notag\\
&+\sin \theta \cos \theta [ \G_D(\e_x,\e_y)+\G_D(\e_y, \e_x)-4h \D(\e_x)-4h \G_Q(\e_x,\e_y,\e_y) \notag \\
&+4h^2 \Q(\e_x,\e_y)   ] \big\}, \label{eqn:image}
\end{align}
where $\G_Q$ is a Stokes quadrupole, $\D$ is a potential (source) dipole, and $\Q$ is a source quadrupole, all of which are given by
\begin{subequations}
\begin{align}
\G_Q(\x; \c, \d, \e)  &= \frac{1}{8 \pi \mu r^3} \left[ (\d \cdot \e) \c + (\c \cdot \e) \d - (\c \cdot \d) \e - \frac{3g_1}{r^2}+ \frac{15(\c \cdot \x)(\d \cdot \x)(\e \cdot \x) \x}{r^4}  \right], \\
\D(\x; \e) &= \frac{1}{8 \pi \mu r^3} \left[ -\e + \frac{3 (\e \cdot \x) \x}{r^2}   \right] ,\\
\Q(\x; \d, \e) &= \frac{-3}{ 8 \pi \mu r^4} \left[ \frac{(\d \cdot \x) \e + (\e \cdot \x) \d + (\d \cdot \e) \x}{r} - \frac{5 (\e \cdot \x)(\d \cdot \x)\x}{r^3}  \right],\\
g_1 &=  \left[ (\d \cdot \e)(\c \cdot \x)+ (\c \cdot \e)(\d \cdot \x)+(\c \cdot \d)(\e \cdot \x)  \right] \x +(\d \cdot \x) (\e \cdot \x) \c \notag \\
& \ \ \ \  + (\c \cdot \x) (\e \cdot \x) \d - (\c \cdot \x) (\d \cdot \x)\e .
\end{align}
\end{subequations}
The flow field generated by the image singularities,  $\G^*_D$, represents the total modification in the flow field due to the presence of the wall.

\subsubsection{Fax\'{e}n's law}
Fax\'{e}n's law provides a way to determine the translational velocity, $\U$, and rotational rate, $\BOmega$,  of a body due to an arbitrary ambient flow, $\v^*$, in Stokes flows. Let us consider a swimmer with the shape of a prolate spheroid with major and minor axis lengths given by $a$ and $b$ respectively, and a body aspect ratio  defined as $\gamma = a/b$. Fax\'{e}n's law in that case is given by \cite{Kim1991}
\begin{subequations}\label{eqn:Faxen}
\begin{align}
\U &= \v^*(\x_0)+ O(a^2 \nabla^2 \v^*|_{\x_0}), \label{eqn:Faxen1} \\
\BOmega &= \frac{1}{2} \bomega^*(\x_0) + \frac{\gamma^2-1}{\gamma^2+1} \e \times \left[ \mathbf{E}^*(\x_0) \cdot \e \right] +O(a^2 \nabla^2\bomega|_{\x_0}), \label{eqn:Faxen2}
\end{align}
\end{subequations}
where $\bomega^* = \nabla \times \v^*$ and $\mathbf{E}^* = (\nabla \v^* + \nabla \v^{*T})/2$ denote, respectively, the vorticity and rate of strain of the  flow. In the case of a swimmer modeled as a singularity near a surface, we take $\v^*$ to be the image system flow field evaluated at the body centroid, $\x_0$. Using the image flow $\v^*(\x)= \G^*_D$ by substituting Eq.~(\ref{eqn:image}) into Eq.~(\ref{eqn:Faxen}), we obtain the wall-induced kinematics 
\begin{subequations}
\begin{align}
\U &= \frac{\alpha}{8\pi\mu} \left[ \frac{3 \sin 2 \theta}{8h^2} \e_x -\frac{3(1-3\cos^2\theta)}{8h^2} \e_y \right], \\
\BOmega &= \frac{\alpha}{8\pi\mu} \left\{- \frac{3 \sin 2\theta}{16h^3} \left[ 1+ \frac{\gamma^2-1}{2(\gamma^2+1)} (1+\cos^2\theta)  \right] \right\} \e_x \times \e_y, 
\end{align}
\end{subequations}
which we now examine. 
\subsubsection{Hydrodynamic attraction/repulsion and re-orientation}
The induced velocity component normal to the boundary is given by
{\begin{align}
U_y (\theta, h) = \U \cdot \e_y = -\frac{3 \alpha}{64\pi\mu h^2} (1-3\cos^2\theta),\label{eq:AttractRepel}
\end{align}
and allows us to answer   the question: does the wall attract or repel the swimming cell? The effect of wall depends on the type of swimmers, with an opposite effect in the case of  pushers ($\alpha>0$) versus pullers ($\alpha<0$). For a pusher ($\alpha >0$), swimming parallel to the wall ($\theta = \pi/2$), we see that $U_y$ is negative, meaning that the cell is attracted to  the wall. Allowing the cell to be tilted, we see that the sign of $U_y$ becomes positive if the angle $\theta < \cos^{-1} 1/\sqrt{3}$ at which point the wall repels the swimmer. Due to the linearity of the wall-induced flow with $\alpha$, the  opposite conclusion holds for a puller and in that case swimming parallel to the wall leads to a repulsion.

In addition to  inducing attraction or repulsion, the wall also hydrodynamically re-orients the swimmer. The induced rotational velocity on the swimmer acts in the $\e_z = \e_x \times \e_y$  direction (the direction perpendicular to the page in  Fig.~\ref{fig:WallFarField}) at a rate
\begin{align}
\BOmega &= - \frac{3 \alpha \cos \theta \sin \theta}{64 \pi \mu h^3} \left[ 1+ \frac{\gamma^2-1}{2(\gamma^2+1)} (1+\cos^2 \theta)\right] \e_z . \label{eq:orientation}
\end{align}
The re-orientation depends on the swimming mechanism $\alpha$. The shape of the cell $\gamma$ does not affect the direction of the induced rotational velocity, since the quantity in the square bracket in Eq.~(\ref{eq:orientation}) is always positive. To focus on a specific example, \textit{E. coli} bacteria are pushers ($\alpha>0$) and prolate cells ($\gamma \gg 1$). For \textit{E. coli}, the sign of the rotational velocity is therefore given by the sign of $-\cos \theta \sin \theta$ (see Eq.~\ref{eq:orientation}). We thus get that an  \textit{E. coli} cell is always  re-oriented in the direction parallel to the wall \cite{Berke2008} (Fig.~\ref{fig:WallFarField}$\b$). When $0 \le \theta \le \pi/2$ (resp.~$\pi/2 \le \theta \le \pi$), the rotational velocity is negative (resp.~positive), bringing the cell back to an orientation  parallel to the wall. On the other hand, hydrodynamic interactions are expected to re-orient a puller in the direction perpendicular to the surface (Fig.~\ref{fig:WallFarField}$\c$). Berke \textit{et al.} \cite{Berke2008} investigated these hydrodynamic interactions and proposed that  they are  responsible for the experimentally-observed accumulation of swimming bacteria near surfaces.

\subsubsection{Interaction between swimmers}

\begin{figure}[t]
\begin{center}
\includegraphics[width=0.8\textwidth]{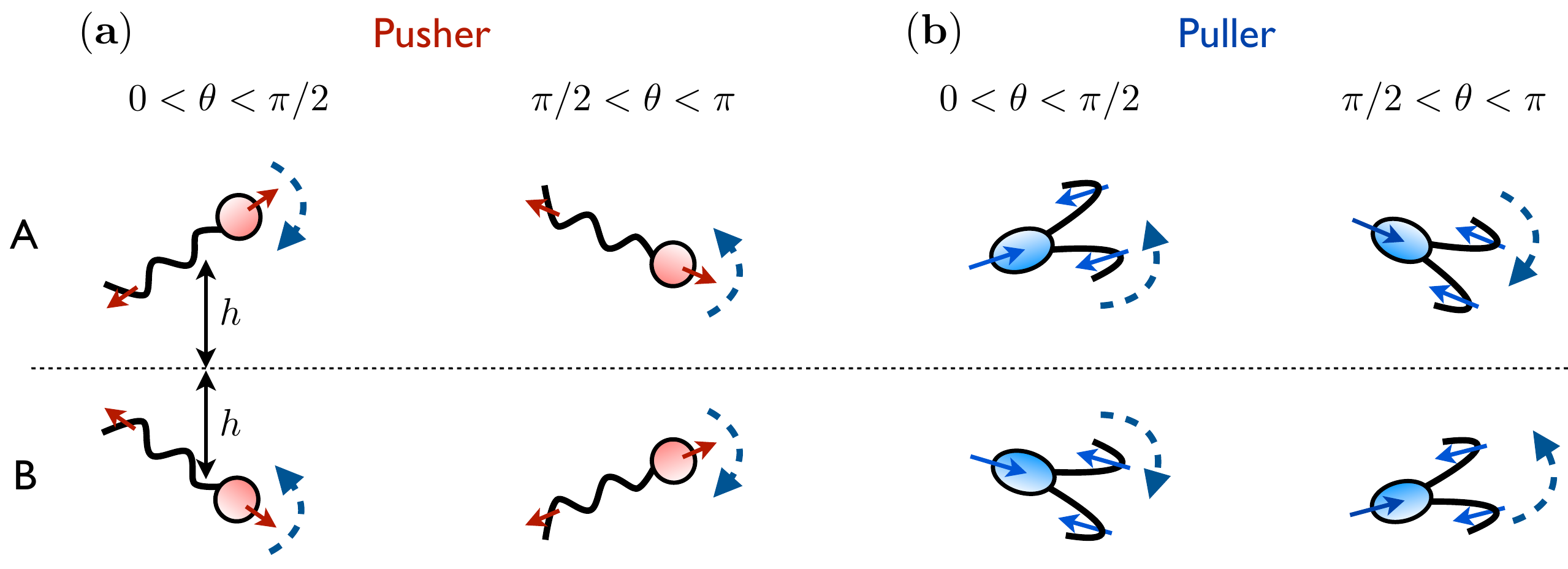}
\caption{Hydrodynamic interactions between two identical swimmers oriented side-by-side. ($\mathbf{a}$): Pushers are reoriented  to be perpendicular to their line of centers.   ($\mathbf{b}$): In contrast pullers are reoriented to be parallel to their line of centers. }
\label{fig:TwoSwimmers}
\end{center}
\end{figure}

We can use  similar calculations  to draw conclusions on the hydrodynamic interactions between two identical swimmers. Consider swimmer A and its mirror-image swimmer B, separated by a distance $2h$ (Fig.~\ref{fig:TwoSwimmers}). By symmetry, the effect of swimmer A on swimmer B is  identical to  that from  B on  A. The  translational and rotational velocities induced on swimmer A by  the flow created by swimmer B are found using  Fax\'{e}n's law applied to the flow generated by swimmer B, which is simply the mirror image of a Stokes dipole, leading to
\begin{subequations}
\begin{align}
\U &= -\frac{\alpha}{32\pi\mu h^2} (1-3\cos^2 \theta) \e_y, \\
\BOmega &= -\frac{3 \alpha \sin \theta \cos \theta}{256 \pi \mu h^3} \left( 1+ \frac{\gamma^2-1}{\gamma^2+1} \cos^2 \theta \right) \e_z . \label{eq:CellsReorient}
\end{align}
\end{subequations}
We obtain effects qualitatively similar  to the case of swimming near a solid boundary.  When two cells swim side-by-side ($\theta = \pi/2$), the induced migration velocity is given by $-\alpha/32\pi\mu h^2 \e_y$, meaning that   hydrodynamic interactions acts to attract  two pushers ($\alpha>0$) and repel two pullers ($\alpha<0$). The opposite holds for the case where two cells swim head on ($\theta = 0$), where the induced velocity changes sign and becomes $\alpha/16 \pi \mu h^2 \e_y$. Regarding the hydrodynamic reorientation, and focusing on prolate cells for simplicity ($\gamma >1$),  we get that  two pushers are reoriented to be perpendicular to their line of centers, and thus end up swimming  parallel to each other (Fig.~\ref{fig:TwoSwimmers}$\a$). On the other hand, two pullers are reoriented so as to be parallel to their line of centers (Fig.~\ref{fig:TwoSwimmers}$\b$).  As a final remark, we note that the configuration in Fig.~\ref{fig:TwoSwimmers}  is physically equivalent to swimmer A swimming  at a distance $h$ from  a flat stress-free surface, since in that case   the image singularity required to satisfy the surface condition is simply the mirror image of the Stokes dipole.

\subsection{Flagellar synchronization}\label{sec:Syn}

Continuing on topics involving hydrodynamic interactions, we investigate in this section the  experimental observation of  flagellar  synchronization for cells swimming in close proximity \cite{Woolley2009,Hayashi1998,Riedel2005,Yang2008} (Fig.~\ref{fig:SynExp}). As a first modelling approach to the problem, Taylor \cite{Taylor1951} studied two swimming sheets with identical, prescribed, waveforms and showed that the energy dissipated between the sheets (and equal  to the rate of working of the swimmers)  is minimized when the two sheets swim in phase. Recently, Elfring and Lauga \cite{Elfring2009} revisited the dynamics of this problem and showed that front-back asymmetry of the flagellar waveform is required for synchronization to dynamically occur. 

This geometrical requirement can be shown using a combination of symmetry arguments and kinematic reversibility. Consider two identical  sheets swimming using  travelling waves of deformation (the usual Taylor model). Assume that their waveforms have  up-down and front-back symmetry  (such as a pure sinewave) and that they are are positioned with respect to each other so as to have a finite phase difference, as shown in Fig.~\ref{fig:Syn}$\a$. Without loss of generality, suppose that a pair of nonzero stabilizing forces, $f$ and $-f$,  act on each swimmer in the direction  to bring the phase difference to zero. One reflection about the vertical axis leads to the configuration shown in Fig.~\ref{fig:Syn}$\b$, and a second reflection about the horizontal axis leads to the configuration shown in Fig.~\ref{fig:Syn}$\c$. By kinematic reversibility (see Sec.~\ref{sec:KinRev}), we can reverse the direction of the wave propagation in each swimmer, which reverses the sign of the pair of forces to $-f$ and $f$, leading to the configuration shown in Fig.~\ref{fig:Syn}$\mathbf{d}$. The setups in Figs.~\ref{fig:Syn}$\a$ and $\mathbf{d}$ are identical but are subject to equal and opposite forces. We therefore conclude that these forces cannot exist, and thus $f=0$ for doubly-symmetric  waveforms. Thus, only waveforms with broken geometrical symmetries have any hope of synchronizing.  We reproduce below  the lubrication analysis by Elfring and Lauga \cite{Elfring2009} detailing the synchronization dynamics.

\begin{figure}[t]
\begin{center}
\includegraphics[width=0.7\textwidth]{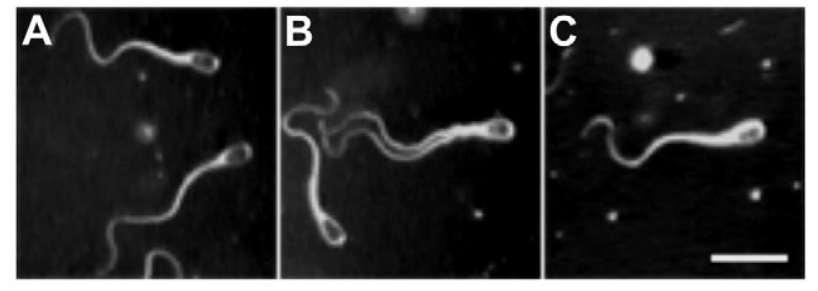}
\caption{Time sequence of the synchronization of two swimming bull spermatozoa.  Scale bar, $25\mu$m. Reprinted with permission from Woolley \textit{et al.} \cite{Woolley2009}. Copyright \copyright 2009 The Company of Biologists Ltd.}
\label{fig:SynExp}
\end{center}
\end{figure}

\subsubsection{Formulation}

We consider two identical swimming sheets whose waveform is described by the function $ag(z)$, where $a$ is the wave amplitude and $g$ its  dimensionless shape (Fig.~\ref{fig:Syn2}, left). In order to focus on cells swimming along straight trajectories, we adopt waveforms possessing reflection symmetry about the horizontal axis $g(z+\pi) = -g(z)$ \cite{Goldstein1977}, but not front-back symmetry. Suppose the bottom sheet (\# 1) swims with a velocity $-U$ (thus to the left of Fig.~\ref{fig:Syn2}) and the top sheet  (\# 2) swims with a different velocity $-U+U_\Delta$. In a frame moving with the  sheet \# 1, its position  is given by $Y_1 = ag(kz-\omega t)$, propagating a wave to the right at a phase speed $c=\omega/k$. Sheet  \# 2, situated at a mean distance $\bar{h}$ above and parallel to the  sheet  \# 1, moves at a velocity $U_\Delta$ to the right relative to the bottom sheet. Its instantaneous position is given by $Y_2 = \bar{h} + ag(kz-\omega t+ \phi)$, where $\phi(t)$ is the phase difference between the two waveforms. Note the explicit time dependence in this waveform that arises from the difference in swimming velocities of the two sheets, $U_\Delta$. Geometrically, the phase is related to the history of the difference in swimming as  $\phi(t) = \tilde{\phi} - k\int_0^t U_\Delta(t') dt'$, where $\tilde{\phi}$ is the initial phase difference. The cases $\phi = 0$ and $\phi = \pi$ are referred as the in-phase and opposite-phase configurations respectively (Fig.~\ref{fig:Syn2}, right). 

The key question to address  is whether or not the two sheets can attain synchronized states \textit{i.e.}~if  the phase difference, $\phi(t)$, dynamically reaches  a steady value. The evolution of the phase difference respects the overall force-free condition. If  however the sheets were not allowed to undergo relative motion, they would be subject to hydrodynamic forces, whose signs and magnitudes would govern  the physics of synchronization. We  therefore first focus on the values of the  hydrodynamic force acting on the sheets. We analyze the problem in the lubrication limit, where the distance between the two sheets is small compared with the wavelength, $\delta = k \bar{h} \ll 1$.

\begin{figure}[t]
\begin{center}
\includegraphics[width=0.9\textwidth]{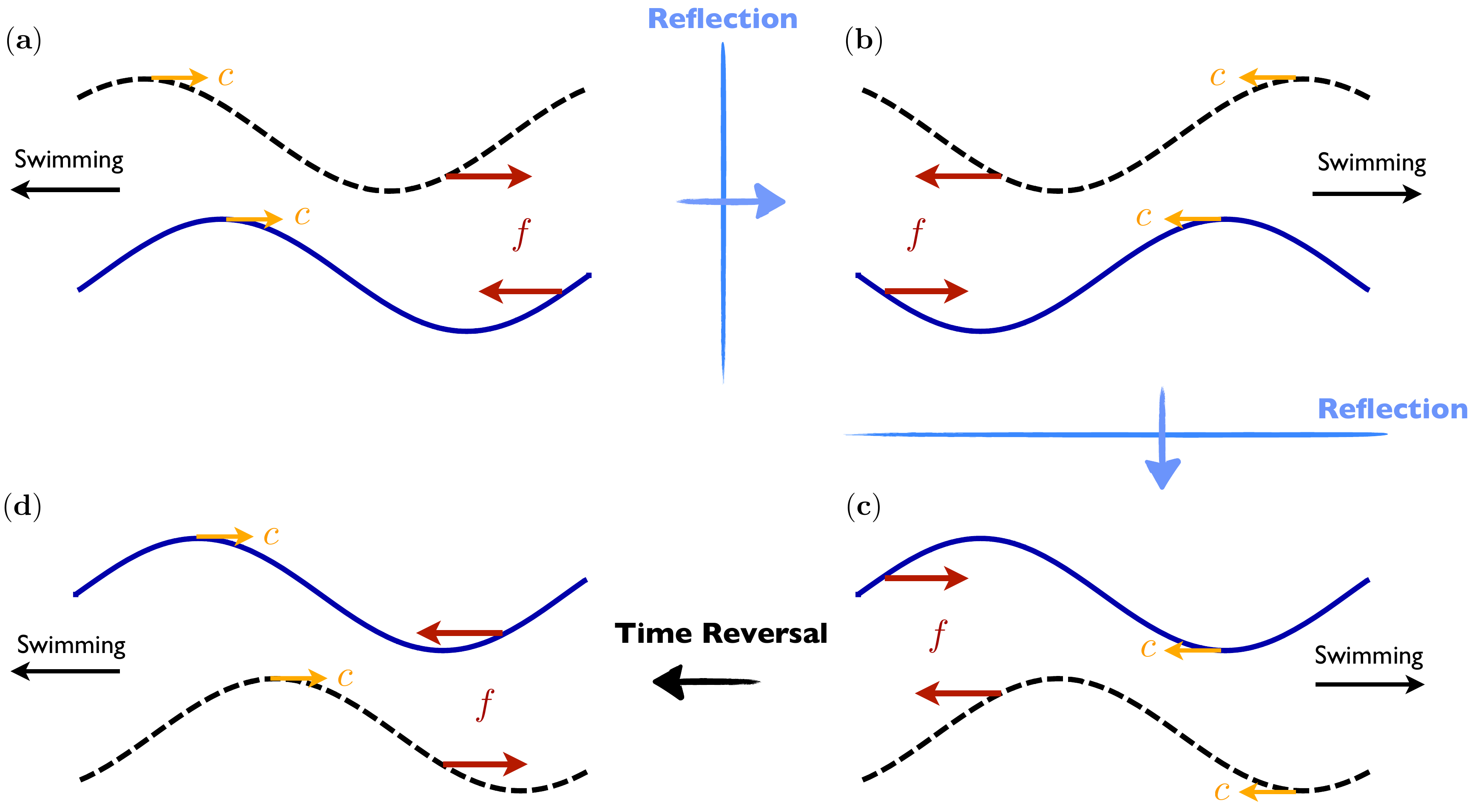}
\caption{Swimmers whose waveforms  have two reflection symmetries cannot synchronize ($f=0$).}
\label{fig:Syn}
\end{center}
\end{figure}

\subsubsection{Non-dimensionalization}
As in Sec.~\ref{sec:lubrication}, we non-dimensionalize time as $t^* = t\omega$, the horizontal length as $z^* = zk$, and the horizontal velocity as $u^* = u/c$. The only difference is that here we scale the vertical length by the mean separation distance between the two sheets, $y^* = y/\bar{h}$. The vertical velocity is then non-dimensionalized as $v^* = v/(\delta c)$. The dimensionless instantaneous positions of the sheets are thus now  $Y_1^* =a^*g^*(x^*)$ and $Y^*_2 = 1+a^*g^*(x^*+\phi^*)$, where $x^*= z^*-t^*$ is the wave variable and  $\phi^* = \tilde{\phi}^* - \int_0^{t^*} U_\Delta^*(t^{'*})dt^{'*}$.The stars represent dimensionless variables and are dropped for convenience hereafter. All the variables below are dimensionless unless otherwise stated.  As a general feature of lubrication theory, forces on the sheet increase as inverse powers of $\delta$, and  dominate the forces from the fluids located on the other side of the sheets. We therefore only consider the flow between the two sheets and ignore the outer problem in the analysis below. Similarly to Sec.~\ref{sec:lubrication}, the governing equations in this lubrication limit are  given by Eq.~(\ref{eqn:lubGov}) in dimensionless variables.

\subsubsection{Lubrication analysis}

\begin{figure}[t]
\begin{center}
\includegraphics[width=0.9\textwidth]{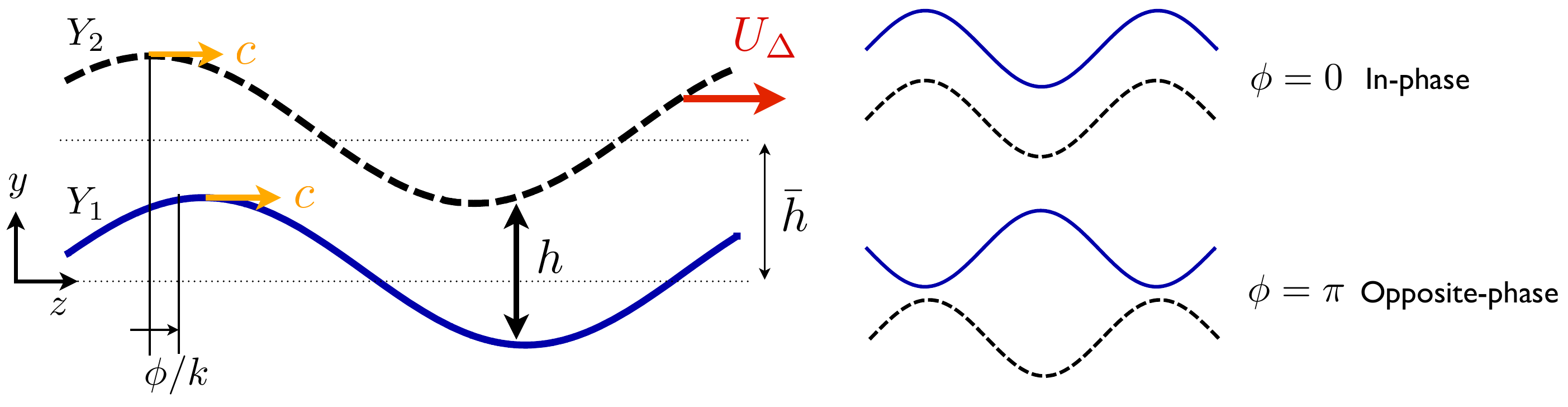}
\caption{Geometrical setup for the synchronization of two swimming sheets \cite{Elfring2009}.}
\label{fig:Syn2}
\end{center}
\end{figure}

The boundary conditions require some thought since we are in a frame moving at a velocity $-U+c$, where the shape of the bottom sheet appears to be stationary.  Under this frame, the material points on the bottom and top sheets have a horizontal velocity $-c$ and $-c + U_\Delta$ respectively. The boundary conditions in dimensionless variables are therefore given by
\begin{subequations} \label{eqn:SynBC}
\begin{align}
u(x, y=Y_1) &= -1, \label{eqn:SynBC1}\\
v(x, y=Y_1) &= \frac{\partial Y_1}{\partial t} = - \frac{dY_1}{dx},\label{eqn:SynBC2} \\
u(x, y=Y_2) &= -1 + U_\Delta, \label{eqn:SynBC3}\\
v(x, y=Y_2) &= \frac{\partial Y_2}{\partial t}=  - \frac{dY_2}{dx}\cdot \label{eqn:SynBC4}
\end{align}
\end{subequations}
We then proceed by the standard procedures in lubrication theory (recall Sec.~\ref{sec:lubrication}) to obtain the velocity field $u(x,y)$ by integrating the $x$-momentum equation, Eq.~(\ref{eqn:lubMomX}), twice with the boundary conditions, Eqs.~(\ref{eqn:SynBC1}) and (\ref{eqn:SynBC3})
\begin{align}
u(x,y) = \frac{1}{2} \frac{d p}{d x} (y-Y_1)(y-Y_2) + U_\Delta \frac{y-Y_1}{Y_2-Y_1}-1 . \label{eqn:synVelocity}
\end{align}
We then integrate the continuity equation, Eq.~(\ref{eqn:lubContinuity}), in $y$ between the sheets to obtain
\begin{align}
\int_{Y_1}^{Y_2} \frac{\partial u}{\partial x} dy + v(Y_2) - v(Y_1) = 0 .
\end{align}
Applying the Leibniz's rule and the boundary conditions, Eq.~(\ref{eqn:SynBC}), we have
\begin{align}
\frac{d}{dx} \left(  \int_{Y_1}^{Y_2} u dy \right) = U_\Delta \frac{dY_2}{dx} \cdot \label{eqn:synReynolds}
\end{align} 
The integral can be evaluated with Eq.~(\ref{eqn:synVelocity}) to give
\begin{align}
\int_{Y_1}^{Y_2} u dy = - \frac{1}{12} \frac{dp}{dx} h^3-h \left(1 - \frac{U_\Delta}{2} \right) \label{eqn:synFlowRate},
\end{align}
where $h(x)= Y_2-Y_1$. Substituting Eq.~(\ref{eqn:synFlowRate}) into Eq.~(\ref{eqn:synReynolds}), we obtain the Reynolds equation
\begin{align}
- \frac{1}{12} \frac{d}{dx} \left( \frac{dp}{dx}h^3 \right)= \frac{d}{dx} \left[h+ U_\Delta \left(Y_2 - \frac{h}{2} \right) \right],  \label{eqn:synReynolds2}
\end{align}
which we will solve to obtain the pressure gradient necessary to compute the hydrodynamic force acting on the sheets.

\subsubsection{Possible synchronized states}
Let us  focus on  possible synchronized states and their stability. A
 true synchronized state, $\phi= \phi_0$, should be a fixed point of the dynamics. We should thus get that $d\phi/dt = 0$,   meaning no relative motion between the sheets, $U_\Delta = 0$, and zero net force.  We  enforce the condition $U_\Delta =0$ and determine the value(s) of $\phi_0$ satisfying the overall force-free condition.  With $U_\Delta = 0$, an integration of  the Reynolds equation, Eq.~(\ref{eqn:synReynolds2}), leads to the pressure gradient
\begin{align}
 \frac{dp}{dx}= -12 \left(\frac{q}{h^3}+\frac{1}{h^2} \right),
\end{align}
where $q$ is a numerical constant. We determine this unknown constant $q$ by enforcing the periodicity of the pressure field $\int_0^{2\pi} dp/dx  \ dx =0$, which implies $q=-I_2/I_3$, where $I_n = \int_{0}^{2\pi} h^{-n} dx$. The pressure gradient is therefore given by
\begin{align}
\frac{dp}{dx} = 12 \left( \frac{I_2}{h^3 I_3} - \frac{1}{h^2}  \right) \cdot \label{eqn:synPressure}
\end{align}
Similarly to Eq.~(\ref{eqn:lubForceFree}), the leading-order hydrodynamic force on the top sheet is given by
\begin{align}
f_x = \int_0^{2\pi} \left( Y_2 \frac{dp}{dx} - \frac{\partial u}{\partial y} \right) \bigg|_{y=Y_2} dx. \label{eqn:synForceGeneral}
\end{align}
Substituting the velocity field, Eq.~(\ref{eqn:synVelocity}), and the value for the  pressure gradient, Eq.~(\ref{eqn:synPressure}), into Eq.~(\ref{eqn:synForceGeneral}), we arrive at the expression of the hydrodynamic force acting on the top sheet
\begin{align}
f_x = 6 a \int_0^{2\pi} \left( \frac{I_2}{h^3 I_3} - \frac{1}{h^2}  \right) \left [ g(x+\phi) + g(x) \right] dx, \label{eqn:synForce0}
\end{align}
when there is no relative motion between the sheets ($U_\Delta = 0$).
From Eq.~(\ref{eqn:synForce0}), we can see that the force vanishes at $\phi =0$ and $\phi=\pi$. For $\phi =0$ (in-phase configuration), the separation distance $h$ becomes a constant, leading to $I_2/(h^3I_3) - 1/h^2 = 0$. For $\phi = \pi$ (opposite-phase configuration), $g(x+\pi)+g(x) = 0$ due to the symmetry required for cells swimming along straight lines. The two possible synchronized states are therefore the in-phase ($\phi_0 =0$) and opposite-phase ($\phi_0=\pi$) configurations.

\subsubsection{Stability of synchronized states}
We proceed to evaluate the stability of these fixed points by expanding the hydrodynamic force, Eq.~(\ref{eqn:synForce0}), about the fixed points as: $\phi = \phi_0 + \phi'$, where $\phi' \ll 1$ represents a small perturbation. Expanding about the in-phase fixed point ($\phi_0 = 0$), we obtain the force
\begin{align}
f_{0} \approx -72 a^4  \phi'^3 \int_0^{2\pi} g(x) g'(x) ^3 dx . \label{syn:FixedStab1}
\end{align}
Similarly, expanding the force about the opposite-phase fixed point ($\phi_0 = \pi$) and assuming a small amplitude wave ($a \ll 1$), the force is given by
\begin{align}
f_{\pi} \approx 72 a^4 \phi'^3 \int_0^{2\pi} g(x) g'(x) ^3 dx . \label{syn:FixedStab2}
\end{align}
We can see that the integral 
\begin{align}
A = \int_0^{2\pi} g(x) g'(x)^3 dx \label{A}
\end{align}
in Eqs.~\ref{syn:FixedStab1} and \ref{syn:FixedStab2}, which depends only on the geometry of the waveform, solely dictates the stability of the fixed points. Furthermore, for a given waveform, the two fixed points have always opposite stability as shown by the difference in sign between Eq.~(\ref{syn:FixedStab1}) and Eq.~(\ref{syn:FixedStab2}). If $A<0$ (resp.~$A>0$),  then the  in-phase (resp.~opposite-phase) configuration is stable while the other one is unstable.

\subsubsection{Evolution of the phase difference}

Although the  synchronized states have been determined, we have yet to solve for the dynamic  evolution of the phase difference, $\phi(t)$, towards the synchronized state. This requires relaxing the condition $U_\Delta =0$ and instead solving for the value of $U_\Delta$ leading to free swimming. We thenÊ geometrically integrate $ d\phi(t)/dt=-U_\Delta $ in a quasi-static fashion. To proceed, we integrate Eq.~(\ref{eqn:synReynolds2}) to obtain the pressure gradient as
\begin{align}
\frac{dp}{dx} = \frac{6 U_\Delta -12}{h^2} - \frac{12 U_\Delta Y_2 +C}{h^3}, \label{eqn:synPressureFree}
\end{align}
where $C$ is an integration constant. By enforcing the periodicity of the pressure field, the constant is determined to be $C=\left[ 6 U_\Delta (I_2-2K)-12 I_2  \right]/I_3$, where $K = \int_0^{2\pi} Y_2 h^{-3} dx$. 
\begin{figure}[t]
\begin{center}
\includegraphics[width=0.8\textwidth]{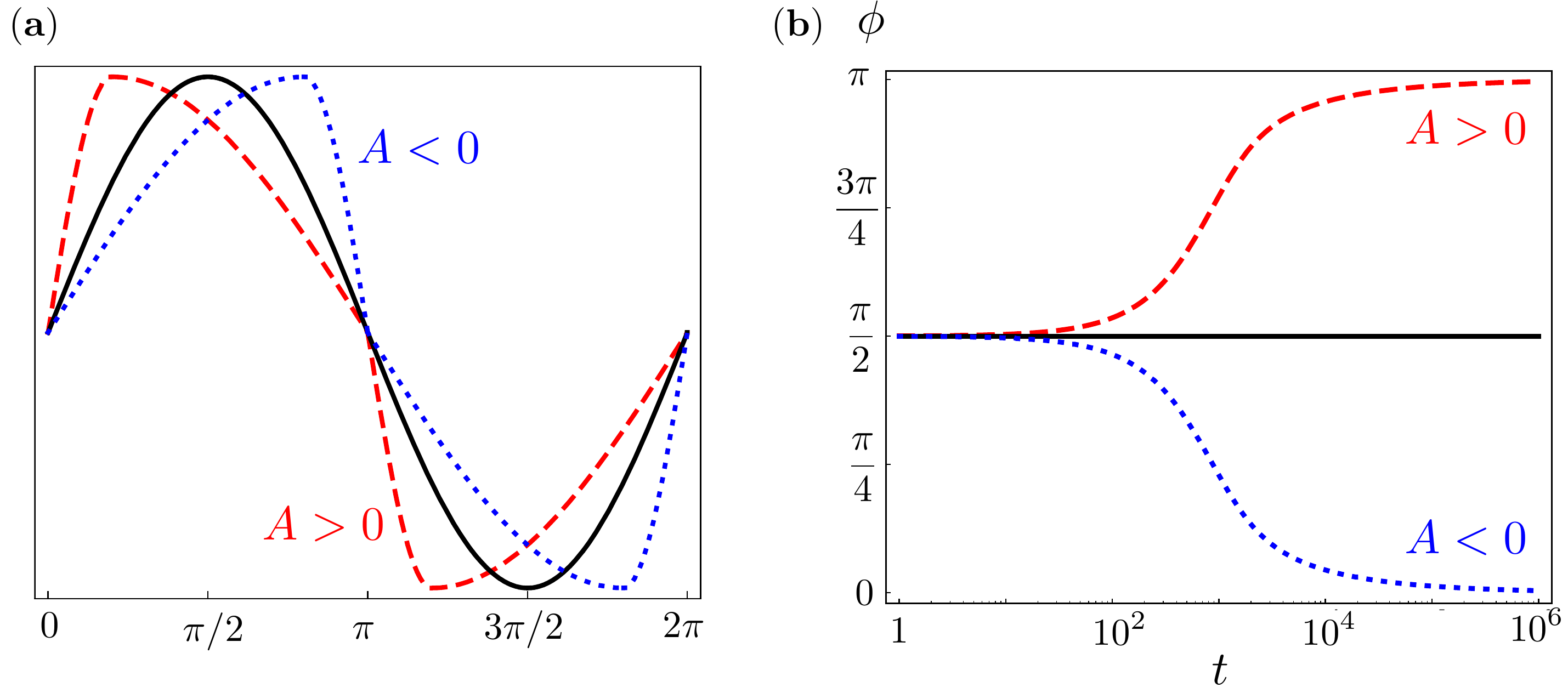}
\caption{The synchronization of two identical swimming sheets. 
($\mathbf{a}$): Skewed sinusoidal waveforms corresponding to $A>0$ (red dashed line) and $A<0$ (blue dotted line). The black line corresponds to an un-skewed sinewave. ($\mathbf{b}$) Evolution of the phase difference, $\phi(t)$, for different flagellar waveforms.}
\label{fig:Syn3}
\end{center}
\end{figure}
Substituting the new expression of the pressure gradient, Eq.~(\ref{eqn:synPressureFree}), and the velocity field, Eq.~(\ref{eqn:synVelocity}),  into Eq.~(\ref{eqn:synForceGeneral}) to compute the leading-order horizontal hydrodynamic force on the top sheet, we have
\begin{align}
f = \int_0^{2\pi} \left[ \frac{1}{2}  \left(   \frac{6 U_\Delta -12}{h^2} - \frac{12 U_\Delta Y_2 +C}{h^3} \right) \left(Y_2 + Y_1 \right) - \frac{U_\Delta}{h}  \right] dx. \label{eqn:SynBig}
\end{align}
Finally, by enforcing the force-free condition, $f=0$, we can solve Eq.~(\ref{eqn:SynBig}) for $U_\Delta$ as
\begin{align}
U_\Delta = - \frac{d \phi}{dt}= 6 \int_0^{2\pi} \left( \frac{I_2}{h^3I_3 }-\frac{1}{h^2} \right) dx \bigg/ \int_{0}^{2\pi} \left[ \frac{1}{h}-3 \left( \frac{1}{h^2}- \frac{2 Y_2 I_3 + I_2-2K}{I_3 h^3}  \right) (Y_1+ Y_2) \right] dx . \label{eqn:SynPhi}
\end{align}
Notice that, as  expected,  Eq.~(\ref{eqn:SynPhi}) reduces to Eq.~(\ref{eqn:synForce0}) in steady state ($U_\Delta =0$). Supplying an initial phase difference $\phi(t=0) = \tilde{\phi}$, Eq.~(\ref{eqn:SynPhi}) can be computed  to obtain the time-evolution of the phase difference $\phi(t)$. For illustration, Elfring and Lauga \cite{Elfring2009} considered two skewed sinusoidal waveforms shown in  Fig.~\ref{fig:Syn3}$\a$, which correspond to opposite  signs for the constant $A$ (Eq.~\ref{A}). By numerically integrating Eq.~(\ref{eqn:SynPhi}) with the initial condition $\tilde{\phi}=\pi/2$, the phase $\phi(t)$ is shown in Fig.~\ref{fig:Syn3}$\b$, where the skewed sine function corresponding to $A>0$ (red dashed line) attains an opposite-phase synchronized state, whereas the other skewed sine function having $A<0$ (blue dotted line) ends up an in-phase synchronized state. As expected from the analysis, an un-skewed sinusoidal function (black solid line) maintains its initial phase difference for all time.

\subsubsection{Energy dissipation}
We close by briefly remarking on  energy dissipation  \cite{Taylor1951,Elfring2009}. As first considered by Taylor\cite{Taylor1951}, one can calculate the rate of energy dissipation in the volume of fluid between the sheets at the synchronized states.  The global maximum and minimum of energy dissipation occur  respectively at $\phi = 0$ and $\pi$, independently of the waveform. In contrast, as shown in the  analysis above, the stability of the fixed points are dictated by the waveform geometry through the integral $A$ (Eq.~\ref{A}). Therefore, there is no relationship  between the configuration yielding the minimum energy dissipation and the location of a stable fixed point. In particular, two sheets can be forced into a stable configuration at which the energy dissipation is indeed the maximum  -- specifically, all geometries such that  $A>0$ which  lead to $\phi = \pi$. 
In other words, the system here does not always adopt a state with minimized energy dissipation.

\section{Swimming with Elasticity}
In this section, we consider the effects of  elasticity on inertialess locomotion. Elasticity can be present in the deforming body (\textit{e.g.}~flexible flagellum) or in the fluid medium (\textit{e.g.}~polymeric fluids). We first illustrate the roles of elasticity of the body in generating propulsive thrust at low Reynolds number (Sec.~\ref{sec:ElasticityBody}) and then turn our attention to the first effects of viscoelasticity on microscopic locomotion (Sec.~\ref{sec:ElasticityFluid}).

\subsection{Flagellar elasticity}\label{sec:ElasticityBody}

In the previous sections, we assumed the flagellar waveforms were prescribed and computed the resulting swimming kinematics. However, exactly how the flagellar waveform is actuated and maintained is an important question to address \cite{Machin1958}. Several actuation mechanisms have been elucidated, and they can be categorized into boundary and distributed actuations \cite{Lauga2009}. For boundary actuation, the flagellum is driven by a motor at its base and the rest of the flagellum is passive; this is the situation arising in bacterial flagella.  In the case of distributed actuation relevant to eukaryotic flagella and cilia, there are molecular (dynein) motors distributed all along the flagellum that cause  microtubules  to slide, resulting in bending and propulsion. Extensive theoretical studies have been carried out to determine  the flagellar waveform and propulsion velocity resulting from   distributed actuation  \cite{Camalet2000,Fu2007,Fu2008,Kruse2007,Evans2010, Gadelha2010}. Here we focus on a classical calculation of boundary actuation to illustrate the role of flagellar elasticity in enabling locomotion \cite{WigginsGoldstein}.

Specifically, we consider an elastic filament  wiggled periodically at one end, and we investigate how the  filament flexibility enables  the development of non-reciprocal kinematics  and  propulsion. The elastic filament has a  cross-sectional radius $r$ and total length $L$ described by the position vector $\r(s,t) = [x(s, t),y(s, t),z(s, t)]$, where $s$ denotes the arc-length along the filament (see Fig.~\ref{fig:RFT}$\a$). The filament is actuated on one end harmonically with an angular frequency $\omega$. In contrast to previous sections, where $\r(s,t)$ is prescribed, here the deformation is an unknown function of space and time to be determined by balancing viscous and elastic forces on the filament. In order to write down the force balance and determine the shape, we first need descriptions of the viscous and the elastic forces, which are given, respectively, by slender body theory from hydrodynamics and Euler-Bernoulli beam theory from elasticity.

\subsubsection{Hydrodynamics: Slender body theory}
We assume the filament  is sufficiently slender ($L \gg r$) that we can apply  slender body theory (Sec.~\ref{sec:RFT}) to describe  hydrodynamic forces. Similarly to Sec.~\ref{sec:RFT}, we use only the leading-order theory (resistive force theory) and ignore hydrodynamic interactions between distinct parts of the filament. This local theory was shown to be quantitatively correct for gentle distortions of the filament shape \cite{Gray1955, Wiggins1998, Yu2006, Kruse2007}. The local viscous force  per unit length acting on the filament is thus given by (Eq.~\ref{eqn:RFT})
{\begin{align}
\f_{\text{vis}}(s)  &= - \left[ \xi_{\parallel} \t\t+\xi_{\perp} (\I-\t\t) \right] \cdot \v = - \left[ \xi_\perp \I + (\xi_\parallel - \xi_\perp) \t \t \right] \cdot \v, \label{eqn:ElastoHydroHydro}
\end{align}
where $\t(s,t) \equiv  \r_s (s,t)$ is the local tangent vector to the filament and $\v(s) \equiv \r_t (s,t)$  the local velocity of the filament (assuming no background flow). }

\subsubsection{Elasticity: Euler-Bernoulli beam theory}
{We consider here an elastic and inextensible filament. When the filament is deformed, elastic bending and tensile forces arise trying to minimize the energy and restore the filament to its undeformed shape. These elastic forces can be obtained by taking a variational derivative of the energy functional, $\mathcal{E}=\frac{1}{2}\int_0^L \left[ A  \left( \r_{ss} \cdot \r_{ss} \right) + \sigma (\r_s \cdot \r_s) \right] ds$, where $A=EI$ is  the filament bending stiffness (the product of Young modulus $E$ and the second moment of area $I$), and $\sigma(s, t)$ is the Lagrange multiplier (tension) enforcing local inextensibility, $(\r_s \cdot \r_s)_t = 0$. The elastic force, per unit length, is then given by
\begin{align}
 \f_{\text{elastic}}(s) = - A \r_{ssss} + \left(\sigma \r_s \right)_s, \label{eqn:ElastoHydroElasto}
\end{align}
and the reader is referred to classical monographs for its derivation \cite{landau_lifshitz_elas}. Note that here we study a planar problem, where no twisting along the filament occurs. In three-dimensional problems, the energy cost due to twisting can enter the energy functional and contribute another restoring force \cite{Wolgemuth2000, Powers2010}.}

\subsubsection{Elastohydrodynamics}
Since we are in the low Reynolds number regime, inertial forces are negligible. Hence, the local viscous  force density, Eq.~(\ref{eqn:ElastoHydroHydro}), balances the local elastic force, Eq.~(\ref{eqn:ElastoHydroElasto}), $\f_{\text{vis}} + \f_{\text{elastic}}=\mathbf{0}$, yielding the equation governing the elastohydrodynamics of the filament
\begin{align}
 \left[ \xi_\perp \I + (\xi_\parallel - \xi_\perp) \r_s \r_s \right] \cdot \r_t = - A \r_{ssss} + \left(\sigma \r_s \right)_s. \label{eqn:FullElastohydro}
\end{align}
Upon the supply of appropriate boundary conditions, the equation can be solved to obtain the resulting deformation, $\r(s,t)$, along the filament.

\subsubsection{Non-dimensionalization}

We now non-dimensionalize the variables and equations in order to identify the relevant dimensionless parameters governing the physics of this problem. We scale lengths by $L$, times by the actuation frequency $\omega^{-1}$,  velocities by $L \omega$, {and forces by $A/L^2$}. Using the same symbols for simplicity, the dimensionless elastohydrodynamic equation now reads
\begin{align}
\left[ \I + \left( \gamma^{-1}-1 \right) \r_s \r_s \right] \cdot \r_t =  \Sp^{-4} \left[- \r_{ssss} + \left( \sigma \r_s\right)_s  \right], \label{eqn:ElastoHydroNon}
\end{align}
and two dimensionless groups appear: the drag anisotropy ratio $\gamma = \xi_{\perp}/\xi_{\parallel}$ we encountered in Sec.~\ref{sec:RFT}, and the so-called Sperm number, $\Sp = L \left( \xi_{\perp} \omega / A\right)^{1/4}$, which characterizes the relative influence of the viscous and bending forces \cite{WigginsGoldstein,Lowe2003}. All  variables hereafter are dimensionless unless otherwise stated.

{There are two unknowns in Eq.~(\ref{eqn:ElastoHydroNon}), namely $\r(s,t)$ and $\sigma(s,t)$. The equation for $\sigma(s,t)$ is obtained from the inextensibility condition, $(\r_s \cdot \r_s)_t = 0$, which implies $\r_s \cdot \r_{ts} = 0$. In order to apply this condition, we first invert Eq.~(\ref{eqn:ElastoHydroNon}) to obtain
\begin{align}
\r_t = \Sp^{-4}\left[ \I + \left(\gamma -1 \right) \r_s \r_s  \right] \left[ -\r_{ssss} + \left( \sigma \r_s\right)_s  \right],
\end{align}
which can be simplified as 
\begin{align}
\r_t = -\Sp^{-4}\left[ -\r_{ssss} + \sigma_s \r_s + \sigma \r_{ss} + \left(\gamma -1 \right) \r_s \left( -\r_s \cdot \r_{ssss} +\sigma_s  \right) \right], \label{eqn:ObtainTension}
\end{align}
noting that $\r_s \cdot \r_{ss} = 0 $ by differentiating the relation $\r_{s} \cdot \r_{s} = 1$ once. We then differentiate Eq.~(\ref{eqn:ObtainTension}) and take the inner product with $\r_s$ to apply the condition $\r_s \cdot \r_{ts} = 0$, resulting in the equation for the Lagrange multiplier 
\begin{align}
\gamma \sigma_{ss} - \sigma (\r_{ss} \cdot \r_{ss}) + 7 \r_{ss} \cdot \r_{ssss} + 6 \r_{sss} \cdot \r_{sss} = 0, \label{tension}
\end{align}
while the following identities (obtained by repeated differentiation of the identity $\r_s \cdot \r_s = 1$)  have been used for simplification: $\r_s \cdot \r_{ss} = 0$, $\r_s \cdot \r_{sss} = - \r_{ss} \cdot \r_{ss}$, $\r_s \cdot \r_{ssss} = -3 \r_{ss} \cdot \r_{sss}$, and $\r_{s} \cdot \r_{sssss} = -4 \r_{ss} \cdot \r_{ssss} - 3 \r_{sss} \cdot \r_{sss}$. Eq.~(\ref{tension}) is then a second-order differential equation for the Lagrange multiplier $\sigma(s,t)$, together with Eq.~(\ref{eqn:ElastoHydroNon}), forming a coupled system of equations for the elastohydrodynamic problem.}

\subsubsection{Deformation of the filament}
In order to model the boundary actuation by a motor at the base of the flagellum, we assume for simplicity that  we sinusoidally oscillate vertically one end of the filament, $s=0$, with a dimensionless amplitude $\epsilon$ while the  other end of the filament, $s=L$, is free. Given this actuation, the motion of the filament is confined to the $x-y$ plane, $\r(s,t) = [x(s,t), y(s,t), 0]$. {The coupled nonlinear elastohydrodynamic equations for the shape $\r(s,t)$ and $\sigma(s,t)$, Eqs.~(\ref{eqn:ElastoHydroNon}) and (\ref{tension}), can be solved numerically with prescribed boundary conditions. In order to make analytical progress, we assume that the amplitude of actuation is small $\epsilon \ll 1$. With this approximation, we have  $s \approx x$ and the position vector can be approximated as $\r \approx x \ \e_x + y(x,t) \ \e_y$. The leading-order local velocity of the filament is then given by $\v = d\r/dt \approx [0, \partial y/ \partial t]$. Since the boundary actuation is $O(\epsilon)$, we expect $y \sim \epsilon$. From Eq.~(\ref{tension}), we then see that $\sigma \sim \epsilon^2$. As a result, the leading-order balance in the elastohydrodynamic equation, Eq.~(\ref{eqn:ElastoHydroNon}), comes from the $y$-direction and is given by
\begin{align}
\frac{\partial y}{\partial t} = - \Sp^{-4} \frac{\partial^4 y}{\partial x^4}, \label{eqn:ElastoHydroLinear}
\end{align}
a hyper-diffusion equation \cite{Wiggins1998}. Note that the higher-order term $\sigma$ does not appear in the dynamic balance to leading order, significantly simplifying the analysis.} 

At the actuated end, $x = 0$, the vertical displacement is given by $y\mid_{x=0} = \epsilon \cos t$. This end is hinged and hence is free of bending moment, $y_{xx}\mid_{x=0} = 0$.  At the other end of the filament, $x=1$, there is no  force, $y_{xxx}\mid_{x=1} = 0$, and no bending moment, $y_{xx} \mid_{x=1} = 0$. The boundary conditions and their nature (dynamics and kinematic) are summarized in Table \ref{tbl:ElastohydroBC}.

\begin{table}
\small
  \caption{\ Boundary conditions for the boundary-actuated filament}
  \label{tbl:ElastohydroBC}
  \centering
  \begin{tabular*}{0.8\textwidth}{@{\extracolsep{\fill}}ccl}
    \hline
    Location & Boundary conditions & Physical meaning (nature) \\
    \hline
    $x=0$ & $y = \epsilon \cos t$ \ or \  $h = 1$  & Driven (wiggling) end (kinematic) \\
    $x=0$ & $y_{xx} = 0$ \ or \ $h_{xx} = 0$ & Bending moment free (dynamic) \\
    $x=1$ & $y_{xx}=0$ \ or \ $h_{xx}=0$ & Bending moment free (dynamic)\\
    $x=1$ & $y_{xxx}=0$ \ or \ $h_{xxx}=0$ & Force free (dynamic)\\
    \hline
  \end{tabular*}
\end{table}

Because of the linearity of Eq.~(\ref{eqn:ElastoHydroLinear}) and the oscillatory boundary condition at $x=0$, we expect a harmonic solution in time and thus assume a separable solution of the form.
\begin{align}
y = \epsilon \mathcal{R}\left[e^{it} h(x)\right],
\end{align}
where $\mathcal{R}$ denotes taking the real part of a complex number. This allows us to reduce Eq.~(\ref{eqn:ElastoHydroLinear}) into an ordinary differential equation with constant coefficients for $h(x)$ as
\begin{align}
i h = - \Sp^{-4} \frac{d^4h}{dx^4} \cdot
\end{align}
Assuming a solution of the form $h = c e^{k x}$, where $c$ is a constant, we determine that
\begin{align}
k^4 = -i \Sp^4,
\end{align}
which has four roots  as
\begin{align}
k_n = i^n e^{- i\pi/8} \Sp,\quad n = 1,2,3,4 . \label{eqn:kn}
\end{align}
Superimposing these modes, the general solution is given by
\begin{align}
h(x) = \sum_{n=1}^4 c_n e^{k_n x}, \label{eqn:hx}
\end{align}
where $k_n$'s are given by Eq.~(\ref{eqn:kn}), and the constants $c_n$'s are determined from the boundary conditions for $h(x)$ summarized in Table~\ref{tbl:ElastohydroBC} and satisfy the 4-by-4 linear system  
\begin{subequations}
\begin{align}
\sum_{n=1}^4 c_n &= 1,\\
\sum_{n=1}^4 k_n c_n &= 0,\\
\sum_{n=1}^4 k_n^2 e^{k_n} c_n  &= 0,\\
\sum_{n=1}^4 k_n^3 e^{k_n} c_n  &= 0 .
\end{align}
\end{subequations}
The leading-order deformation of the filament, $y(x,t) = \epsilon \mathcal{R}[e^{it} h(x)] = \epsilon \mathcal{R} \left[ \sum_{i=1}^4 c_n e^{k_n x+it} \right]$, is then completely determined upon solving the above simultaneous equations for $c_n$.

\begin{figure}[t]
\begin{center}
\includegraphics[width=1\textwidth]{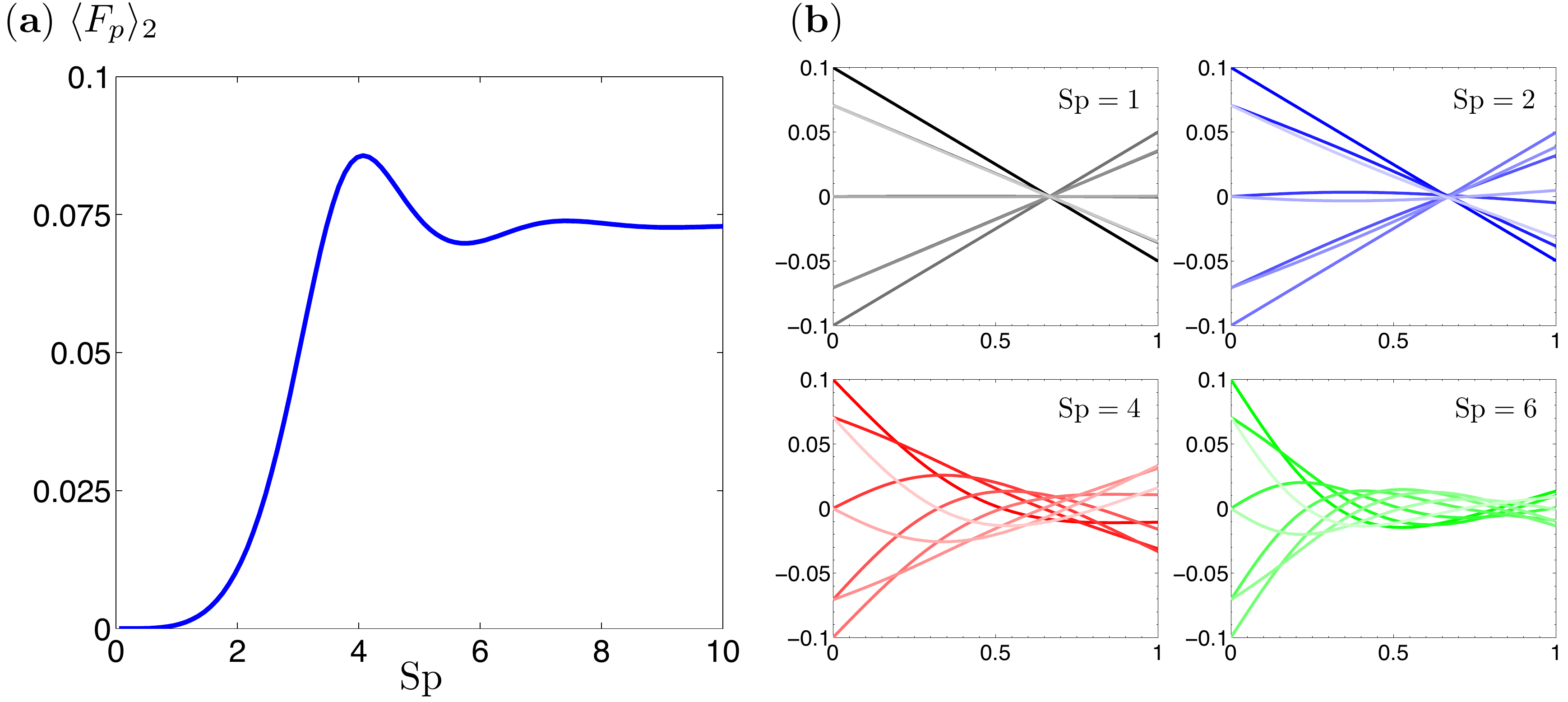}
\caption{Propulsion by the periodic actuation of a flexible filament. 
($\mathbf{a}$) Average dimensionless propulsive force as a function of the sperm number, Sp. ($\mathbf{b}$) Deformations of the filament  over one period ($T=2\pi$) at different times $n\pi/4$, where $n=[1,2, ...,7]$ and different fixed values of sperm numbers ($\rm Sp= 1$, 2, 4, and 6); the intensity of color decreases as time advances and the amplitude is $\epsilon = 0.1$. }
\label{fig:SolidElas}
\end{center}
\end{figure}

\subsubsection{Propulsive force}
We now calculate the propulsive force developed by the actuation at one end of the filament. The hydrodynamic force acting on the filament is given by Eq.~(\ref{eqn:ElastoHydroHydro}), which has the dimensionless form (scaled by $\xi_\perp L \omega$)
\begin{align}
\f_{\text{vis}} &= -\left[\I + \left(\gamma^{-1}-1\right) \t\t  \right] \cdot \v . \label{eqn:ForceDimensionless}
\end{align}
Using the small-amplitude approximation, $\epsilon \ll 1$, we have the leading-order  velocity $\v \approx [0 \ \ y_t]^T$  and tangent $\t \approx [1\ \  y_x]^T$, noting that $y$, $y_x$, and $y_t$ are all $O(\epsilon)$. Substituting these approximations into Eq.~(\ref{eqn:ForceDimensionless}) and keeping only the leading-order terms for each component, we obtain the viscous force as
\begin{align}
\f_{\text{vis}}&\approx -
\begin{pmatrix}
 (\gamma^{-1}-1) y_x y_t \\
y_t
\end{pmatrix}, 
\end{align} and  we note that the $x$ and $y$-components scale as $O(\epsilon^2)$ and $O(\epsilon)$ respectively. These results should be compared with the dimensional form of the viscous force in the swimming problem, Eq.~(\ref{eqn:Similar}). Note also that the $y$-component of the viscous force has been used in the leading-order force balance with the elastic force, Eq.~(\ref{eqn:ElastoHydroLinear}). The leading-order viscous force in the $x$-direction can be rewritten as 
\begin{align}
f_x = - (\gamma^{-1}-1) y_x y_t =  \Sp^{-4 }(\gamma^{-1}-1) y_x y_{xxxx},
\end{align}
where we have used   Eq.~(\ref{eqn:ElastoHydroLinear}). The total viscous force acting  along the filament is therefore given by
\begin{align}
F_x =\int_0^1 f_xdx&=\Sp^{-4 }(\gamma^{-1}-1)  \int_0^{1}  y_x y_{xxxx} dx = \Sp^{-4 }(\gamma^{-1}-1) \left( - y_{x} y_{xxx} + \frac{1}{2} y_{xx}^2 \right)_{x=0}
\end{align}
upon integration by parts and using the boundary conditions, $y_{xxx}\mid_{x=1}=y_{xx}\mid_{x=1} = 0$. Note that this is the force acting on the filament by the fluid. The force required to hold the filament in place, or the propulsive thrust $F_p$, has the same magnitude but opposite direction: $F_p = - F_x$. Averaging the propulsive force over a period of oscillation gives the mean propulsive force
\begin{align}
\langle F_p \rangle =  \frac{(\gamma^{-1}-1)}{2\pi \Sp^4} \int_0^{2 \pi}  \left( y_{x} y_{xxx} - \frac{1}{2} y_{xx}^2 \right)_{x=0} dt = \epsilon^2\langle F_p \rangle_2, 
\end{align}
which is of order $\epsilon^2$ (as indicated by our use of the subscript 2).

The scaled average dimensionless propulsive force, $\langle F_p \rangle_2$, is plotted in Fig.~\ref{fig:SolidElas}$\a$ as a function of the sperm number $\text{Sp}$. We observe that propulsion goes to zero  in the limit $\text{Sp}\ll1$. 
The low-$\text{Sp}$ limit corresponds to the situation where  elastic forces dominate and the filament takes the shape of a pivoting rigid rod (Fig.~\ref{fig:SolidElas}$\b$, $\text{Sp}=1$),  a reciprocal motion unable to  generate  propulsion according to the scallop theorem (Sec.~\ref{sec:scallop}). For non-zero values of Sp, flexibility allows the propagation of waves along the filament, breaking the kinematic reversibility and enabling propulsion \cite{WigginsGoldstein,Wiggins1998,Yu2006,Lauga2007F}. For large values of $\text{Sp}$, viscous forces dominate. The bending of the filament is localized near the actuation point, which can be mathematically shown by the exponential decay of the solution amplitude, Eq.~(\ref{eqn:hx}), when Sp is large \cite{Wiggins1998} (Fig.~\ref{fig:SolidElas}$\b$). {The portion where bending is small contributes little propulsive thrust. We therefore expect a plateau as Sp increases, as seen in Fig.~\ref{fig:SolidElas}.} The optimal value of Sp leading to a maximum propulsive force occurs around $\text{Sp}\approx4$.  It should be remarked that the filament discussed here is held fixed in the horizontal direction and not allowed to move; only the force  required to hold the filament is calculated. Should the filament be free to move, the swimming velocity can be determined by enforcing the overall force-free and torque-free conditions \cite{Lauga2007F}.

\subsection{Fluid Elasticity}\label{sec:ElasticityFluid}

We now turn our attention to elasticity in the fluid medium and its impact on locomotion. Many biological fluids are polymeric and display non-Newtonian rheological properties \cite{Katz1978, Katz1981, Suarez1992, Wilking2011, Fauci2006, Lauga2009}. As a result, the locomotion of microorganisms in viscoelastic fluids  has {recently} attracted considerable interest. Fundamental questions, such as whether fluid elasticity enhances or hinders propulsion, remain under debate \cite{Lauga2007, Fu2009, Teran2010, Shen11, Liu2011, Spagnolie2013}. In this section we illustrate an extension of Taylor's classical swimming sheet calculation, Sec.~\ref{sec:TaylorSheet},  in a viscoelastic fluid. Lauga\cite{Lauga2007} considered various non-Newtonian constitutive models, including the Oldroyd-B, FENE-P, Johnson-Segalman-Oldroyd, Giesekus models, and obtained a seemingly universal conclusion regarding the effect of viscoelasticity on small-amplitude,  inertialess swimming. The origin and limitations of different constitutive models are beyond the scope of this chapter and the reader is  referred  to classical books  on the subject \cite{Bird1987, Bird1987b}. Here we focus on the Oldroyd-B model,  arguably the most popular constitutive equation, both because of its simplicity and the fact that it can be derived exactly from  kinetic theory by modelling the polymeric fluid as a dilute solution of elastic  dumbbells \cite{Bird1987, Bird1987b}.

\subsubsection{Governing equations}
Since we now consider  a non-Newtonian problem,  we no longer have the Stokes equations but the general Cauchy's equation of motion without inertial terms, 
\begin{align}
\nabla p = \nabla \cdot \btau, \label{eqn:LaugaMomentum}
\end{align}
where $\btau$ is the deviatoric viscoelastic stress tensor. The continuity equation, 
\begin{align}
\nabla \cdot \v = 0,
\end{align}
remains in effect for incompressible flows. 

\subsubsection{Constitutive equation}
We require constitutive equations, which relate stresses and kinematics of the flow, in order to close the system of equations. For polymeric fluids, the deviatoric stress may be decomposed into two components, $\btau = \btau^s + \btau^p$, where $\btau^s$ is the Newtonian contribution from the solvent, and $\btau^p$ is the polymeric contribution to the stress. For the Newtonian contribution, the constitutive equation is given simply by $\btau^s = \mu_s \gamd$, where $\mu_s$ is the Newtonian contribution to the viscosity and $\gamd = \nabla \v + \nabla \v^T$. The polymeric  contribution is described by  the classical Oldroyd-B model, where the polymeric stress, $\btau^p$, satisfied the upper-convected Maxwell equation
\begin{align}
\btau^p + \lambda \stackrel{\triangledown}{\btau^p} = \mu_p \gamd, \label{eq:upperMaxwell}
\end{align}
where $\mu_p$ is the polymer contribution to the viscosity and $\lambda$ is the polymeric relaxation time. In Eq.~(\ref{eq:upperMaxwell}), the upper-convected derivative for a tensor $\mathbf{A}$ is defined as 
\begin{align}
\stackrel{\triangledown}{\mathbf{A}} = \frac{\partial \mathbf{A}}{\partial t} +\v \cdot   \nabla \mathbf{A}  - \nabla \v^T \cdot \mathbf{A} - \mathbf{A} \cdot \nabla \v ,
\end{align}
and represents the rate of change of $\mathbf{A}$ in the frame translating, rotating, and deforming with the fluid. From Eq.~(\ref{eq:upperMaxwell}), we can obtain the Oldroyd-B constitutive equation for the total stress, $\btau$, as given by
\begin{align}
\btau + \lambda_1 \dbtau = \eta \left( \gamd + \lambda_2 \dgamd \right),
\end{align}
where $\mu=\mu_s+\mu_p$, $\lambda_1=\lambda$, and $\lambda_2 = \mu_s \lambda / \mu$. Here, $\lambda_1$ and $\lambda_2$ are, respectively, the relaxation and retardation times of the fluid. The relaxation time is the typical decay rate of stress when the fluid is at rest, and the retardation time measures the decay rate of residual rate of strain when the fluid is stress-free \cite{Bird1987, Bird1987b}. It can be noted that $\lambda_2 < \lambda_1$, and both are exactly zero in the  Newtonian case.

\subsubsection{Non-dimensionalization}
We adopt the same non-dimensionalizations as in Taylor's original calculations (Sec.~\ref{sec:TaylorSheet}). We scale times as $1/\omega$, lengths as $1/k$, and hence speed as by the phase speed of the wave $c = \omega/k$. Shear rates and stresses are hence non-dimensionalized as $\omega$  and $\mu\omega$ respectively. The dimensionless equations are then given by
\begin{subequations}\label{eq:govern}
\begin{align}
\nabla \cdot \v &=0,\\
\nabla p&=\nabla \cdot \btau, \label{eq:mecheq}\\  
\btau + \De_1 \dbtau &= \gamd + \De_2 \dgamd, \label{eqn:LaugaOB}
\end{align} 
\end{subequations}
where $\De_1=\lambda_1 \omega$ and $\De_2 = \lambda_2 \omega$ are defined as the two Deborah numbers for the flow and we have adopted the same symbols for convenience.

\subsubsection{Boundary Conditions}
The geometry and boundary conditions of this problem remain unchanged compared with the Newtonian calculations and are therefore given by Eqs.~(\ref{eqn:TaylorBCFar}) and (\ref{eqn:TaylorBC}).

\subsubsection{Asymptotic Expansions}

In the spirit of Taylor's original calculations, we consider the small-amplitude limit $\epsilon \ll1$ and look for regular perturbation expansions for the stream function, the swimming speed, and the deviatoric stress as
\begin{subequations}
\begin{align}
\psi &= \epsilon \psi_1+\epsilon^2 \psi_2 + ...,\\
U &= \epsilon U_1+\epsilon^2 U_2 + ...,\\
\btau &= \epsilon \btau_1 + \epsilon^2 \btau_2+... \cdot 
\end{align}
\end{subequations}
Substituting the expansion for stress into the constitutive equation, Eq.~(\ref{eqn:LaugaOB}), we obtain
\begin{align}
\left( \epsilon \btau_1 + \epsilon^2 \btau_2 +... \right) + \De_1 \left( \epsilon \frac{\partial \btau_1}{\partial t} +  \epsilon^2 \frac{\partial \btau_2}{\partial t} + \epsilon^2 \v_1 \cdot \nabla \btau_1- \epsilon^2 \nabla \v_1^T \cdot \btau_1 - \epsilon^2 \btau_1 \cdot \nabla \v_1+...  \right)  \notag \\
=\left( \epsilon \gamd_1 + \epsilon^2 \gamd_2 +... \right) + \De_2 \left( \epsilon \frac{\partial \gamd_1}{\partial t} +  \epsilon^2 \frac{\partial \gamd_2}{\partial t} + \epsilon^2 \v_1 \cdot \nabla \gamd_1- \epsilon^2 \nabla \v_1^T \cdot \gamd_1 - \epsilon^2 \gamd_1 \cdot \nabla \v_1+...  \right).
\end{align}
The boundary conditions are expanded similarly to  the Newtonian case, see Eqs.~(\ref{eqn:TaylorExpandBegin}) and (\ref{eqn:TaylorExpandEnd}).

\subsubsection{First-order solution}
Collecting the terms of the same order, we have the $O(\epsilon)$ constitutive equation given by
\begin{align}
\btau_1 + \De_1 \frac{\partial \btau_1}{\partial t} = \gamd_1 + \De_2 \frac{\partial \gamd_1}{\partial t} \cdot \label{eqn:LaugaOBFirst}
\end{align}
The equation of mechanical equilibrium,  Eq.~(\ref{eqn:LaugaMomentum}), gives at first order $\nabla p_1 = \nabla \cdot \btau_1$. We proceed to take the divergence of the constitutive equation, Eq.~(\ref{eqn:LaugaOBFirst}), and  relate the divergence of stress to the gradient of pressure. We then eliminate the pressure by taking the curl of that equation, resulting in
\begin{align}
\left( 1+\De_2 \frac{\partial}{\partial t} \right) \nabla^4 \psi_1 =0, \label{eqn:LaugaGovernFirst}
\end{align}
where we have used the kinematic relation $\nabla \times \nabla \cdot \gamd_1 = - \nabla^4 \psi_1 \e_z$. Eq.~(\ref{eqn:LaugaGovernFirst}) is subject to the same $O(\epsilon)$ boundary conditions as in the Newtonian case, Eq.~(\ref{eqn:TaylorBCFirst}). It is straightforward to see that the solution satisfying the biharmonic equation, $\nabla^4 \psi_1 = 0$, with the same boundary conditions in the Newtonian case will also satisfy Eq.~(\ref{eqn:LaugaGovernFirst}). Therefore, after a possible transient,  the first-order harmonic solution is given by 
\begin{align}
\psi_1  = (1+y) e^{-y} \sin(x-t),
\end{align}
 the same as the Newtonian first-order solution, Eq.~(\ref{eqn:Taylor1stSol}). Note that although the flow field and swimming kinematics remain unchanged by the presence of viscoelastic stresses at leading order, the rate of work of the sheet is however modified. Readers interested in this calculation are referred to the original paper \cite{Lauga2007}.

\subsubsection{Second-order solution}
The second-order constitutive equation is given by 
\begin{align}
\left( 1+ \De_1 \frac{\partial}{\partial t}\right) \btau_2 - \left( 1+ \De_1 \frac{\partial}{\partial t}\right) \gamd_2 = & \ \De_1 \left( \nabla \v_1^T \cdot \btau_1 + \btau_1 \cdot \nabla \v_1 - \v_1 \cdot \nabla \btau_1 \right) \notag \\
&- \De_2 \left( \nabla \v_1^T \cdot \gamd_1 + \gamd_1 \cdot \nabla \v_1 - \v_1 \cdot \nabla \gamd_1 \right), \label{eqn:LaugaOBSecond}
\end{align}
where the right-hand side can be computed explicitly using the first-order solution. For convenience, we write the first-order solution in Fourier notations as
\begin{align}
\psi_1 = \mathcal{R}( \tilde{\psi}_1 e^{it}) =  \frac{\tilde{\psi}_1 e^{it} + \tilde{\psi}_1^* e^{-it}}{2}, \ \ \ \ \ \tilde{\psi}_1 = i(1+y) e^{-y} e^{-ix}, \label{eqn:LaugaFirstOrderSolution}
\end{align}
where $\mathcal{R}$ denotes taking the real part of the quantity and the star denotes the complex conjugate in this section. Using similar notations for the stress and rate-of-strain tensors, we write variables on the right-hand side of Eq.~(\ref{eqn:LaugaOBSecond}) as
\begin{align}
\v_1 = \frac{\tilde{\v}_1 e^{it} + \tilde{\v}_1^* e^{-it}}{2},\quad  \btau_1 = \frac{\tilde{\btau}_1 e^{it} + \tilde{\btau}_1^* e^{-it}}{2}, \quad \gamd_1 = \frac{\tilde{\gamd}_1 e^{it} + \tilde{\gamd}_1^* e^{-it}}{2} \cdot   \label{eqn:LaugaReal}     
\end{align}
We further relate $\tilde{\btau}_1$ to $\tilde{\gamd}_1$ by rewriting Eq.~(\ref{eqn:LaugaOBFirst}) in Fourier notations as
\begin{align}
\tilde{\btau}_1 = \frac{1+i\De_2}{1+i\De_1} \tilde{\gamd}_1 . \label{eqn:LaugaOBFirstReal}
\end{align}
By substituting the above expressions, Eqs.~(\ref{eqn:LaugaReal}) and (\ref{eqn:LaugaOBFirstReal}), into the second-order constitutive equation, Eq.~(\ref{eqn:LaugaOBSecond}), and averaging in time, we end up with \begin{align}
\langle \btau_2 \rangle-\langle \gamd_2 \rangle = \mathcal{R} \left[ \frac{\De_1-\De_2}{2(1+i\De_1)} \left( \nabla \tilde{\v}_1^{T*} \cdot \tilde{\gamd}_1 + \tilde{\gamd}_1 \cdot \nabla \tilde{\v}_1^{*}-\v_1^* \cdot \nabla \tilde{\gamd}_1  \right) \right], \label{eqn:LaugaGovern2}
\end{align}
where $\langle ... \rangle$ denotes  time-averaging over a period of oscillation of the wave. The right-hand side of Eq.~(\ref{eqn:LaugaGovern2}) can be calculated using the first-order solution, Eq.~(\ref{eqn:LaugaFirstOrderSolution}), which leads
\begin{align}
\tilde{\v}_1 = 
\begin{pmatrix}
-iy\\ -(1+y)
\end{pmatrix} e^{-y}e^{-ix}, \ \ 
\nabla \tilde{\v}_1=
\begin{pmatrix}
-y & i(1+y) \\
i(y-1) & y
\end{pmatrix} e^{-y} e^{-ix}, \ \
\nabla \tilde{\gamd}_1=
\begin{pmatrix}
-2y & 2iy \\
2iy & 2y
\end{pmatrix} e^{-y} e^{-ix} .
\end{align}
With the first-order solution, Eq.~(\ref{eqn:LaugaGovern2}) then becomes in a matrix form
\begin{align}
\langle \btau_2 \rangle-\langle \gamd_2 \rangle = \frac{\De_1 - \De_2}{1+\De_1^2} e^{-2y} 
\begin{pmatrix}
6y^2-2y-1 & \De_1(1+2y-y^2) \\
\De_1(1+2y-2y^2) & 2y^2 +2y+1
\end{pmatrix}. \label{eqn:LaugaOBAveraged}
\end{align}
Upon taking the divergence of Eq.~(\ref{eqn:LaugaOBAveraged}) to relate the divergence of the stress tensor to the pressure gradient, and then taking the curl to eliminate the pressure, we obtain the equation for the time-averaged second-order stream function as 
\begin{align}
\nabla^4 \langle \psi_2  \rangle  = \frac{8\De_1(\De_1 - \De_2)}{1+\De_1^2} (1-3y+y^2) e^{-2y} .
\end{align}
This equation is subject to the time-averaged  Newtonian boundary conditions (Eq.~\ref{eqn:TaylorBCSecondb}),
\begin{subequations}
\begin{align}
\frac{\partial \langle \psi_2 \rangle}{\partial y} \bigg|_{x, y \rightarrow \infty} &= U_2, \label{eqn:TaylorBCSecond1av}\\
\frac{\partial \langle \psi_2 \rangle }{\partial x}\bigg|_{x, y \rightarrow \infty} &= 0, \label{eqn:TaylorBCSecond2av}\\
\frac{\partial \langle \psi_2 \rangle}{\partial y}\bigg|_{x, y = 0} &= \frac{1}{2}, \label{eqn:TaylorBCSecond3av}\\
\frac{\partial \langle \psi_2 \rangle}{\partial x}\bigg|_{x, y = 0} &= 0. \label{eqn:TaylorBCSecond4av}
\end{align}
\end{subequations}
Due to the presence of viscoelastic stresses, the equation for the second-order stream function, Eq.~(\ref{eqn:LaugaOBAveraged}), is an inhomogenous biharmonic equation, whose solution is a superposition of the time-averaged  homogenous solution, Eq.~(\ref{eqn:TaylorBiharmonicHomo}), 
\begin{align}
\langle \psi_{2} \rangle_h = Ax + B y,
\end{align}
and a particular solution of the form
\begin{align}
\langle \psi_{2} \rangle_p = \left(a+b y+ cy^2 \right) e^{-2y}.
\end{align}
The unknown coefficients are determined by substituting the particular solution into Eq.~(\ref{eqn:LaugaOBAveraged}) and we obtain
\begin{align}
a= 0, \ \ \ b=c= \frac{\De_1(\De_1-\De_2)}{2(1+\De_1^2)} \cdot
\end{align}
The admissible solutions are therefore given by
\begin{align}
\langle \psi_2 \rangle = \langle \psi_{2} \rangle_h + \langle \psi_{2} \rangle_p = Ax + By +\frac{\De_1 (\De_1-\De_2)}{2(1+\De_1^2)}y(1+y) e^{-2y}, \label{eqn:LaugaSolutionSecond}
\end{align}
and Eqs.~(\ref{eqn:TaylorBCSecond2av}) or (\ref{eqn:TaylorBCSecond4av}) give $A=0$. Finally,  Eq.~(\ref{eqn:TaylorBCSecond3av}) leads to
\begin{align} 
B + \frac{\De_1(\De_1-\De_2)}{2(1+\De_1^2)} = \frac{1}{2} \ \ \  \Rightarrow \ \ \ B = \frac{1+\De_1 \De_2}{2(1+\De_1^2)},
\end{align}
which is then used with Eq.~(\ref{eqn:TaylorBCSecond1av}) to lead to the swimming speed
\begin{align}
U_2 = B = \frac{1+\De_1 \De_2}{2(1+\De_1^2)} \cdot
\end{align}
In the  Newtonian limit, we have $\De_1 = \De_2 = 0$, and the solution  reduces to $U_2 = U_N = 1/2$, which is the Newtonian swimming speed first   obtained by Taylor, Eq.~(\ref{eqn:TaylorSpeed}). We may then compare the viscoelastic to the Newtonian swimming speeds and find that
\begin{align}
\frac{U_2}{U_N} = \frac{1+\De_1 \De_2}{1+ \De_1^2} \cdot
\end{align}

Since $\De_2 \le \De_1$, the waving sheet always swims slower in a viscoelastic fluid compared with a Newtonian fluid, \textit{i.e.~}$U_2 \le U_N$. This relationship  continues to hold in the case of a cylindrical filament propagating a travelling wave \cite{Fu2009}. Further numerical simulations of a swimming sheet in a complex fluid  recovered the asymptotic results presented in this section in the limit of small  wave amplitude  \cite{Teran2010}, while suggesting  that fluid viscoelasticity can increase swimming speeds in the case of finite-amplitude swimmers. 

Recent experiments on the locomotion of \textit{Caenorhabditis elegans} in synthetic polymeric solutions showed quantitative agreement with the asymptotic analysis  \cite{Shen11}. In contrast,  experiments on rotating helices in viscoelastic fluids reported more complex results where  decrease of the swimming speed is seen  for small-amplitude motion while a modest increase of the swimming speed is obtained  for larger amplitudes  \cite{Liu2011}. A recent numerical study on the locomotion of helices in viscoelastic fluids connects results from small-amplitude theories to large-amplitude experimental measurements. Further work will be needed to fully unravel to the  role of fluid elasticity in small-scale locomotion.

\section{Synthetic micro-propellers} \label{sec:Synthetic}

In the previous sections, we  reviewed classical theoretical models addressing the  swimming of  microorganisms. Beyond the biological realm, similar concepts and ideas may be applied to analyzing and designing synthetic micro-swimmers. These are of current interest for potential   biomedical applications such as micro-surgery and targeted drug delivery \cite{Nelson2010}. Thanks in part to our improved   understanding of low-Reynolds-number hydrodynamics and to advances in micro- and nano-fabrication, a variety of synthetic propelling devices have been proposed. Some are biomimetic and use biology as an inspiration while others take advantage of  different mechanisms offered by physics in order to achieve micro-propulsion.  In this section, we very briefly introduce several of these mechanisms together with their basic physical principles. Interested readers are referred to comprehensive reviews of recent progress on the design of synthetic micro-swimmers \cite{Walther2008, Wang2009, Lauga2009, Nelson2010, Ebbens2010, Lauga2011}.

\begin{figure}[t]
\begin{center}
\includegraphics[width=1\textwidth]{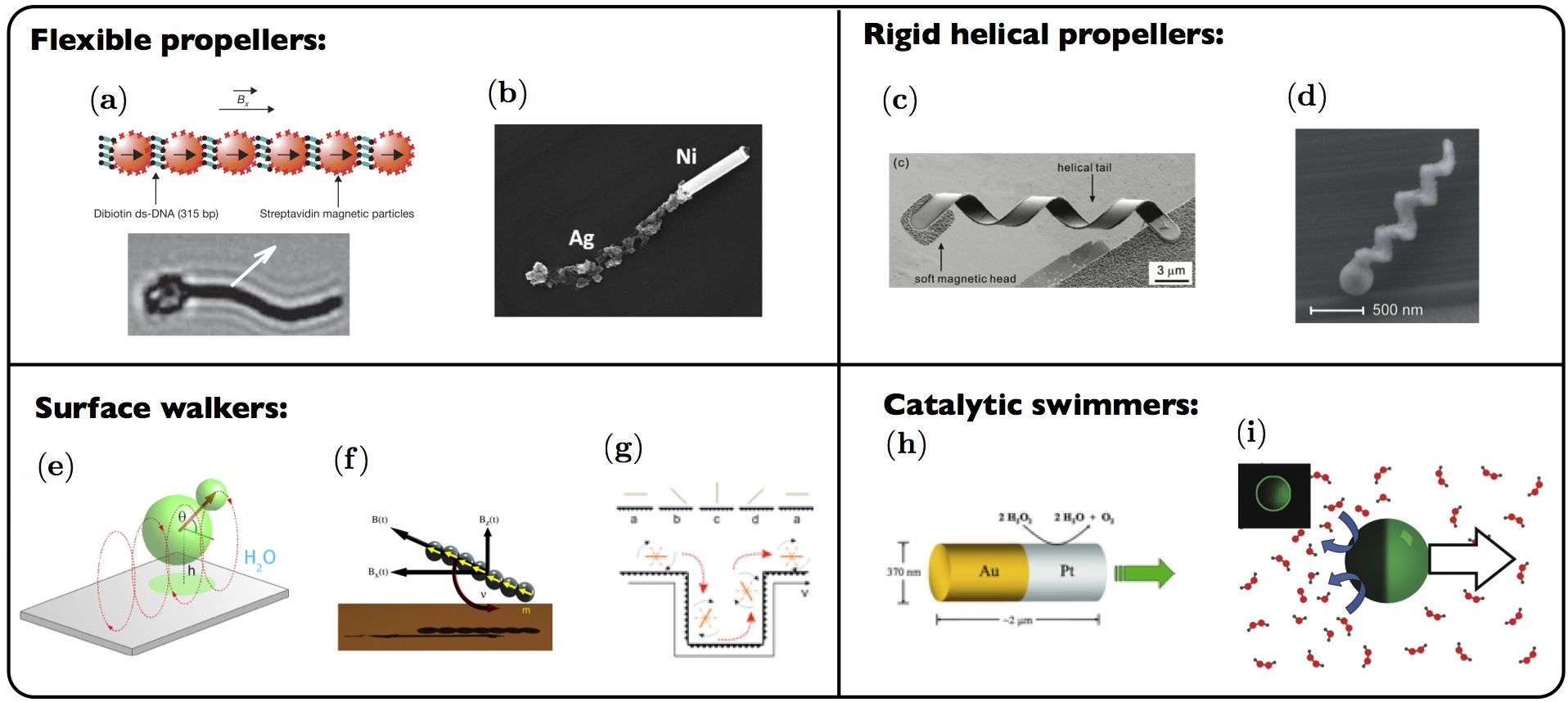}
\caption{Different designs of synthetic micro-propellers. Flexible propellers: ($\mathbf{a}$) \cite{Dreyfus2005} and ($\mathbf{b}$) \cite{Pak2011}; rigid helical propellers: ($\mathbf{c}$)\cite{Zhang2009b} and ($\mathbf{d}$) \cite{Ghosh2009}; surface walkers: ($\mathbf{e}$)\cite{Tierno2008}, ($\mathbf{f}$) \cite{Zhang2010} and ($\mathbf{g}$)\cite{Sing2010} ; catalytic swimmers: ($\mathbf{h}$)\cite{Paxton2004} and ($\mathbf{i}$)\cite{Ebbens2011}. All images were reprinted with permission: $(\mathbf{a})$ from Dreyfus \textit{et al.}\cite{Dreyfus2005}. Copyright \copyright 2005 Nature Publishing Group; $(\mathbf{b})$ from Pak \textit{et al.}\cite{Pak2011} with permission from The Royal Society of Chemistry;  $(\mathbf{c})$ from Zhang \textit{et al.} \cite{Zhang2009b} \copyright 2009 American Chemical Society; $(\mathbf{d})$  from Ghosh and Fischer \cite{Ghosh2009} \copyright 2009 American Chemical Society; $(\mathbf{e})$  from Tierno \textit{et al.} \cite{Tierno2008} \copyright 2008 American Physical Society; $(\mathbf{f})$ from Zhang \textit{et al.} \cite{Zhang2010} \copyright 2010 American Chemical Society;  $(\mathbf{g})$ from Sing \textit{et al.} \cite{Sing2010} \copyright 2008 National Academy of Sciences, USA; $(\mathbf{h})$ from Paxton \textit{et al.} \cite{Paxton2004} \copyright 2004 American Chemical Society; $(\mathbf{i})$ from Ebbens and Howse \cite{Ebbens2011} \copyright 2011 American Chemical Society.}
\label{fig:Synthetic}
\end{center}
\end{figure}

We categorize the design of different synthetic micro-propellers  as shown in Fig.~\ref{fig:Synthetic}.  As a preliminary remark, we note that many propellers driven by external fields are often referred to as swimmers in the literature. Although they are force-free, strictly speaking, they do not represent true self-propulsion like that of  swimming microorganisms because of the presence of non-zero external moments. This is why we use the generic term propellers  instead of swimmers. 

The first propeller category  is that of flexible propellers, which exploit the flexibility of a body -- typically a slender filament -- to develop non-reciprocal deformation for propulsion. The underlying physical principle is similar to that discussed in Sec.~\ref{sec:ElasticityBody}. Dreyfus \textit{et al.}\cite{Dreyfus2005} realized the idea experimentally by fabricating a 24 mm long flexible filament composed of paramagnetic beads linked by DNA, and the filament was attached to a red blood cell (Fig.~\ref{fig:Synthetic}$\a$). Different from the boundary actuation discussed in Sec.~\ref{sec:ElasticityBody}, actuation in this propeller was distributed along the filament by the paramagnetic beads, driven by an external, transverse, planar magnetic field. The presence of the red blood cell broke the frontÐback symmetry of the device, and allowed the propagation of a travelling wave along the filament and propulsion. Recently, metallic nanowires have been also used to fabricate flexible swimmers  \cite{Gao2010, Pak2011}. A typical nanowire motor consists of two segments, silver (Ag) and nickel (Ni) (Fig.~\ref{fig:Synthetic}$\b$). The ferromagnetic nickel segment is driven by a rotating magnetic field, and the flexibility of the silver segment allows chiral deformations to develop along the filament, leading to propulsion. The importance of chiral deformation for micro-propulsion has been illustrated in the second example discussed in Sec.~\ref{sec:KinRev} (Fig.~\ref{fig:SA}$\b$). In contrast to the propeller proposed by Dreyfus \text{et al.}\cite{Dreyfus2005}, the actuation in these nanowire motors is acting solely on the rigid magnetic nickel portion of the filament, while the flexible silver portion is passive.

The second group of swimmers is that of  rigid helical propellers, similar to the helical bacterial flagella discussed in Sec.~\ref{sec:RFThelical}. These rigid helices are rotated by external magnetic fields and achieve translation thanks  to the chirality of their shapes \cite{Zhang2009b, Ghosh2009}(Figs.~\ref{fig:Synthetic}$\c$ and $\mathbf{d}$). Comparing the flexible nanowire motors with  rigid helical propellers, the former develops  chirality dynamically due to the balance between viscous and elastic forces acting on the filament and the chiral deformation changes with the actuation, while the latter has the chirality already built in the rigid structure and does not change with the actuation. However, they typically require more complex fabrication techniques.

The third group of swimmers is composed of surface walkers, which rely on the presence of a rigid surface to break the spatial symmetries  enabling propulsion. These surface walkers typically utilize a rotating magnetic field and exploit the fact that viscous drag varies spatially at different phases of rotation due to the presence of a nearby boundary. Consider the doublet \cite{Tierno2008,Tierno2010} shown in Fig.~\ref{fig:Synthetic}$\mathbf{e}$ as an example: the viscous drag is larger when the smaller particle in the doublet is closer to the surface than when it is farther away from the surface. Averaging the viscous drag over one period of rotation, there is hence a net force in the lateral direction, leading to a lateral translation in order to satisfy the force-free condition. The same principle  applies to other objects externally driven to rotate near a surface, including a single sphere, a chain of superparamagnetic beads \cite{Sing2010} or nanowires \cite{Zhang2010}.

Another type of swimmers are called catalytic swimmers, which rely on chemical reactions between the swimmer and a fuel  in the surrounding fluid for propulsion, \textit{e.g.}~hydrogen peroxide. These swimmers usually consist of two different materials (janus) so that the chemical reaction occurs only with one half (typically platinum) of the swimmers \cite{Walther2008}.  The asymmetric distribution of reaction products hence drives the self-diffusiophoretic motion of the swimmer \cite{Golestanian2005, Golestanian2007,Howse2007}, which can also be understood as an osmotic propulsion mechanism \cite{Cordova2008}. The creation of bubbles in this setup can also be exploited for  propulsion \cite{Gibbs2009, Gao2012}.

Besides the types of swimmers discussed above,  other interesting designs have been proposed, including Purcell's three-link swimmer, two-\cite{Avron2005} or three-sphere\cite{Najafi2004} swimmers, and the surface-treadmilling swimmer\cite{Taylor1952}.

\section{Concluding remarks}

In this chapter, we have presented an extensive catalog of   theoretical models and analytical techniques employed in the studies of low-Reynolds-number locomotion. Useful exact solution methods such as Lamb's general solution (Sec.~\ref{sec:Lamb}) and  the reciprocal theorem (Sec.~\ref{sec:reciprocal}) have been introduced. In many problems, however, in order to make analytic progress one has to focus on certain asymptotic limits, such as small-amplitude analysis (Secs.~\ref{sec:TaylorSheet} and \ref{sec:ElasticityFluid}), slender body theory (Secs.~\ref{sec:RFT} and \ref{sec:ElasticityBody}), lubrication theory (Secs.~\ref{sec:lubrication} and \ref{sec:Syn}), and far-field approximations (Secs.~\ref{sec:FarField} and \ref{sec:WallSingularities}). On the other hand, different numerical methods, such as the boundary element method \cite{Pozrikidis1992} and the method of regularized Stokeslets \cite{Cortez2001}, have also been developed to address the subject, but a discussion of these approaches is beyond the scope of this chapter.

{The statement of Purcell's scallop theorem in Sec.~\ref{sec:scallop}  appears to be simple but it has far-reaching importance in locomotion at small scales. The theorem holds for reciprocal motion in Newtonian fluids at zero Reynolds number. An examination of these assumptions reveals different ways around the theorem to produce micro-propulsion \cite{Lauga2011}, including generating non-reciprocal motion (propagation of flagellar waves) and exploiting non-Newtonian rheological properties \cite{Normand2008, Keim2012, Pak2012}. Another way to escape from the constraints of the scallop theorem is via inertia. For small-scale locomotion, the Reynolds number is small but cannot be exactly zero, unless no motion occurs. A fundamentally interesting question is then, how much inertial force is necessary to break the constraints of the scallop theorem \cite{Childress2004}? Is the breakdown continuous or discontinuous? This topic has recently been studied extensively  
\cite{Vandenberghe2004,Alben2005, Lu2006,Vandenberghe2006,Lauga2007Con}.

Finally, we remark on another  physical process at small scales not taken into account in this chapter -- namely, the presence of noise and fluctuations. Not noticeable in the macroscopic life, on very small scales the effects of  Brownian motion can be dramatic, akin to  walking in a hurricane in our world \cite{Astumian2001}. The deterministic approach outlined in this chapter will then be valid only on short time scales. On longer time scales, a motile cell will typically always undergo effective diffusion. Interested readers are referred to a biophysical introduction to this  topic \cite{Berg1993}.

This chapter was designed to  serve as a pedagogical introduction to the theoretical modelling of low-Reynolds-number locomotion, and we hope that it will inspire many to contribute to this active and exciting field. There is still plenty of room at the bottom!

\footnotesize{
\bibliography{biblio.bib}
}

\end{document}